%% file: main.tex
\documentclass[a4paper,12pt,pdftex]{these}

\usepackage{these}
\usepackage{thesetitle}
\usepackage{thesefonts}
\usepackage{natbib}

\titre{Détections de pulsars milliseconde avec le FERMI Large Area Telescope}
\nom{Guillemot}
\prenom{Lucas} 
\ed{Sciences Physiques et de l'Ingénieur}					
\specialite{Astrophysique, Plasmas et Corpuscules}
\date{24 Septembre 2009}                          
\annee{2009}                                      
\labo{CENBG}
\logolabo{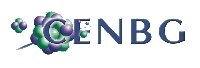}
\ordre{3841}

\dominitoc

\begin{document}

\maketitle

\include{frontmatter/dedEpi}  
 
\include{frontmatter/avPropos}

\tableofcontents             

\renewcommand{\thechapter}{\Roman{chapter}}

\include{introduction/introduction}

\partie{Astronomie des pulsars}

\include{part1/chap1}
\include{part1/chap2}

\partie{Instrumentation radio, X et $\gamma$}

\include{part2/chap3}
\include{part2/chap4}

\partie{Observation de pulsars milliseconde avec le LAT}

\include{part3/chap5}

\include{part3/chap6}

\include{conclusion/conclusion}

\begin{appendix}

\include{annexes/annexeA} 
\include{annexes/annexeB}
\include{annexes/annexeC}

\end{appendix} 

\partie*{Références}

\bibliographystyle{aa}

\bibliography{main}

\listoffigures 
               
\listoftables   

\include{frontmatter/acronymes}

\include{frontmatter/resume}

\end{document}

%% file: frontmatter/dedEpi.tex
\cleardoublepage 

\dedicace{\`{A} Doroté, à ma famille, à mes amis}

\cleardoublepage 



%% file: frontmatter/avPropos.tex
\chapter*{Remerciements}

Cette thèse a été une belle aventure scientifique mais aussi et surtout une aventure humaine, et de bonnes relations avec ses encadrants et collègues sont plus que primordiales. Tout en haut de la montagne de remerciements que je m'apprête à distribuer, mes premières pensées vont ainsi à David Smith et Denis Dumora, mes directeur et co-directeur de thèse. Au cours de ces trois années passionnantes, j'ai bénéficié de leur clairvoyance, leur rigueur et leur disponibilité. Je les remercie sincèrement pour m'avoir permis de mener ces travaux à bien avec opiniâtreté, et j'espère avoir absorbé un peu de leur compétence. Qu'ils soient convaincus de mon amitié et de ma reconnaissance.\\ 

Un grand merci également aux autres membres de l'équipe Astroparticules du CENBG : Marianne Lemoine-Goumard, Thierry Reposeur, Benoît Lott et les doctorants Marie-Hélène Grondin, Lise Escande et Damien Parent. Encore une fois, de bonnes relations dans le travail, agrémentées de discussions enrichissantes, de gaité, de cappuccino et de chocolat ont rendu le quotidien des plus agréables. J'exprime ma reconnaissance au directeur du CENBG, Bernard Haas, pour son accueil et son soutien. Merci aux doctorants du Solarium, anciens et actuels, Rémi Huguet, Jérémie Messud, Julien Le Bloas et Hao Tran Viet Nhan, pour toutes ces conversations autour d'un café. Plus généralement, j'ai bien entendu une pensée pour l'ensemble du personnel du CENBG, si efficace et sympathique.\\

Je tiens à remercier Isabelle Grenier et Denis Bernard, qui m'ont fait l'honneur d'être rapporteurs de cette thèse, qui a largement bénéficié de leurs remarques constructives et leurs discussions enrichissantes. Merci à Jean-Marc Huré qui m'a fait l'honneur de présider le jury de thèse, jury complété par David Smith, Denis Dumora et Isma\"el Cognard, que je remercie chaleureusement pour leur aide précieuse aux différentes étapes de la préparation du manuscrit et de la soutenance.\\

Au cours de cette thèse j'ai eu la chance de travailler au sein de la collaboration Fermi tout en étant en contact avec la communauté des pulsaristes radio et X. De nombreuses personnes ont directement contribué à ces travaux, que ce soit dans la préparation des observations de pulsars en rayons $\gamma$ avec le Fermi LAT, dans l'analyse des données observationnelles, l'interprétation ou la rédaction des articles scientifiques qui en ont découlé. Je remercie donc Dave Thompson, Alice Harding, Roger Romani, Eric Grove, Isabelle Grenier, Tyrel Johnson, Matthew Kerr, Christo Venter, Natalie Webb, J\"urgen Kn\"odlseder, et j'en oublie certainement pour ce qui est de la collaboration Fermi. Je remercie également Isma\"el Cognard pour son aide ô combien précieuse pour les questions de chronométrie des pulsars mais aussi Gilles Theureau, Grégory Desvignes, Michael Kramer, Simon Johnston, Dick Manchester, Patrick Weltevrede et Aristeidis Noutsos.\\

Je termine en exprimant ma sincère reconnaissance à l'ensemble de mes proches qui ont évidemment contribué au bon déroulement de cette thèse. En premier lieu, un immense merci à Doroté pour sa présence et son soutien indéfectible. Je remercie mes parents et mes frères pour leur confiance et pour avoir parcouru de nombreux kilomètres pour assister à ma soutenance de thèse, comme l'ont fait Lise et Jean-Claude Schiff (très bon le kouglopff !). Merci également à tous mes amis. Pour reprendre une pensée de Julio Iglesias : \emph{« l'amitié, c'est la fidélité, et si on me demandait qu'est-ce que la fidélité ? Je répondrais : c'est l'amitié ! »} \reflectbox{?}\footnote{Ce point d'ironie est un clin d'oeil à Thierry !} Que les personnes dont les noms suivent soient convaincus de mon amitié fidèle (l'ordre ne suit aucune logique) : Benjo, Nico, Momo, Bouc, Marguerite, Julie, Mathieu P., Erwan G., Erwan J., Eddy, Simon, Florian, Fanny, Yoann, Agnès, \'Emeline, Youenn, Dorian, Loïc, \'Elie, Brice et Hélène ! J'ajoute, un peu en vrac : Michel Gestin, Tony Vairelles, Billy Corgan et la galette saucisse.\\

Voilà, j'espère avoir correctement couvert l'ensemble des personnes qui méritent d'être remerciées. Merci tout de même à ceux que j'ai eu l'inconscience d'oublier !

%% file: introduction/introduction.tex
\partie*{Introduction} 


\paragraph{Motivations et plan de la thèse\newline}

Si l'astronomie dans le domaine visible est millénaire, l'astronomie $\gamma$ n'est que cinquantenaire et n'en était qu'à ses balbutiements il y a peu de temps encore. L'atmosphère terrestre nous protège en grande partie de ces photons très énergétiques, ainsi l'Homme ne peut observer le ciel dans ce domaine d'énergie que depuis le vingtième siècle, en plaçant ses instruments sur des ballons ou des satellites en orbite autour de la Terre, ou en ayant recours à des télescopes à effet Tcherenkov au sol pour ce qui est des photons de très haute énergie ($E > 30$ GeV). L'astronomie $\gamma$ renseigne sur des phénomènes violents, capables d'accélérer des particules chargées vers des énergies extrêmes : noyaux actifs de galaxie, vestiges de supernova, sursauts $\gamma$, pulsars, \emph{etc}. Le \emph{Large Area Telescope} (LAT), à bord du satellite Fermi, appartient à la troisième génération de télescopes $\gamma$ placés en orbite. Le satellite Fermi a été lancé le 11 juin 2008, et le LAT a débuté ses observations entre 20 MeV et plus de 300 GeV quelques jours après. Sa surface effective de collection de photons $\gamma$ est considérablement meilleure que celle de son prédécesseur, EGRET (actif de 1991 à 2000). Le champ de vue du LAT est plus large, et sa résolution angulaire plus fine. Mais le LAT ne se résume pas à une évolution d'EGRET : il observe le ciel $\gamma$ entre 10 et 100 GeV, domaine du spectre électromagnétique largement inexploré jusqu'alors. Ainsi, le LAT observe désormais plusieurs centaines de sources, connues ou inconnues auparavant, avec un niveau de sensibilité sans précédent. De plus, le LAT balaye le ciel presque uniformément toutes les trois heures, à la différence d'EGRET qui opérait principalement en pointé, ce qui fait du LAT un instrument de choix pour la découverte de phénomènes transitoires.\newline

L'astronomie des pulsars est plus jeune encore que l'astronomie $\gamma$, le premier pulsar ayant été découvert en 1967 par J. Bell \& A. Hewish à Cambridge. On connaît à présent 1900 de ces objets environ. Les pulsars sont des étoiles à neutrons fortement magnétisées et en rotation rapide, émettant un rayonnement sous forme de faisceaux qui balayent le ciel au cours de leur rotation. Pour une vaste majorité, les pulsars sont observés dans le domaine radio. Leur étude à ces longueurs d'onde touche à une large gamme de domaines de la physique, de la relativité générale à la mesure du champ magnétique dans la Galaxie ou la recherche d'ondes gravitationnelles. D'autres pulsars sont également détectés en optique, en rayons X ou en rayons $\gamma$. Cette catégorie de pulsars demeure tout de même largement minoritaire : seuls neuf pulsars avaient par exemple été détectés dans le domaine $\gamma$ au début de l'activité de Fermi, grâce à SAS--2, COS--B, COMPTEL, EGRET et AGILE. Ce faible nombre de détections a laissé de nombreuses interrogations, notamment en ce qui concerne le mécanisme de production de rayonnement $\gamma$ dans leur magnétosphère. En outre, ces objets ont en commun d'être relativement jeunes : aucun pulsar milliseconde n'a été détecté avec certitude par EGRET ou AGILE. Des pulsations faiblement significatives ont néanmoins été observées pour deux de ces objets, PSR J0218+4232 et J1824$-$2452A. La question de l'émission dans le domaine $\gamma$ par les pulsars milliseconde, étoiles à neutrons en rotation très rapide, représentant 10\% des pulsars connus et pouvant disposer de réservoirs d'énergie rotationnelle colossaux (typiquement, 10$^{50}$ à 10$^{52}$ erg, ce qui est comparable à certains sursauts $\gamma$), restait donc en suspens. Les observations du ciel $\gamma$ par le LAT offraient la possibilité d'éclaircir certaines zones d'ombre : les pulsars milliseconde peuvent-ils accélérer des particules chargées vers les très hautes énergies, afin que celles-ci  produisent un rayonnement $\gamma$ détectable à distance ? Si oui, quelles sont les particularités de ce rayonnement $\gamma$, par rapport à l'émission des pulsars milliseconde dans les autres longueurs d'onde d'une part, et par rapport à celui des pulsars ordinaires ?\newline

Cette thèse, consacrée à la recherche de pulsars milliseconde en rayons $\gamma$ avec le LAT, est à la jonction de ces deux astronomies. L'observation des pulsars à haute énergie renseigne sur les mécanismes opérant dans la magnétosphère de ces objets aux propriétés singulières, d'autant plus que pour les pulsars détectés en rayons $\gamma$, on s'aperçoit qu'une bien plus grande part de l'énergie rotationnelle est convertie en rayonnement $\gamma$ plutôt que dans le domaine radio. Plus généralement, l'étude de l'émission $\gamma$ des pulsars se place dans le contexte de l'évolution des étoiles massives. Par ailleurs, les premiers résultats du LAT indiquent qu'en majorité, les sources galactiques sont des pulsars, rendant ces objets incontournables pour toute étude de sources galactiques. Il a également été proposé que les pulsars proches soient les principaux responsables de l'émission des électrons et positrons détectés autour de la Terre. Ces différentes raisons motivent l'étude des pulsars à haute énergie. \newline

La problématique peut se résumer de façon relativement simple. Aucun pulsar milliseconde n'a été fermement détecté dans les données d'EGRET et d'AGILE, ce qui laisse à penser que si ces objets émettent en rayons $\gamma$, ils doivent être peu lumineux. Par conséquent, les photons éventuellement émis sont espacés dans le temps, de sorte que la construction de leur courbe de lumière (c'est-à-dire l'intensité du signal en fonction de la fraction de tour du pulsar sur lui-même) requiert une connaissance très précise de leur période de rotation et de leur orbite, pour les objets appartenant à des systèmes binaires. Ces différents paramètres sont connus \emph{via} la chronométrie, en particulier dans le domaine radio. La découverte éventuelle de pulsations en rayons $\gamma$ pour ces pulsars à rotation très rapide repose sur leur chronométrie, en plus des performances du télescope. La datation des pulsars a donc été un premier objectif de cette thèse. Une fois les paramètres de chronométrie acquis, la recherche de pulsations dans les données du LAT a été la suite logique du travail préparatoire.\newline

Dans la première partie de ce manuscrit nous rappelons un certain nombre de propriétés des pulsars, notamment au travers d'une description simplifiée consistant à les assimiler à des dipôles magnétiques tournants. Nous introduisons ensuite les pulsars milliseconde, objets étudiés dans cette thèse. Leur scénario de formation et certaines de leurs propriétés se distinguent de ceux des pulsars communs. Dans un second volet de cette partie introductive nous établissons un état des lieux des connaissances sur les pulsars dans le domaine $\gamma$, du point de vue théorique et observationnel. \newline

La seconde partie est consacrée aux aspects instrumentaux concernant la préparation et l'analyse des données du LAT. Dans un premier temps, nous introduisons le concept de chronométrie des pulsars, puis nous décrivons la campagne de chronométrie des pulsars d'intérêt pour le LAT, dont la coordination a été l'un des objectifs au cours de cette thèse. Dans un second temps, nous rappelons le principe de fonctionnement et les caractéristiques du \emph{Large Area Telescope} et de ses composants principaux. \newline

Dans la troisième et dernière partie, nous présentons la recherche de pulsations pour des pulsars milliseconde dans les données enregistrées par le LAT. Comme nous le verrons, huit de ces objets ont été détectés, les établissant pour la première fois en tant que puissants accélérateurs de particules chargées. Les courbes de lumière de ces huit pulsars sont présentées, ainsi que les résultats spectraux. Les différentes caractéristiques sont interprétées en terme de modèle d'émission dans la magnétosphère. Quelques pulsars milliseconde semblent être de bons candidats à la détection par le LAT au cours des mois à venir. 

\paragraph{Publications\newline}

Ce manuscrit porte essentiellement sur la découverte de pulsations pour huit pulsars milliseconde en rayons $\gamma$. Ce résultat est dû aux bonnes performances du LAT, mais aussi au succès de la campagne de suivi des pulsars dans les domaines radio et X, à laquelle j'ai activement contribué. Grâce à l'expérience concernant le problème des calculs de phases rotationnelles acquise au travers de la campagne de chronométrie, j'ai également contribué à l'analyse de pulsars jeunes observés par le LAT, Vela et PSR J2021+3651. Plus généralement, l'ensemble des analyses de pulsars émetteurs radio et X avec le LAT ont jusqu'à présent bénéficié de la chronométrie des pulsars. Mes différents travaux ont trouvé leur concrétisation dans les articles suivants :\newline

\begin{itemize}
\item J.~T. O'Brien, S. Johnston, M. Kramer, \dots, \textbf{L. Guillemot}, \emph{PSR J1410$-$6132: a young, energetic pulsar associated with the EGRET source 3EG J1410$-$6147}, MNRAS 388, 1 (2008)\\
\item D.~A. Smith, \textbf{L. Guillemot}, F. Camilo, \emph{et al.}, \emph{Pulsar timing for the Fermi gamma-ray space telescope}, A\&A 492, 923 (2008)\\
\item A.~A. Abdo, \emph{et al.} (\emph{Fermi}-LAT Collaboration), \emph{Pulsed gamma-rays from the millisecond pulsar J0030+0451 with the Fermi Large Area Telescope}, ApJ 699, 1171 (2009)\\
Corresponding authors: \textbf{L. Guillemot}, T.~J. Johnson, M. Kerr\\
\item A.~A. Abdo, \emph{et al.} (\emph{Fermi}-LAT Collaboration), \emph{A population of gamma-ray millisecond pulsars seen with the Fermi Large Area Telescope}, Science 325, 848 (2009)\\
Corresponding authors: \textbf{L. Guillemot}, T.~J. Johnson, M. Kerr, D.~A. Smith\\
\item A.~A. Abdo, \emph{et al.} (\emph{Fermi}-LAT Collaboration), \emph{Pulsed Gamma-rays from PSR J2021+3651 with the Fermi Large Area Telescope}, ApJ 700, 1059 (2009)\\
Corresponding authors:  D.~A. Smith, \textbf{L. Guillemot}, M. Kerr\\
\item A.~A. Abdo, \emph{et al.} (\emph{Fermi}-LAT Collaboration), \emph{Fermi Large Area Telescope Observations of the Vela Pulsar}, ApJ 696, 1084 (2009)
\end{itemize}

%% file: part1/chap1.tex
\chapter[Introduction aux pulsars]{Introduction aux pulsars} 

\minitoc

\chaptabst{L}{es}{pulsars sont des « étoiles à neutrons hautement magnétisées et en rotation rapide ». Dans ce premier chapitre nous rappelons les grandes étapes de la découverte des pulsars. On peut les assimiler en première approximation à des dipôles magnétiques tournants, entourés d'une magnétosphère en co-rotation qui est le siège de l'émission de rayonnement à différentes énergies. On observe que la rotation des pulsars sur eux-mêmes est pseudo-périodique, la rotation ralentissant progressivement. L'hypothèse du dipôle magnétique tournant permet d'estimer la quantité d'énergie cinétique rotationnelle perdue au cours du temps par les pulsars, ainsi que leur âge ou l'intensité de leur champ magnétique, à partir de la période de rotation $P$ et de sa diminution $\dot P$. On s'aperçoit de plus que différents types de pulsars existent, en fonction de $P$ et de $\dot P$. Dans la deuxième moitié de ce chapitre nous nous concentrons sur une des catégories, les pulsars milliseconde, et rappelons leur spécificité. Ceux-ci ont des périodes de rotation comprises entre 1 et 30 ms, avec des taux de ralentissement extrêmement faibles. Ils appartiennent généralement à des systèmes binaires, une conséquence de leur formation et leur évolution. Bien que représentant environ 10\% des pulsars connus, ils ont une grande importance en astronomie et en physique en général, de par leur nature extrême.}

\section{Historique}

Deux ans seulement après la découverte du neutron par James Chadwick, en 1932, Walter Baade et Fritz Zwicky proposaient que des étoiles à neutrons, c'est-à-dire des étoiles composées presque exclusivement de neutrons, pourraient naître au coeur des supernovae. Et quelques années plus tard, en 1939, Oppenheimer et Volkoff entreprirent même les premiers modèles de structure des étoiles à neutrons \citep{Oppenheimer1939}. Malgré ces travaux pionniers, la communauté astronomique n'a guère accordé d'intérêt à la recherche de ces astres ; on n'avait pas entrevu alors que les étoiles à neutrons pourraient être, par exemple, détectées \emph{via} la recherche de fluctuations rapides dans le signal radio. Si bien qu'il a fallu attendre plus de 25 ans avant que l'on ne découvre une étoile à neutrons... par hasard. 

En 1967, Jocelyn Bell et son directeur de thèse Anthony Hewish scrutaient le ciel avec le radiotélescope de Cambridge. Une instrumentation spéciale avait été conçue dans le but de détecter des scintillations brèves du signal radio, une signature caractéristique des quasars. Le 6 août 1967, des trains d'ondes radio périodiques furent remarqués à 19$^h$19$^m$ d'ascension droite et +21$^\circ$ de déclinaison. Le signal fut confirmé les jours suivants, au transit de cette région du ciel au-dessus du télescope. La structure des bouffées d'émission radio restait inaccessible cependant, la résolution en temps étant trop grossière. Une nouvelle instrumentation fut alors mise au point pour sonder le signal en deçà de la seconde. Le 28 novembre 1967, on s'aperçut que les trains d'ondes étaient composés de pulsations espacées de 1,337 s environ. Cet astre, source de signal radio pulsé, a été nommée \emph{pulsar}. Ce pulsar est maintenant connu sous le nom de B1919+21\footnote{La convention de nommage des pulsars est la suivante : les pulsars découverts de 1967 jusqu'au milieu des années 90 sont nommés par leur position dans le ciel dans le système de coordonnées équatoriales besselien (B1950) précédée d'un B : Bxxxx+yy. Ceux découverts par la suite sont nommés suivant leur position dans le système julien (J2000) précédée d'un J : Jxxxx+yyyy. Lorsque deux objets ont le même nom suivant cette convention, ce qui est le cas pour certains pulsars des amas globulaires, on les distingue en ajoutant une lettre : par exemple J1824$-$2452A.} \citep{hewish1968}.

Le lien entre les pulsars et les étoiles à neutrons avait été fait quelques temps avant la découverte de B1919+21. Pacini avait en effet postulé que la source d'énergie alimentant la nébuleuse du Crabe était une étoile à neutrons fortement magnétisée et en rotation rapide \citep{Pacini1967}. Indépendamment, Gold introduisait le concept de « pulsars alimentés par leur rotation » : à savoir que des étoiles à neutrons en rotation étaient analogues à des dipôles magnétiques tournant, donc perdant de l'énergie par rayonnement électromagnétique et par émission de particules relativistes \citep{Gold1968}. Ce faisant, la réserve d'énergie (rotationnelle en majorité) diminuait, se traduisant par une diminution progressive de la période de rotation. Mais il demeurait la possibilité que le signal radio, de période 1,337 s, provienne en réalité de l'oscillation résonnante de naines blanches \citep{Meltzer1966}, ce modèle expliquant naturellement des périodicités de l'ordre de la seconde. Les découvertes successives en 1968 d'un pulsar de période 89 ms au coeur du reste de supernova de Vela \citep{Large1968}, et d'un pulsar de période 33 ms au coeur de la nébuleuse du Crabe \citep{Staelin1968} établissaient l'identité des pulsars en tant qu'étoiles à neutrons de façon définitive.

\section{Pulsars : propriétés générales}

\subsection{\'Etoiles à neutrons}

Les étoiles à neutrons sont les résidus compacts des explosions de supernovae de type II, engendrées par l'effondrement gravitationnel du coeur des étoiles massives (typiquement $>$ 8 \msol) en fin de vie, lorsque la pression de radiation ne suffit plus à compenser leur contraction. Tandis qu'un objet très dense se forme au coeur de l'étoile progénitrice, les couches externes de l'atmosphère stellaire sont éjectées vers le milieu ambiant, et forment par la suite le vestige de supernova (\emph{supernova remnant}, SNR). 

Les modèles de structure des étoiles à neutrons placent typiquement leur masse entre 0,5 \msol\ et 2,5 \msol\ (pour une revue détaillée, voir \citet{LattimerPrakash2004}). Un certain nombre de masses de pulsars ont été mesurées (\emph{cf.} Figure 3 de \citet{LattimerPrakash2007}), les mesures les plus précises provenant de la chronométrie de pulsars binaires. Les masses relevées sont généralement compatibles avec 1,4 \msol, valeur de masse la plus couramment utilisée pour les étoiles à neutrons. En revanche, on ne dispose à ce jour d'aucune mesure précise de rayon d'étoile à neutrons. Les modèles d'équation d'état de la matière, qui décrivent la relation entre densité et pression au sein de l'étoile, placent généralement ce rayon entre 10 et 12 km, comme le montre la Figure \ref{eos} (A). La mesure simultanée du rayon et de la masse d'une étoile à neutrons placerait des contraintes fortes sur les différents modèles.

\sfig[Diagrammes masse-rayon des étoiles à neutrons]{chap1/masserayon.tex}{(A) : diagramme masse-rayon pour certains modèles d'équation d'état de la matière des étoiles à neutrons. Voir \citet{LattimerPrakash2001} et \citet{LattimerPrakash2007} pour la nomenclature des différents modèles représentés et des zones d'exclusion. Figure issue de \citet{LattimerPrakash2007}. (B) : courbes représentant $I = \frac{2}{5} M R^2$, pour différentes valeurs du moment d'inertie $I$ autour de $I_0 = 10^{45}$ g cm$^2$.}{eos}

L'équation d'état de la matière définit le profil de densité radial de l'étoile, et donc le moment d'inertie $I = k M R^2$. Pour une sphère de densité homogène, on a $k = \frac{2}{5}$, alors que les modèles prédisent des valeurs variant entre 0,3 et 0,45 lorsque le rapport $M/R$ est compris entre 0,1 et 0,2 \msol\ km$^{-1}$ \citep{LattimerPrakash2001}. En utilisant $k =$ 0,4, $M =$ 1,4 \msol\ et $R =$ 10 km, on obtient $I \simeq 10^{45}$ g cm$^2$, valeur que l'on utilisera dans la suite. Pour fixer les idées, la Figure \ref{eos} (B) montre quelques exemples de relations masse-rayon où l'on utilise $k =$ 0,4 et l'on fait varier $I$ entre 0,5 $I_0$ et 1,5 $I_0$, avec $I_0 = 10^{45}$ g cm$^2$. Bien que les différents modèles d'équations d'état de la matière vont au-delà de l'hypothèse de la sphère de densité uniforme (\emph{cf.} Figure 6 de \citet{LattimerPrakash2001}), on s'aperçoit que la valeur du moment d'inertie de $10^{45}$ g cm$^2$ est très incertaine, ce que nous garderons à l'esprit dans la suite de ce manuscrit. Elle pourrait varier fortement d'un pulsar à un autre, en particulier s'il y a accrétion de matière au cours de la vie de l'étoile à neutrons, comme c'est \emph{a priori} le cas pour les pulsars milliseconde, comme nous le verrons par la suite. 

\subsection{Description simplifiée : modèle du dipôle}

Un certain nombre de propriétés essentielles des pulsars peuvent être comprises en les assimilant à des dipôles magnétiques tournants (voir Figure \ref{Magnetosphere}). 

Le champ magnétique dipolaire, dont l'axe n'est pas nécessairement aligné avec l'axe de rotation, induit en tournant un champ électrique $\vec{E} \propto \left( \vec{\Omega} \times \vec{r} \right) \times \vec{B}$, où $\vec{\Omega}$ est la vitesse angulaire et $\vec{B}$ est le champ magnétique régnant à la distance $\vec{r}$ du dipôle. Dans ces conditions, Goldreich \& Julian ont montré que l'étoile à neutrons ne peut être entourée de vide : en effet la composante de champ électrique parallèle au champ magnétique $E_\parallel$ à la surface de l'étoile est intense, et arrache des particules chargées vers l'extérieur, peuplant ainsi la magnétosphère \citep{Goldreich1969}. \`{A} l'équilibre, la magnétosphère est remplie d'une distribution de charges $\rho = - \vec{\Omega} . \vec{B} / \left( 2 \pi c \right)$, et la composante $E_\parallel$ du champ électrique est écrantée. Les particules chargées et les champs électrique et magnétique de la magnétosphère entrent en co-rotation avec l'étoile. Un pulsar de période de rotation $P$ est entouré d'un « cylindre de lumière » imaginaire, de rayon $c P/(2 \pi)$, qui est en co-rotation avec le pulsar, et dont la surface se déplace à la vitesse de lumière. Comme le montre la Figure \ref{Magnetosphere}, les lignes de champ magnétique comprises dans ce cylindre de lumière peuvent se refermer normalement (ce sont les lignes de champ « fermées »), tandis que celles s'étendant au-delà du cylindre ne peuvent boucler (lignes de champ « ouvertes »). Une répartition de charges existe donc là où les lignes de champ peuvent se refermer. Au contraire, les lignes de champs ouvertes laissent s'échapper les particules chargées au-dessus des pôles magnétiques. Par cascades électromagnétiques, les particules chargées engendrent l'émission de photons dont les directions sont délimitées par les dernières lignes de champ ouvertes, celles qui interceptent le cylindre de lumière. Ainsi un faisceau d'émission centré autour de l'axe magnétique du pulsar se forme. Il balaye l'environnement du pulsar au cours de la rotation. Si le faisceau intercepte la trajectoire de la Terre, l'observateur terrestre peut détecter un signal périodique, se répétant à chaque rotation du pulsar \citep{Handbook}.

\fig[Schéma simplifié de la magnétosphère des pulsars]{width=8cm}{chap1/magnetosphere.pdf}{Description simplifiée de la magnétosphère des pulsars. Figure issue de \citet{Handbook}}{Magnetosphere}

Ce modèle donne une interprétation simple de l'origine de l'émission des pulsars dans le domaine radio, comme créée par les particules chargées se déplaçant le long des lignes de champs ouvertes. Comme nous le verrons par la suite, l'émission de haute énergie des pulsars n'est pas nécessairement produite au-dessus des pôles. Néanmoins l'idée d'une zone d'émission située dans la magnétosphère, visible de façon périodique par l'observateur terrestre si la configuration géométrique le permet, est toujours applicable.

\subsection{Période, ralentissement et quantités dérivées}

La description faite ci-dessus de la magnétosphère des pulsars, bien que très incomplète, indique qu'il existe au moins deux sources de perte d'énergie pour le pulsar et sa magnétosphère : l'émission de rayonnement dipolaire à basse fréquence, due à la rotation du moment magnétique, et l'échappement de particules chargées le long des lignes de champ ouvertes, aboutissant à un faisceau d'émission le long de l'axe magnétique. L'énergie cinétique rotationnelle $E = \frac{1}{2}I \Omega^2$, où $I$ est le moment d'inertie de l'étoile à neutrons et $\Omega = 2 \pi / P$ la vitesse angulaire, est le principal réservoir d'énergie du pulsar. La diminution de $E$ avec un taux $\dot{E} = - I \Omega \dot{\Omega}$ entraîne par conséquent le ralentissement de la rotation du pulsar, avec un taux $\dot{P} = - 2 \pi \dot{\Omega} / \Omega^2 > 0$. La quantité $\dot{E}$ est la perte d'énergie due au freinage (\emph{Spin-down power}). En fonction des paramètres $P$ et $\dot{P}$, et en prenant $I = 10^{45}$ g cm$^2$, on a :

\begin{eqnarray}
\dot{E} = 4 \pi^2 I \dfrac{\dot{P}}{P^3} \simeq 3,95 \times 10^{31} \mathrm{erg\ s}^{-1} \left(\dfrac{P}{\mathrm{1\ s}}\right) \left(\dfrac{\dot{P}}{10^{-15}}\right)^{-3}
\label{spindown}
\end{eqnarray}

Ces pulsars dont la période augmente au cours du temps en raison de leur perte d'énergie rotationnelle forment la catégorie des pulsars « alimentés par rotation » (\emph{Rotation-Powered}). On connaît à ce jour environ 1900 de ces astres, parmi lesquels B1919+21, Vela et le Crabe, cités auparavant. Le catalogue ATNF\footnote{Disponible en ligne : http://www.atnf.csiro.au/research/pulsar/psrcat/expert.html} répertorie ces pulsars, et certaines de leurs propriétés \citep{CatalogueATNF}. C'est cette catégorie de pulsars qui est étudiée dans cette thèse. Une autre catégorie d'objet est celle des pulsars « alimentés par accrétion » (\emph{Accretion-Powered}). Le rayonnement de ces derniers est dû à la chute de matière provenant d'une étoile compagnon. La forte température acquise par la matière dans sa chute crée des points chauds à la surface de l'étoile à neutrons. Ces zones sont émettrices de rayons X et sont en rotation avec le pulsar, ce qui crée un signal pulsé pour un observateur fixe. On comprend cependant que le mécanisme de rayonnement de ces objets est très différent de celui des pulsars alimentés par leur rotation.

Si l'on suppose que la perte d'énergie cinétique rotationnelle du pulsar est due exclusivement au rayonnement dipolaire magnétique, on a \citep{Jackson1962} :

\begin{eqnarray}
\dot{E} = - I \Omega \dot{\Omega} = \frac{2}{3 c^3} \vec{\mu}^2 \Omega^4 \sin^2(\alpha)
\label{nrjdipole}
\end{eqnarray}

Dans cette équation, $\vec{\mu}$ désigne le moment dipolaire magnétique du pulsar, et $\alpha$ est l'angle que forment l'axe magnétique et l'axe de rotation. D'après \ref{nrjdipole}, il vient :

\begin{eqnarray}
\dot{\Omega} = - \left( \frac{2 \vec{\mu}^2 \sin^2(\alpha)}{3 I c^3} \right) \Omega^3
\label{pertedipolaire}
\end{eqnarray}

Cette dernière équation est le cas particulier de $\dot{\Omega} \propto \Omega^n$ pour une perte d'énergie purement dipolaire, soit $n = 3$ d'après \ref{pertedipolaire}. La quantité $n$ est appelée « indice de freinage ». Cet indice peut se mesurer : en effet $\dot{\Omega} \propto \Omega^n$ implique $n = \Omega \ddot{\Omega} / \dot{\Omega}^2$, et comme nous le verrons dans la suite, $\Omega$, $\dot \Omega$ et $\ddot \Omega$ sont accessibles \emph{via} la chronométrie des pulsars. En pratique les mesures de $n$ sont rares, car $\ddot \Omega$ est souvent contaminé par des instabilités de la rotation (le \emph{timing noise}, \emph{cf.} \ref{labeltimingnoise}), dont il est nécessaire de s'affranchir pour accéder au comportement intrinsèque du pulsar. Les mesures disponibles sont en contradiction avec l'hypothèse $n = 3$, par exemple le pulsar du Crabe : $n = 2,51 \pm 0,01$ \citep{Lyne1993}, ou Vela : $n = 1,4 \pm 0,2$ \citep{Lyne1996}. Il est généralement admis que l'écart entre les indices de freinage mesurés et la valeur théorique de 3 est une preuve qu'une part importante de l'énergie cinétique est dissipée \emph{via} un vent de particules chargées et de champ magnétique émis par les pulsars.

La relation de proportionnalité $\dot{\Omega} \propto \Omega^n$ peut s'intégrer simplement, à condition que $n \ne 1$, pour estimer l'âge du pulsar :

\begin{eqnarray}
T = \dfrac{P}{(n-1) \dot{P}} \left[ 1 - \left( \dfrac{P_0}{P} \right)^{n-1} \right]
\end{eqnarray}

Dans cette équation, $T$ désigne l'âge du pulsar, $P$ et $\dot{P}$ sont les valeurs actuelles de la période et du ralentissement, et $P_0$ est la période du pulsar à sa naissance. En faisant l'hypothèse que la période initiale $P_0$ est négligeable devant la période actuelle, et que le ralentissement est d'origine dipolaire uniquement (soit $n = 3$), on a : 

\begin{eqnarray}
T = \tau = \frac{P}{2 \dot{P}}
\label{age}
\end{eqnarray}

La quantité $\tau$ est nommée « âge caractéristique », et représente une estimation raisonnable de l'âge véritable du pulsar. Le pulsar du Crabe a par exemple un âge caractéristique de 1240 ans, alors que l'association avec la supernova historique indique un âge de 950 ans, soit un écart de 25\% environ. Pour la quasi-totalité des pulsars, l'âge historique est inconnu, ainsi l'âge caractéristique $\tau$ est le seul estimateur dont on dispose.

Enfin, les valeurs de $P$ et $\dot{P}$ et le modèle dipolaire de champ magnétique des pulsars permettent d'estimer le champ magnétique régnant en différents endroits de la magnétosphère, et en particulier à la surface du pulsar et au cylindre de lumière. On peut montrer que l'intensité du champ magnétique induit par le moment dipolaire à la distance $r$ est donné par :

\begin{eqnarray}
B(r) & = & \dfrac{1}{r^3} \sqrt{ - \dfrac{3 I c^3}{2 \sin^2 \alpha} \dfrac{\dot{\Omega}}{\Omega^3} } \nonumber \\ 
& = & \dfrac{1}{r^3} \sqrt{ \dfrac{3 I c^3}{8 \pi^2 \sin^2 \alpha} P \dot{P} }
\end{eqnarray}

Soit $B_S$ le champ magnétique en surface, dit « champ magnétique caractéristique », et $B_{LC}$ le champ magnétique au cylindre de lumière, de rayon $R_L = c P / (2 \pi)$. Pour un pulsar canonique, de rayon $R$ = 10 km, de moment d'inertie $I = 10^{45}$ g cm$^2$, et en choisissant un angle d'inclinaison magnétique $\alpha$ égal à 90$^\circ$ (rotateur orthogonal), on a :

\begin{eqnarray}
B_S = B(R) & \simeq & 3,2 \times 10^{19} \mathrm{G} \ \sqrt{P \dot{P}} \\
B_{LC} & = & B_S \left( \dfrac{2 \pi R}{c P} \right)^3
\label{champs}
\end{eqnarray}

Il faut souligner que ces deux dernières équations utilisent une description dipolaire du champ, alors que les mesures d'indices de freinage disponibles indiquent un mécanisme de perte d'énergie rotationnelle plus complexe. Les incertitudes existant sur $R$ et $I$ sont importantes, et enfin une valeur de 90$^\circ$ pour l'inclinaison magnétique a été utilisée, alors que certaines valeurs de $\alpha$ mesurées sont très différentes : citons par exemple le cas de PSR J0218+4232, pour lequel cet angle est proche de 0$^\circ$ \citep{Stairs1999}. Il en va de même pour $\dot{E}$ et l'âge caractéristique $\tau$ : les paramètres sont mal connus, ainsi les incertitudes peuvent être du même ordre de grandeur que les valeurs elles-mêmes. Mais à défaut d'informations plus précises, les équations \ref{spindown}, \ref{age} et \ref{champs} fournissent des renseignements utiles sur les pulsars et leur magnétosphère. 

Remarquons que des calculs alternatifs des propriétés énergétiques et magnétiques des pulsars existent. Par exemple, Spitkovsky a appliqué les équations de la MHD à la magnétosphère des étoiles à neutrons dans l'hypothèse d'un plasma libre et obtenu une solution numérique à ces équations \citep{Spitkovsky2006}. La perte d'énergie par rayonnement électromagnétique ainsi calculée est : 

\begin{eqnarray}
\dot{E}_\mathrm{Spitkovsky} = k_1 \frac{\vec{\mu}^2 \Omega^4}{c^3} \left( 1 + k_2 \sin^2\alpha \right)
\end{eqnarray}

Les coefficients $k_1$ et $k_2$ valent respectivement (1 $\pm$ 0,05) et (1 $\pm$ 0,1). Le rapport entre $\dot{E}_\mathrm{Spitkovsky}$ et la perte d'énergie cinétique rotationnelle par rayonnement dipolaire (\emph{cf.} \ref{nrjdipole}) est donc :

\begin{eqnarray}
\frac{\dot{E}_\mathrm{Spitkovsky}}{\dot{E}_\mathrm{Dip\hat{o}le}} = \frac{3}{2} k_1 \frac{1 + k_2 \sin^2\alpha}{\sin^2\alpha}
\end{eqnarray}

\`{A} l'inverse, le champ magnétique surfacique calculé numériquement est proportionnel à\linebreak $\left(  1 + k_2 \sin^2\alpha \right)^{-1/2}$ tandis que le champ magnétique de surface dans l'hypothèse du dipôle tournant est proportionnel à $\left(\sin^2\alpha \right)^{-1/2}$. Les formules de Spitkovsky fournissent donc une description plus réaliste des propriétés des pulsars, en particulier lorsque $\alpha$ est faible. En effet : pour un angle $\alpha$ proche de 0, $\dot{E}_\mathrm{Dip\hat{o}le}$ tend vers 0 et le champ magnétique surfacique correspondant diverge tandis que les quantités calculées numériquement prennent des valeurs finies. Néanmoins, l'angle d'inclinaison magnétique $\alpha$ est inconnu pour la plupart des pulsars, aussi nous utiliserons les équations \ref{spindown}, \ref{age} et \ref{champs} avec $\alpha = 90^\circ$ dans la suite de ce manuscrit. Notons tout de même qu'à mesure que des valeurs de l'angle $\alpha$ seront obtenues, l'emploi de tels résultats deviendra plus approprié.

\subsection{Ralentissements apparents et intrinsèques}

Les pulsars se déplacent par rapport au système solaire. Ils sont donc animés d'un mouvement apparent, le mouvement propre, détectable si leur déplacement est suffisamment rapide et leur distance suffisamment faible. Dans le plan du ciel, un pulsar d'ascension droite $\alpha$ et de déclinaison $\delta$ possède un mouvement propre transverse égal à $\mu_T = \sqrt{\mu_\alpha^2 + \mu_\delta^2}$, où $\mu_\alpha = \dot{\alpha}\ \cos(\delta)$ et $\mu_\delta = \dot \delta$.

Une conséquence de ce mouvement propre transverse est l'effet \emph{Shklovskii} \citep{Shklovskii1970} : la distance projetée du pulsar par rapport au système solaire croît, ce qui a pour conséquence le fait que le taux de ralentissement apparent $\dot P$ est supérieur au taux intrinsèque, d'un facteur :

\begin{eqnarray}
\Delta \dot P_\mathrm{Shk} = \frac{v_T^2}{c d} \times P
\label{eqShklovskii1}
\end{eqnarray}

où $v_T$ est la vitesse transverse, donnée par $v_T = \mu_T \times d$ et $d$ est la distance de l'objet. L'équation \ref{eqShklovskii1} peut donc se réécrire en :

\begin{eqnarray}
\Delta \dot P_\mathrm{Shk} = \frac{\mu_T^2 d}{c} \times P \simeq 2,43 \times 10^{-21} \left( \frac{\mu_T}{\mathrm{mas\ yr^{-1}}} \right)^2 \left( \frac{d}{\mathrm{kpc}} \right) \left( \frac{P}{\mathrm{s}} \right)
\label{eqShklovskii2}
\end{eqnarray}

Pour obtenir le taux de ralentissement intrinsèque $\dot P_i$ du pulsar, il faut donc tenir compte de l'effet Shklovskii en soustrayant $\Delta \dot P_\mathrm{Shk}$ au taux apparent $\dot P$. Comme le montre l'équation \ref{eqShklovskii2}, l'effet n'est visible que pour les pulsars proches et à faible valeur de $\dot P$. Pour les pulsars concernés, il peut avoir des conséquences importantes sur les quantités dérivées données par \ref{spindown}, \ref{age} et \ref{champs}. Par exemple, PSR J0437$-$4715 ($P = 5,757$ ms, $\dot P = 5,73 \times 10^{-20}$, $d = 0,16$ kpc, $\mu_T = 141$ mas yr$^{-1}$) a un taux de ralentissement intrinsèque $\dot P_i = 1,39 \times 10^{-20}$ inférieur à $\dot P$ d'un facteur quatre, diminuant d'autant sa perte d'énergie due au freinage $\dot E$.

Plus généralement, toute accélération $a$ le long de la ligne de visée conduit l'observateur à observer un ralentissement $\dot P$ différent du taux intrinsèque d'un facteur $\Delta \dot P = a P / c$ ; c'est notamment le cas pour les pulsars appartenant à des amas globulaires, soumis à des forces gravitationnelles locales. Pour ces objets il faut donc prendre les effets d'accélération locaux dans le calcul de $\dot E$, $\tau$, $B_S$ et $B_{LC}$.

\section{Pulsars milliseconde}

\subsection{Différents types de pulsar}

On a vu dans les paragraphes précédents que les pulsars perdent de l'énergie au cours du temps, ce qui se traduit par une diminution de leur période de rotation. La Figure \ref{PPdot} montre le diagramme $P$ - $\dot P$, représentant le taux de ralentissement des pulsars en fonction de leur période, pour ceux dont on dispose d'une mesure de $\dot P$ fiable, parmi les quelques 1900 recensés par le catalogue ATNF.

\fig[Diagramme $P$ - $\dot P$]{width=14cm}{chap1/PdotP.pdf}{Ralentissement $\dot{P}$ en fonction de la période $P$ pour les pulsars du catalogue ATNF en dehors d'amas globulaires et avec $\dot P$ positif. Les valeurs de $\dot P$ sont corrigées de l'effet Shklovskii lorsque possible. Les pulsars isolés sont représentés par des points tandis que les pulsars appartenant à des systèmes multiples sont représentés entourés d'un cercle. Des lignes d'âge caractéristique $\tau$ constant sont représentées, ainsi que des lignes de champ magnétique caractéristique $B_S$ constant et des lignes de $\dot{E}$ constant. La majorité des pulsars connus sont les pulsars « ordinaires », au centre de ce graphe. En bas à gauche du graphe on trouve les pulsars « milliseconde ». En haut à droite, on trouve les « magnetars », des pulsars au champ magnétique très intense. Pour les six pulsars dont l'indice de freinage $n = \Omega \ddot \Omega / \dot \Omega^2$ a été mesuré, des traits pleins montrent le comportement simulé de $\dot P$ en fonction de $P$ sur $10^5$ ans, dans l'hypothèse où $n$ est constant.}{PPdot}

Comme le montre la Figure \ref{PPdot}, la majorité des pulsars connus ont une période de rotation $P$ comprise entre 0,1 et quelques secondes, et un ralentissement $\dot P$ entre 10$^{-17}$ et 10$^{-13}$. Ce sont les pulsars dits « normaux », ou « ordinaires ». On estime qu'ils naissent avec une période comprise entre 14 et 140 ms \citep{Kramer2003}. La majorité des pulsars normaux est relativement jeune : de quelques milliers à quelques dizaines de millions d'années typiquement. On constate l'existence, dans le coin inférieur droit du graphe, d'une zone dépeuplée de pulsars, nommée  « cimetière ». Le faible nombre de pulsars détectés dans cette gamme de ($P$, $\dot P$) amène à penser qu'il doit exister une limite à partir de laquelle les pulsars, après avoir vu leur période de rotation $P$ croître et leur ralentissement $\dot P$ diminuer au cours de leur vie, cessent d'émettre un rayonnement détectable à grande distance. 

Pour les six pulsars jeunes pour lesquels une mesure de l'indice de freinage $n$ est disponible (\emph{cf.} \citet{KaspiHelfand2002} pour une revue des résultats), la Figure \ref{PPdot} montre le comportement de $\dot P$ en fonction de $P$, simulé sur $10^5$ ans, à partir de l'équation différentielle $n = \Omega \ddot \Omega / \dot \Omega^2 = 2 - P \ddot P / \dot P^2$ où l'on a supposé que $n$ est constant. Les parcours ainsi calculés sont quasi-linéaires, avec une pente de $-1$ si l'indice de freinage est de 3. On constate que les six vecteurs ne sont pas dirigés vers la région la plus densément peuplée en pulsars, signifiant vraisemblablement que l'indice de freinage $n$ évolue au cours du temps afin que les pulsars jeunes viennent peupler cette région. On remarque que l'un des pulsars voit son ralentissement $\dot P$ augmenter, alors que ce taux diminue chez les cinq autres pulsars. Il s'agit du pulsar de Vela, dont l'indice de freinage est inférieur à 2, ce qui implique $\ddot P > 0$ \citep{Lyne1996}. De tels objets pourraient être à l'origine des « magnetars », pulsars ayant les champs magnétiques les plus intenses \citep{Lorimer2009}.

Enfin, on trouve dans la partie inférieure gauche du graphe les pulsars « milliseconde » (MSPs), qui se démarquent de la majorité des pulsars ordinaires par une rotation extrêmement rapide ($P \leq 30$ ms) et stable ($\dot P \leq 10^{-17}$). Le plus rapide des MSPs connus à ce jour est PSR J1748$-$2446ad, découvert dans l'amas globulaire Terzan 5, ayant une période $P \simeq 1,396$ ms \citep{Hessels2006} ! En outre, on remarque que les MSPs appartiennent à des systèmes multiples pour la plupart, au contraire des pulsars ordinaires.

\subsection{Formation}

Le modèle du dipôle tournant conduit à des âges caractéristiques pour les MSPs de l'ordre de 10$^9$ à 10$^{11}$ années, soit supérieurs à l'âge de l'univers dans certains cas. Par conséquent, le schéma d'évolution décrit précédemment pour les pulsars normaux ne permet pas d'expliquer l'existence des MSPs. La Figure \ref{PPdot} indique par un cercle les pulsars appartenant à des systèmes multiples. Comme on l'a déjà évoqué, il est frappant de constater que moins d'un pourcent des pulsars normaux appartiennent à des systèmes multiples, alors que plus de 80\% des pulsars milliseconde sont concernés, et tandis que ce taux est bien supérieur à 1\% pour les étoiles en général. Cette faible valeur s'explique par l'importante probabilité de rupture du système, du fait de la violence du phénomène de supernova. Pour les systèmes qui survivent à la supernova, il a été proposé que si l'étoile compagnon vient à déborder du lobe de Roche du pulsar, celui-ci peut acquérir une partie de la masse et du moment cinétique, accélérant ainsi sa rotation \citep{Alpar1982}. La découverte récente d'un pulsar milliseconde qui a été entouré d'un disque d'accrétion dans un passé proche tend à confirmer ce scénario \citep{Archibald2009}. Pour cette raison, on parle également de pulsar « recyclés ». 

Pendant la phase d'accrétion, il y a production de rayonnement X thermique par la matière tombant vers l'étoile à neutrons, ce qui rend le système détectable en tant que « binaire à rayons X » (\emph{X-ray Binary}). On en distingue deux catégories, selon la masse de l'étoile compagnon: les systèmes à haute masse (HMXBs) et à faible masse (LMXBs). Dans le premier cas, il se peut que l'étoile compagnon soit suffisamment massive pour évoluer en supernova, puis en étoile à neutrons ; si le système survit à la seconde explosion, on obtient un double pulsar. On connaît neuf de ces systèmes, le premier découvert étant PSR B1913+16 \citep{Hulse1975}. Plus généralement, la phase d'accrétion dans les HMXBs peut être interrompue par l'explosion du compagnon, ce qui produit des pulsars « modérément recyclés » (\emph{mildly recycled}). Pour ce qui est des LMXBs, où cette fois le compagnon est peu massif, la période d'accrétion n'est pas interrompue par une supernova, ainsi la phase d'accrétion est beaucoup plus longue. Il en résulte généralement un pulsar milliseconde à très courte période (typiquement, quelques ms), en orbite avec une naine blanche.

Il existe néanmoins des MSPs isolés et à très courte période, pour lesquels le recyclage n'a pourtant pas été interrompu. Dans les amas globulaires, ce phénomène peut s'expliquer par la forte densité d'étoiles : les interactions entre étoiles y sont plus nombreuses et les systèmes multiples peuvent se défaire. En revanche la cause est moins bien connue pour les pulsars du champ galactique. Il a été proposé que l'étoile compagnon puisse être évaporée par l'intense vent de particules produit par le MSP \citep{Ruderman1989}.

\subsection{Nombre et répartition}

La catalogue ATNF recense 169 pulsars\footnote{En juin 2009.} ayant une période inférieure à 30 ms\footnote{Cette limite supérieure sur la période des pulsars milliseconde est une des valeurs couramment utilisées dans la littérature. La condition $P \leq 50$ ms est également utilisée. La limite de 30 ms, bien qu'arbitraire, présente l'avantage d'exclure le pulsar du Crabe ($P \simeq 33$ ms) aux propriétés très différentes de celles des MSPs, en particulier son taux de ralentissement $\dot P$ beaucoup plus grand.}. Parmi ces 169, 97 appartiennent à des amas globulaires, les autres sont dans le « champ galactique ». En réalité, si l'on tient compte des pulsars découverts récemment et dont les découvertes n'ont pas été publiées, le nombre total de MSPs connus à ce jour s'élève à un peu plus de 200. En particulier, bien que quelques pulsars soient en attente de confirmation, de nombreux pulsars milliseconde des amas globulaires ne sont pas recensés par le catalogue ATNF\footnote{\emph{cf.} http://www.naic.edu/~pfreire/GCpsr.html}. Les listes des pulsars milliseconde du champ galactique et des amas globulaires et certaines de leurs caractéristiques sont données en Annexes \ref{tableauMSPsGAL} et \ref{tableauMSPsGC}.

En moyenne, les MSPs connus sont moins lumineux dans le domaine radio que leurs alter-ego normaux \citep{Kramer1998,Bailes1997}. Il en résulte par construction que les MSPs détectés dans le champ galactique sont plus proches que les pulsars normaux en moyenne. \`{A} cause de leur proximité, de nombreux MSPs paraissent être situés en dehors du plan galactique (soit $|b| > 10^\circ$). Les amas globulaires se situant dans le halo de la Galaxie, ils apparaissent également en dehors du plan, en coordonnées galactiques. Pour ces différentes raisons les MSPs sont majoritairement situés à haute latitude galactique, comme le montrent les Figures \ref{CarteGalactique} (A) et (B).

\sfig[Répartition des pulsars en coordonnées galactiques]{chap1/GL_GB_Distance.tex}{(A) : latitude galactique $b$ en fonction de la distance pour les pulsars du catalogue ATNF. (B) : Carte du ciel en coordonnées galactiques. Les pulsars normaux sont représentés par des points, les pulsars milliseconde sont entourés d'un cercle.}{CarteGalactique}

Moins d'une centaine de pulsars milliseconde galactiques sont donc recensés. Les études de populations de MSPs émetteurs radio et des LMXBs prédisent des taux de naissance de pulsars milliseconde compris entre 3 et 6,5 $\times 10^{-6}$ par an \citep{Ferrario2007,Story2007,Kiel2006}. En prenant l'âge de la Galaxie égal à 13,6 milliards d'années, on obtient une population galactique de MSPs comprise entre 40000 et 90000 objets.

\subsection{Profil d'émission}

La Figure \ref{ProfilsMSPs} montre quelques exemples de profils radio de pulsars milliseconde. On constate une certaine diversité parmi ces courbes de lumière, dans la multiplicité des pics, la complexité des structures, leur largeur et l'écartement des pics (pour les cas où le nombre de pics est supérieur à un). Les courbes de lumière aux différentes longueurs d'onde représentent les empreintes des pulsars, et donnent une idée de la configuration géométrique (inclinaison de l'axe magnétique par rapport à l'axe de rotation, angle de visée par rapport à ces axes) que peuvent compléter des études de la polarisation de l'émission \citep{Ord2004,Stairs1999,Xilouris1998}. Par exemple, le pulsar J1939+2134 présente deux pics, séparés d'environ 180$^\circ$. On peut en déduire que l'on observe les deux pôles magnétiques au cours d'une seule rotation, ce qui correspond à une configuration où la ligne de visée et l'axe magnétique sont proches, et sont orientés orthogonalement à l'axe de rotation.

\fig[Exemples de profils radio de pulsars milliseconde]{width=14cm}{chap1/ProfilsMSPs.pdf}{Exemples de profils radio de pulsars milliseconde. Ces profils ont été enregistrés à Nançay, à la fréquence de 1400 MHz.}{ProfilsMSPs}

Les profils radio des MSPs sont en règle générale semblables à ceux des pulsars normaux. Comme on l'a déjà évoqué, l'émission radio est supposée produite par les particules chargées en déplacement le long des lignes de champ ouvertes. Le cylindre de lumière, de rayon $c P / \left( 2 \pi \right)$ est beaucoup plus petit chez les MSPs, ce qui se traduit par une ouverture de faisceau plus large en moyenne, par rapport aux pulsars normaux. Différents modèles phénoménologiques ont été proposés pour expliquer la morphologie des profils radio des pulsars en général, et la diversité de ces profils, dans un schéma simple de dipôle magnétique tournant. D'après le modèle « coeur et cône », le faisceau radio possède un coeur entouré d'un ou plusieurs cônes évidés, ces différentes composantes étant centrées sur l'axe magnétique \citep{Rankin1983}. L'alternative principale est le modèle « à taches », pour lequel le cône délimité par les lignes de champ ouvertes est peuplé de taches d'émission, distribuées de façon non uniforme \citep{Lyne1988}. Selon des travaux récents, les données sont le mieux reproduites par un modèle hybride utilisant des taches d'émission dans une structure ayant un coeur et des cônes \citep{Karastergiou2007}. On constate cependant d'après la Figure \ref{ProfilsMSPs} que certains profils de MSPs sont extrêmes de par leur complexité, par exemple celui de J2124$-$3358. Cette complexité est parfois attribuée à l'existence d'un champ multipolaire à proximité de la surface \citep{Kramer1997}.

\subsection{Les pulsars milliseconde en physique et en astronomie}

Pour terminer ce chapitre introductif sur les pulsars en général et les pulsars milliseconde en particulier, replaçons ces objets dans un contexte plus général que l'étude de leur émission en rayons $\gamma$, qui sera abordée dans la suite. L'étude des étoiles à neutrons touche un panel large de domaines de la physique fondamentale et de l'astrophysique. De par leur nature extrême, les MSPs ont une place toute particulière dans les différentes découvertes issues de l'astronomie des pulsars. Citons quelques exemples d'avancées réalisées grâce à l'étude des pulsars (liste non exhaustive) :

\begin{itemize}
\item Les premières exoplanètes ont été découvertes \emph{via} la chronométrie du pulsar milliseconde B1257+12, en 1990 à Arecibo \citep{Wolszczan1992}.
\item Des tests fins de la théorie de la relativité générale ont pu être réalisés grâce aux doubles pulsars tels que B1913+16, l'orbite observée étant remarquablement conforme aux prédictions \citep{Hulse1994}. Un prix Nobel a été décerné à Hulse et Taylor en 1993 en récompense de ces travaux. Plus récemment, l'observation du système double J0737$-$3039 dans lequel les deux pulsars sont visibles a fourni les contraintes les plus fortes dont on dispose à ce jour sur la théorie de la relativité :  par exemple la valeur du paramètre orbital « post-képlérien » $s$ mesurée est conforme à celle prédite par la relativité générale, avec une précision de 0,05\% \citep{Kramer2006}.
\item La dispersion du signal radio émanant des pulsars donne la possibilité de cartographier la densité d'électrons dans la Galaxie. De la même manière, l'étude de la polarisation des pulsars permet de sonder le champ magnétique régnant dans la Galaxie \citep{Han2006}. 
\item Une limite supérieure sur la variation temporelle de la constante de gravitation $G$ a pu être déterminée, grâce au pulsar J0437$-$4715 \citep{Verbiest2008}.
\item La chronométrie des pulsars appartenant à des systèmes binaires, pulsars milliseconde pour la plupart, a permis la mesure de nombreuses masses d'étoiles à neutrons \citep{LattimerPrakash2007}. Ces mesures de masses permettent de contraindre l'équation d'état de la matière super-dense des étoiles à neutrons. D'autres contraintes sont fournies par la détection de pulsars milliseconde ultra rapides : le plus rapide connu à présent est J1748$-$2446ad, dont la fréquence de rotation est de 716 Hz \citep{Hessels2006}.
\end{itemize}

Tandis que les télescopes actuels opérant du domaine radio aux rayons $\gamma$ tentent de percer certains mystères sur les pulsars -- combien sont-ils dans la Galaxie, comment se forment-ils, comment évoluent-ils, quels sont leurs mécanismes d'émission -- de grands projets tels que LOFAR et SKA \citep{Leeuwen2008,Smits2009} sont à venir. En plus d'augmenter considérablement le nombre de pulsars connus (plusieurs milliers de détections sont attendues) ce qui permettra une bien meilleure compréhension de ces objets, ils pourraient mettre en application une idée ambitieuse : utiliser un réseau de pulsars milliseconde comme d'un détecteur d'ondes gravitationnelles.

%% file: part1/chap2.tex
\chapter[Les pulsars en rayons $\gamma$]{Les pulsars en rayons $\gamma$}

\minitoc

\chaptabst{N}{ous}{présentons ici les différents modèles de production de rayonnement $\gamma$ par les pulsars, ces modèles se répartissant en deux grandes familles : \emph{Polar Cap} (PC) d'une part, et \emph{Outer Gap} (OG) d'autre part. La différence principale entre les différentes théories d'émission $\gamma$ est le site de production du rayonnement : près des pôles magnétiques ou au contraire en altitude dans la magnétosphère. La géométrie de l'émission ainsi que les mécanismes variant d'un modèle à l'autre, les propriétés prédites sont différentes et le LAT peut permettre de discriminer les modèles. Nous nous penchons par la suite sur le cas, spécifique \emph{a priori}, des pulsars milliseconde. Ceux-ci ont notamment un champ magnétique bien plus faible que celui des pulsars ordinaires. Les modèles théoriques prévoient néanmoins une dizaine de détections de MSPs par le LAT dans sa première année. Dans la suite de ce chapitre, nous passons en revue les propriétés des pulsars $\gamma$ connus. \`{A} ce sujet, EGRET à bord de CGRO (\emph{Compton Gamma Ray Observatory}) a été la principale source de connaissances avant le lancement de Fermi. Des tendances se dégagent parmi les profils d'émission $\gamma$ ainsi que les propriétés spectrales des pulsars détectés par EGRET. Un point commun important est leur grand $\dot E$, et c'est sur ce critère qu'a été basée la campagne de suivi des pulsars par les télescopes radio et X, évoquée dans le chapitre suivant. Dans la fin de ce chapitre nous présentons les motivations de l'étude des pulsars dans cette gamme d'énergie.}

\section{Modèles théoriques}

Les premières sources $\gamma$ identifiées étaient les pulsars du Crabe et de Vela, détectés par les télescopes spatiaux SAS--2 et COS--B dans les années 70, plusieurs décennies avant le lancement de Fermi. Bien que le rayonnement $\gamma$ des pulsars ait été étudié depuis cette époque, le mécanisme d'émission est toujours débattu. 

Les modèles existant actuellement ont en commun de nécessiter une région faiblement peuplée en particules chargées dans la magnétosphère des pulsars. En présence d'un tel déficit de charge par rapport à la distribution de charges à l'équilibre dans la magnétosphère, un champ électrique parallèle $\vec{E_\parallel}$ au champ magnétique se développe, et celui-ci peut accélérer des particules chargées vers les hautes énergies. Ces particules produisent par la suite des photons $\gamma$, par rayonnement de courbure, rayonnement synchrotron ou diffusion Compton inverse. 

Le désaccord entre les différentes théories porte justement sur la région d'accélération des particules chargées. Historiquement, deux grandes familles de modèles se sont confrontées, ceux de la « calotte polaire » (\emph{Polar Cap}) d'une part, pour lesquels l'accélération de charges se produit au-dessus des pôles magnétiques, ceux de la « cavité externe » (\emph{Outer Gap}) d'autre part pour lesquels les photons $\gamma$ sont émis en altitude dans la magnétosphère. Des évolutions de ces théories ont récemment amené à considérer des zones d'émission intermédiaires. Pour plus de détails relatifs aux différents scénarios d'émission $\gamma$, le lecteur pourra par exemple se reporter à \citet{Harding2007,Hirotani2008}.

\fig[Les régions d'accélération dans les différents modèles d'émission $\gamma$]{width=14cm}{chap2/ModelesPulsar.pdf}{Schéma de la magnétosphère des pulsars, représentant les régions d'accélération des particules chargées dans les différents modèles théoriques : calotte polaire (\emph{Polar Cap}), cavité à fentes (\emph{Slot Gap}) et cavité externe (\emph{Outer Gap}). Crédit : Alice Harding.}{ModelesPulsar}

\subsection{Modèles \emph{Polar Cap} et \emph{Slot Gap}}

On a vu dans le chapitre précédent qu'il existe un rayon au-delà duquel les lignes de champ magnétique ne peuvent boucler, c'est le rayon du cylindre de lumière $R_{LC} = c P / (2 \pi)$. Les dernières lignes de champ fermées délimitent, au niveau de la surface de l'étoile, la calotte polaire. Pour les modèles \emph{Polar Cap} (PC), l'accélération des particules chargées se produit sur quelques rayons stellaires au-dessus de cette calotte polaire, le long des lignes de champ magnétique ouvertes. 

Les modèles PC se répartissent eux-mêmes en différentes familles, selon le mode d'approvisionnement en particules chargées à accélérer \citep{Arons1979,Ruderman1975}. Pour les modèles à cavité vide (\emph{vacuum gap}), des particules chargées sont piégées dans les couches superficielles de l'étoile à neutrons, et une zone vide de charge se forme au-dessus de la calotte polaire. Dans les modèles SCLF (\emph{Space-Charge Limited Flow}), des particules chargées sont émises à la surface vers la magnétosphère, où la densité de charges est insuffisante pour annuler le champ électrique parallèle. Dans tous les cas, la région située au-dessus du pôle présente un déficit de charges, induisant l'existence d'un champ électrique intense $E_\parallel$ en son sein, parallèle aux lignes de champ magnétique ouvertes. Des particules chargées y sont donc accélérées parallèlement au champ magnétique, et émettent des rayons $\gamma$ par rayonnement de courbure ou par diffusion Compton inverse. \`{A} proximité de la surface de l'étoile, le champ magnétique est intense. Par conséquent, les photons $\gamma$ créent des paires électron / positron par interaction avec le champ magnétique, dans un zone appelée « front de formation de paires » (\emph{Pair Formation Front}, PFF). Une cascade de paires électron / positron se développe dans cette région de l'espace. Au-delà de l'épaisseur optique du PFF, le champ électrique parallèle $E_\parallel$ est complètement écranté. Ainsi les particules chargées qui s'en échappent ne sont plus accélérées, mais peuvent à leur tour émettre du rayonnement, formant le faisceau d'émission $\gamma$.

\fig[Schéma du front de formation de paires (PFF)]{width=6cm}{chap2/HK_fig1.pdf}{Schéma de la région située au-dessus de la calotte polaire. Le front de formation de paires (PFF) est représenté au-dessus du pôle. L'interstice délimité par le PFF et les dernières lignes de champ magnétique ouvertes est la région d'accélération des particules chargées dans le modèle \emph{Slot Gap}. Figure extraite de \citet{Harding2004}.}{SlotGap}

Des évolutions du modèle PC ont conduit à l'élaboration du modèle « à fentes » (\emph{Slot Gap}, SG) \citep{Arons1983,Muslimov2003}. \`{A} proximité de la surface, la répartition des charges crée un champ électrique $E_\parallel \neq 0$, qui est écranté plus haut dans un front de formation de paires. Cependant la frontière de la calotte polaire est parfaitement conductrice, imposant $E_\parallel = 0$. Par conséquent, le champ électrique au-dessus de cette frontière est moins intense qu'au-dessus du pôle. Les particules chargées ont donc besoin d'une plus grande distance d'accélération pour émettre des photons $\gamma$ énergétiques pouvant induire des cascades de paires. Le PFF est donc recourbé au-dessus des bords de la calotte, comme indiqué par la Figure \ref{SlotGap}. Cette courbure crée des fentes entre le PFF et les dernières lignes de champ magnétique ouvertes, où le champ magnétique $E_\parallel$ n'est pas écranté, et où par conséquent les particules chargées peuvent accélérer jusqu'à haute altitude. Conduites dans la magnétosphère externe, ces charges émettent des photons $\gamma$ par rayonnement de courbure ou synchrotron. 

Dans une approche purement géométrique qui a conduit au modèle \emph{Two-Pole Caustic} (TPC) \citep{Dyks2003}, il a été montré que les profils d'émission des pulsars $\gamma$ connus correspondent relativement bien à un rayonnement produit le long des lignes de champ ouvertes, depuis la surface de l'étoile jusqu'à haute altitude, à l'image de ce qui est prédit par le modèle à fentes. Il est donc intéressant de remarquer que celui-ci pourrait servir de support physique à une description empirique.

\subsection{Modèle \emph{Outer Gap}}

Les modèles à cavité externe originaux \citep{Cheng1986,Romani1996} partent du principe que les charges positives et négatives sont distribuées séparément dans la magnétosphère, suivant le signe de $\vec{\Omega} . \vec{B}$, comme dans le modèle de Goldreich \& Julian \citep{Goldreich1969}. Des cavités pauvres en particules chargées peuvent se former dans la région délimitée par la surface de charge nulle (définie par $\rho \propto \vec{\Omega} . \vec{B} = 0$) et le cylindre de lumière, que les charges provenant de la surface de l'étoile ne peuvent peupler. Dans ces cavités, la composante de champ électrique parallèle au champ magnétique $E_\parallel$ est forte, de sorte que les particules chargées qui s'y trouvent sont accélérées vers les hautes énergies. Des photons $\gamma$ sont ainsi émis, et ceux-ci peuvent former des paires électron / positron par interaction avec les photons X thermiques provenant de la surface de l'étoile à neutrons. Des cascades de paires sont ainsi engendrées : les électrons et positrons sont accélérés, ils émettent des photons, qui eux-mêmes forment des paires, et ainsi de suite. La cavité est donc peuplée de particules chargées, venant écranter le champ électrique le long des lignes de champ magnétique et limiter le volume de la région d'accélération. La taille de la cavité ainsi maintenue, où le champ électrique parallèle $E_\parallel$ n'est pas écranté, définit la géométrie de l'émission. 

\subsection{Différences attendues : géométries et spectres}

Les trois modèles prévoient des zones d'émission des rayons $\gamma$ et des mécanismes de production différents. Les deux observables de l'analyse des pulsars, quelle que soit la longueur d'onde considérée, sont le profil temporel de l'émission, et la forme spectrale du rayonnement. Les observations de pulsars en rayons $\gamma$ permettent de confronter les prédictions morphologiques et spectrales des différents modèles, et par conséquent de les discriminer. 

La Figure \ref{SimVelaGeometry} montre des simulations de profils d'émission $\gamma$ attendus par les différents modèles, pour le pulsar de Vela et pour différentes configurations géométriques (à savoir, $\alpha$ : angle entre l'axe magnétique et l'axe de rotation, et $\zeta$ : angle entre la ligne de visée et l'axe de rotation). Pour le modèle PC, l'accélération de charges se produit au-dessus de la calotte polaire, et est délimitée par les lignes de champ ouvertes. L'émission radio naît également au-dessus des pôles. Par conséquent le modèle PC n'autorise que de faibles écartements radio / $\gamma$. Au contraire, des écartements plus grands sont permis pour des émetteurs de type SG et OG, du fait de la zone d'émission $\gamma$ décentrée. De plus, OG et SG prévoient davantage de pulsars $\gamma$ non détectables en radio en raison du désalignement des faisceaux d'émission aux différentes énergies. La fraction de pulsars détectables en radio parmi les émetteurs $\gamma$ varie donc d'une théorie à l'autre. 

\`{A} ces différences intrinsèques s'ajoutent des effets relativistes qui modifient la morphologie de l'émission. Le temps de parcours des photons $\gamma$ est un premier exemple : des photons émis à une certaine altitude dans la magnétosphère et d'autres photons émis plus tôt ou plus tard à une altitude différente peuvent parvenir à l'observateur simultanément. L'étalement en phase de l'émission $\gamma$ produite en différents endroits de la magnétosphère engendre la formation de caustiques dans les courbes de lumière. \`{A} proximité de la surface de l'étoile, le champ gravitationnel est intense ; ainsi les champs électrique et magnétique sont distordus et le parcours des rayons lumineux courbés. En altitude, à proximité du cylindre de lumière, les champs subissent des retards plus ou moins importants selon la vitesse de rotation. Ces différents effets jouent sur la direction d'émission des photons $\gamma$ voire sur la géométrie des zones d'émission, modifiant le profil d'émission vu par l'observateur \citep{Grenier2006}.

Il est clair d'après la Figure \ref{SimVelaGeometry} que pour un pulsar donné (avec un angle d'inclinaison magnétique $\alpha$ donné), deux observateurs ayant des lignes de visées distinctes par rapport à l'axe de rotation peuvent observer des profils et des intensités lumineuses différentes, voire ne pas détecter les pulsars si l'orientation des faisceaux est défavorable. Ceci conduit à définir un facteur de correction $f_\Omega$, représentant le fait que l'intensité du signal observé sous un certain angle de visée n'est pas nécessairement représentatif de l'émission rayonnée sur le ciel entier \citep{Watters2009} :

\begin{eqnarray}
f_\Omega (\alpha,\zeta) = \frac{\int \int h (\alpha, \zeta^\prime, \Phi) \sin(\zeta^\prime) d\zeta^\prime d\Phi}{2 \int h (\alpha, \zeta, \Phi) d\Phi}
\label{fOmega}
\end{eqnarray}

Dans cette expression, $h$ désigne le flux en énergie émis dans la direction $(\alpha, \zeta, \Phi)$. En pratique, ce facteur est inconnu \emph{a priori}, mais comme le montre la Figure \ref{SimVelaGeometry}, on peut l'estimer par des hypothèses sur la géométrie des faisceaux d'émission. Typiquement, ce facteur de correction est d'environ 1 pour les modèles SG et OG, compris entre 0 et 0,5 pour le modèle PC. 

Du point de vue des spectres d'émission, le modèle PC prévoit une coupure super-exponentielle (c'est-à-dire en $e^{-\left( E / E_c\right)^b}$ où $E_c$ est l'énergie de coupure et $b >$ 1) du fait du mécanisme de production de paires électron / positron par interaction avec le champ magnétique. Pour SG et OG, les coupures sont simplement exponentielles (soit $b$ = 1). On peut donc écrire le flux de photons attendu de la façon suivante :

\begin{eqnarray}
\frac{dN}{dE} = N_0 E^{-\Gamma} e^{-(\frac{E}{E_c})^b}
\label{eqEmission}
\end{eqnarray}

Dans cette expression, $E$ désigne l'énergie du photon, $\Gamma$ est l'indice spectral avant la coupure, $E_c$ est l'énergie de coupure du spectre et $b$ est le paramètre de forme de la coupure. Le terme $N_0$ est un facteur de normalisation. La quantité $dN/dE$ représente le nombre de photons émis par le pulsar et détecté par unité de temps, d'énergie, et de surface de collection. La Figure \ref{SimVelaSpectra} montre, pour le pulsar de Vela, des prédictions faites avant lancement du spectre vu par le LAT au bout d'un an d'activité, pour une émission de type \emph{Polar Cap} ou \emph{Outer Gap}. Contrairement à EGRET, le LAT devrait être en mesure de distinguer entre une coupure simplement ou super-exponentielle. La forme des spectres, et de leur coupure en particulier, sont donc d'autres discriminateurs disponibles. 

\fig[Courbes de lumière $\gamma$ du pulsar de Vela pour différents modèles théoriques]{width=14cm}{chap2/VelaGeometry.pdf}{\emph{En haut à gauche :} courbe de lumière $\gamma$ du pulsar de Vela vue par EGRET entre 30 MeV et 10 GeV. Les trois autres cadrans présentent des simulations de la courbe de lumière attendue pour Vela, dans différents modèles d'émission (\emph{Polar Cap}, \emph{Slot Gap} et \emph{Outer Gap}) et pour différentes configurations géométriques ($\alpha$, $\zeta$), où $\alpha$ est l'angle d'inclinaison de l'axe magnétique par rapport à l'axe de rotation, et $\zeta$ est l'angle que forme la ligne de visée avec l'axe de rotation. Chacun des cadrans est séparé en deux parties. Celle du haut représente l'émission $\gamma$ produite pour les différentes valeurs de phase rotationnelle $\Phi$ (ici entre $-180^\circ$ et $180^\circ$ ou entre $0^\circ$ et $360^\circ$) et d'angle de visée $\zeta$. Pour un observateur fixe, on fait une coupe dans ce plan avec une valeur de $\zeta$ constante, symbolisée par les lignes pointillées. La partie inférieure montre la courbe de lumière prédite pour la valeur de $\zeta$ correspondante. Figure issue de \citet{Harding2007b}.}{SimVelaGeometry}

\fig[Simulations de spectres d'émission $\gamma$ du pulsar de Vela]{width=14cm}{chap2/VelaSpectra.png}{Spectres du pulsar de Vela pour une observation simulée d'un an avec le LAT, dans l'hypothèse des modèles \emph{Polar Cap} et \emph{Outer Gap}. Les données d'EGRET sont superposées. Crédit : Massimiliano Razzano.}{SimVelaSpectra}

\subsection{\'Emission $\gamma$ des pulsars milliseconde}

Un certain nombre de différences importantes existent entre les pulsars normaux et les MSPs : ces derniers ont un champ magnétique en surface typiquement $10^4$ fois inférieur, comme le montre la Figure \ref{PPdot}. Par conséquent, la création de paires électron / positron par interaction des photons avec le champ magnétique à proximité de la calotte polaire est peu favorable. Pour les pulsars milliseconde on s'attend donc à ce que le champ électrique au-dessus des pôles magnétiques ne soit pas écranté, à cause de l'absence de cascades de créations de paires. Selon le modèle PC, les électrons sont donc accélérés depuis la calotte polaire jusqu'à haute altitude et émettent par rayonnement de courbure \citep{Harding2005}. Le modèle prévoit de plus que les électrons puissent être accélérés vers des énergies plus grandes que pour les pulsars jeunes du fait de l'absence de front de formation de paires. De ce fait l'émission de photons peut s'étendre au-delà de 10 GeV, avec une coupure simplement exponentielle (soit $b = 1$ dans l'équation \ref{eqEmission}), contrairement aux pulsars jeunes dans le modèle PC. Une synthèse de population basée sur la théorie de l'émission au-dessus de la calotte polaire a prédit 12 détections en rayons $\gamma$ de MSPs émetteurs radio par le LAT au bout d'un an \citep{Story2007}. Enfin, une conséquence de la non-existence de PFF au-dessus des pôles est l'absence de fentes au contact des dernières lignes de champ ouvertes, requises par le modèle SG pour l'accélération de particules chargées et l'émission de rayons $\gamma$.

L'émission $\gamma$ des MSPs a également été modélisée dans le scénario OG \citep{Zhang2003}. Dans cette étude on fait l'hypothèse d'un champ magnétique multipolaire régnant à proximité de la surface de l'étoile à neutrons. Dans la cavité externe, le champ électrique parallèle aux lignes de champ magnétique permet aux particules chargées de se déplacer, engendrant un courant de retour prenant la direction de la surface de l'étoile. Ces particules interagissent avec l'intense champ magnétique multipolaire et émettent des rayons X, qui viennent alimenter le reste de la magnétosphère et en particulier la cavité externe. L'arrivée de rayonnement X dans la cavité externe recrée les conditions décrites précédemment pour le modèle \emph{Outer Gap} appliqué aux pulsars normaux : les photons X interagissent avec les photons $\gamma$ émis dans la cavité pour engendrer des cascades de paires électron / positron, qui sont accélérées et émettent des photons $\gamma$, \emph{etc}. L'émission de photons $\gamma$ se produisant dans la cavité externe, elle peut donc apparaître décalée du rayonnement radio. Les photons $\gamma$ émis dans ce modèle ne peuvent pas excéder quelques GeV, et la coupure attendue est simplement exponentielle. Le nombre de détections de MSPs par le LAT dans sa première année prédites dans ce modèle est également de l'ordre d'une dizaine \citep{Zhang2007}. 

Bien que les modèles ne s'accordent pas sur les sites et les mécanismes de production des photons $\gamma$, on constate que ceux-ci prédisent un nombre relativement important de détections de MSPs par le LAT.

\section{Observations de pulsars en rayons $\gamma$}

Avant le lancement de Fermi, l'essentiel des connaissances relatives aux pulsars émetteurs $\gamma$ a été acquis grâce au télescope EGRET, à bord de CGRO. Entre 1991 et 1997, EGRET a permis la détection de six pulsars en rayons $\gamma$, parmi lesquels le Crabe et Vela, et quelques candidats sérieux, comme nous le verrons dans la suite de ce chapitre. En ajoutant la détection du pulsar B1509$-$58 jusqu'à 10 MeV par le télescope COMPTEL, également à bord de CGRO, on comptait sept détections certaines de pulsars en rayons $\gamma$ après la mission de CGRO, auxquelles se sont ajoutées récemment les deux détections effectuées par AGILE et de nouveaux candidats.

\subsection{L'héritage de CGRO}

\subsubsection{Pulsars normaux}

L'échantillon de pulsars émetteurs $\gamma$ vus par EGRET et COMPTEL à bord de CGRO était de sept objets et quelques candidats. On ne pouvait donc pas tirer de conclusions définitives à partir de leurs propriétés. Un certain nombre de tendances se dégageaient néanmoins.

Ces sept pulsars sont relativement jeunes et énergétiques \citep{Thompson2004}. Ils sont parmi les pulsars qui perdent le plus d'énergie cinétique rotationnelle par freinage électromagnétique ; disposant ainsi d'un important réservoir à convertir en rayonnement $\gamma$.

La Figure \ref{phasosEGRET} montre les courbes de lumière des pulsars d'EGRET et de COMPTEL dans cinq bandes d'énergie. Le profil d'émission et l'alignement en phase des composantes radio, optiques, X et $\gamma$ varient fortement d'un pulsar à l'autre. On remarque par exemple que le pulsar du Crabe présente deux pics en alignement étroit dans les différentes longueurs d'onde. Le pulsar de Vela n'a qu'un pic visible en radio, mais deux en optique, en X et en rayons $\gamma$. Parmi ces sept pulsars, Geminga est le seul à n'avoir jamais été détecté en radio jusqu'à présent. Malgré ces différences de comportement en fonction de l'énergie, on remarque qu'à l'exception de B1509$-$58, les pulsars présentent deux pics d'émission $\gamma$. Dans certains cas, les plus frappants étant ceux de Vela et de Geminga, des photons sont également détectés dans une zone située entre les deux pics (le \emph{bridge}). Ces observations indiquent que l'émission $\gamma$ est produite dans une zone large, vraisemblablement évidée. 

\fig[Les sept pulsars détectés en rayons $\gamma$ avec EGRET et COMPTEL]{width=16cm}{chap2/phasosEGRET.pdf}{Courbes de lumière dans différentes longueurs d'onde pour les sept pulsars $\gamma$ détectés avec EGRET et COMPTEL. Figure issue de \citet{Thompson2008}.}{phasosEGRET}

Des similitudes se dégagent également parmi les propriétés spectrales des pulsars d'EGRET. Pour un ajustement de leur spectre par une
loi de puissance avec coupure exponentielle (soit $N_0\ E^{-\Gamma} e^{-(E / E_0)}$), les indices spectraux $\Gamma$ sont typiquement inférieurs à 2. L'énergie de coupure $E_0$ quant à elle est généralement comprise entre 1 et 4 GeV \citep{Thompson2004}. On peut écrire la luminosité $\gamma$ de la manière suivante :

\begin{eqnarray}
L_\gamma = 4 \pi f_\Omega h d^2
\label{Lgamma}
\end{eqnarray}

où $h$ est le flux en énergie détecté par le télescope, $d$ est la distance du pulsar et $f_\Omega$ est le facteur de correction géométrique défini précédemment (\emph{cf.} équation \ref{fOmega}). Connaissant la luminosité $\gamma$, on peut calculer l'efficacité de conversion de la perte d'énergie due au freinage en rayonnement $\gamma$, $\eta = L_\gamma / \dot{E}$.

En l'absence d'hypothèse sur la géométrie des faisceaux $\gamma$, il apparaît que la luminosité $L_\gamma (\mathrm{1\ sr}) = h d^2$ au-dessus de 100 MeV varie grossièrement comme $\sqrt{\dot E}$ pour les pulsars d'EGRET, soit $\eta \propto 1 / \sqrt{\dot E}$ \citep{Arons1996}. Cette propriété a été renforcée depuis, sur la base de considérations théoriques \citep{Watters2009}. L'efficacité de rayonnement diminue donc, à mesure que $\dot E$ croit. \'Etant donné que les pulsars ne peuvent convertir plus de 100\% de $\dot E$ en rayonnement $\gamma$, une autre implication de cette loi est que $\eta$ doit saturer à bas $\dot E$. Une autre possibilité est qu'il existe une valeur de seuil pour l'émission $\gamma$, située en deçà de 3 $\times$ 10$^{34}$ erg/s, valeur de $\dot E$ la plus faible parmi les pulsars détectés par EGRET.

\subsubsection{Pulsars milliseconde, et cas de PSR J0218+4232}

Une recherche de pulsations en rayons $\gamma$ a été effectuée dans un lot de données de deux ans d'observations d'EGRET, pour un échantillon de 19 pulsars milliseconde \citep{Fierro1995}. Cette étude n'a permis aucune détection de pulsations en rayons $\gamma$, y compris pour J0437$-$4715, parmi les meilleurs candidats car très proche et de $\dot E$ comparable à ceux des jeunes pulsars $\gamma$ détectés par EGRET\footnote{En réalité, l'effet Shklovskii n'avait pas été pris en compte. Dans le cas de PSR J0437$-$4715, la valeur de $\dot E$ intrinsèque est inférieure à la valeur apparente d'un facteur quatre.}. Cependant, les limites supérieures à 3 $\sigma$ sur les flux observés étaient bien supérieures aux valeurs de flux prédites (\emph{cf.} Table 4 de \citet{Fierro1995}).

En effectuant l'analyse de la distribution spatiale des évènements détectés par EGRET autour de la source du troisième catalogue 3EG J0222+4253, il est apparu que la position de la source entre 100 et 300 MeV était compatible avec celle du pulsar J0218+4232, découvert peu de temps auparavant \citep{Kuiper2000,Navarro1995}, après la fin de la mission EGRET. Au-delà de 1 GeV, le blazar 3C66A situé à une distance angulaire de 1$^\circ$ du pulsar milliseconde est la contrepartie évidente pour l'émission $\gamma$, créant la confusion spatiale. L'analyse temporelle des photons entre 100 MeV et 1 GeV a quant à elle abouti à un signal de significativité 3,5 $\sigma$, avec un profil à deux pics séparés de 0,45, analogues à ceux des pulsars normaux. Plus tard, Kuiper et al. (2002) ont réanalysé les données EGRET conjointement avec des observations aux rayons X du télescope Chandra \citep{Kuiper2002}. Les courbes de lumière radio, X et $\gamma$ ainsi obtenues sont montrées dans la Figure \ref{Kuiper0218}. L'alignement apparent entre les différentes longueurs d'onde leur permit d'accroître la significativité du signal $\gamma$ de 3,5 à 4,9 $\sigma$, grâce à des considérations statistiques basées sur l'alignement apparent entre pics $\gamma$ et X. 

\fig[Le pulsar J0218+4232, observé par Chandra et EGRET]{width=8cm}{chap2/Kuiper0218.pdf}{Courbes de lumière de PSR J0218+4232 à différentes énergies. \textbf{a}: profil radio, à 610 MHz. \textbf{b}: observations en rayons X entre 0,08 et 10 keV, avec Chandra. \textbf{c}: phasogramme $\gamma$ obtenu avec EGRET, entre 100 MeV et 1 GeV. Figure issue de \citet{Kuiper2002}.}{Kuiper0218}

\subsection{Observations de pulsars par le télescope AGILE}

Le télescope AGILE (\emph{Astro-rivelatore Gamma ad Immagini LEggero}) a été mis en orbite en avril 2007, soit un peu plus d'un an avant Fermi. Il a depuis observé le ciel $\gamma$ entre 30 MeV et 30 GeV, avec une sensibilité comparable à celle d'EGRET mais un champ de vue plus large ($\sim$ 2,5 sr) et une meilleure résolution angulaire. AGILE a dans un premier temps observé certains pulsars $\gamma$ connus préalablement : Vela, Crabe, Geminga et B1706$-$44, obtenant ainsi des courbes de lumière $\gamma$ plus détaillées que celles d'EGRET \cite{Pellizzoni2009bis}. Par la suite et pendant les premiers mois d'activité de Fermi, de nouvelles détections fermes de pulsars émetteurs $\gamma$ ont été réalisées grâce à AGILE : PSR J2021+3651 \citep{Halpern2008} et J2229+6114 \citep{Pellizzoni2009}, en plus de quelques détections marginales : J1016$-$5857, J1357$-$6429, J1513$-$5908, J1524$-$5625, J2043+2740, et le pulsar milliseconde appartenant à l'amas globulaire M28, J1824$-$2452A (nous reviendrons sur cette détection marginale à la fin de ce manuscrit). Comme pour EGRET, on s'aperçoit que ces pulsars comptent parmi ceux qui perdent le plus d'énergie cinétique rotationnelle par freinage électromagnétique. 

Aucun pulsar milliseconde ne figurait donc parmi les détections formelles d'EGRET et AGILE aux rayons $\gamma$ au moment des premières observations du ciel par le LAT. Cependant, la détection probable de PSR J0218+4232 par EGRET faisait de ce pulsar une cible prioritaire pour le LAT, tout comme J1824$-$2452A, possible détection d'AGILE. Ces observations suggèrent la possibilité que d'autres pulsars de cette catégorie soient des sources émettrices de rayons $\gamma$.

\subsection{Quels pulsars pour Fermi ?}

Empiriquement, les neuf pulsars détectés par EGRET, COMPTEL et AGILE ainsi que les différentes détections à faible significativité ont en commun de présenter des taux de perte d'énergie par freinage électromagnétique supérieurs à $3 \times 10^{34}$ erg/s \citep{Thompson1999,Pellizzoni2009}. Cependant, la géométrie des faisceaux d'émissions $\gamma$ étant mal connue, la relation entre la valeur de $\dot E$ et le flux $\gamma$ attendu, donc la détectabilité du pulsar, est très incertaine. Pour cette raison, une liste de cibles pour Fermi a été définie en retenant les pulsars vérifiant :

\begin{eqnarray}
\dot E > 10^{34} \mathrm{erg/s}
\end{eqnarray}

Cette coupure conduit à une liste de 230 pulsars\footnote{\`{A} nouveau, ce sont les valeurs apparentes de $\dot P$ qui ont été utilisées.} \citep{PapierTiming}. Les pulsars des amas globulaires pouvant avoir une valeur de $\dot E$ apparente exagérée à cause d'accélérations dans des potentiels gravitationnels locaux, certains ont été exclus, amenant à une liste de 218 pulsars à suivre à l'aide des différents radiotélescopes de par le monde, et les télescopes X pour les pulsars non émetteurs radio (au nombre de cinq dans cette liste : Geminga, J1811$-$1925, J1846$-$0258, J0540$-$6919 et J0537$-$6910).\label{campagne}

Enfin, un critère a été défini dans le but d'accorder une priorité plus ou moins importante au suivi des pulsars. On peut définir un « flux dû au freinage » comme le flux d'une source de luminosité $\dot E$ située à une distance $d$, soit $\dot E / d^2$. Cette quantité représente en quelque sorte le flux maximal pouvant être émis sur l'ensemble du spectre électromagnétique. On a vu précédemment que pour les pulsars d'EGRET, la luminosité $\gamma$ au-dessus de 100 MeV semble croître comme $\sqrt{\dot E}$, soit $\eta = L_\gamma / \dot E \propto \sqrt{\dot E}$. On peut donc pondérer le flux dû au freinage par l'efficacité d'émission en rayons $\gamma$ empirique, et obtenir un flux $\gamma$ effectif : 

\begin{eqnarray}
F_{\mathrm{eff}} = \eta \frac{\dot E}{d^2} = \frac{\sqrt{\dot E}}{d^2}
\label{Feff}
\end{eqnarray}

Pour les pulsars ayant des valeurs de $\dot E$ équivalentes, la priorité est donc donnée aux plus proches. Selon ce classement, les six pulsars d'EGRET se classent parmi les 20 premiers pulsars, intercalés avec quelques objets découverts après la fin de l'activité du télescope $\gamma$. Néanmoins quelques pulsars milliseconde pour lesquels EGRET n'avait pas détecté de pulsations, PSR J0034$-$0534 et J0437$-$4715, se classent parmi les dix premiers, soulignant à nouveau le fait que $F_{\mathrm{eff}}$ ne prend pas en compte tous les paramètres, notamment la géométrie d'émission variable. La Figure \ref{sqrtEdotd2} montre, pour les pulsars recensés par le catalogue ATNF, la valeur de $F_{\mathrm{eff}}$, normalisée à la valeur de Vela, en fonction de leur période de rotation. Les pulsars cibles pour Fermi sont indiqués par des carrés et des triangles. Par construction, ils sont généralement distribués aux grandes valeurs de $\sqrt{\dot E} / d^2$.

\fig[Flux de freinage effectif pour les pulsars du catalogue ATNF]{width=14cm}{chap2/sqrtEdotd2.pdf}{Flux de freinage effectif $\sqrt{\dot E}/d^2$ pour les pulsars du catalogue ATNF. Les flux sont normalisés à celui du pulsar de Vela, 3,13 $\times$ 10$^{19}\ \mathrm{(erg/s)}^{1/2}/d^2$. Les pulsars de valeur de $\dot E$ comprise entre 10$^{34}$ erg/s et 3 $\times 10^{34}$ erg/s sont indiqués par des triangles, ceux vérifiant $\dot E > 3 \times 10^{34}$ erg/s sont montrés par des carrés vides. Les carrés pleins sont les pulsars détectés par EGRET, COMPTEL et AGILE, ainsi que les détections marginales.}{sqrtEdotd2}

\section{Pourquoi étudier les pulsars en rayons $\gamma$ ?}

Dans la première partie de ce chapitre nous avons vu que différents modèles sont en compétition pour l'explication de l'émission $\gamma$ des pulsars : \emph{Polar Cap}, \emph{Slot Gap} et \emph{Outer Gap} notamment. Chacun d'entre eux propose une région particulière de la magnétosphère des pulsars comme point de départ du rayonnement. On a vu par exemple que les modèles prévoient des coupures spectrales de formes différentes, des morphologies d'émission $\gamma$ différentes et des écartements plus ou moins importants avec la composante radio. Avec un échantillon plus large de pulsars émetteurs $\gamma$, on peut établir des tendances dans les profils d'émission ainsi que dans les spectres, et ainsi placer des contraintes sur la géométrie de l'émission dans la magnétosphère. L'augmentation du nombre de pulsars connus permet donc de contraindre, de raffiner voire d'éliminer les théories proposées.

Les télescopes prédécesseurs du LAT, et en particulier EGRET, ont révélé que les pulsars émetteurs $\gamma$ sont ceux qui perdent le plus d'énergie par freinage électromagnétique. Mais EGRET n'a pas permis de comprendre combien l'émission en $\gamma$ est un phénomène général pour les pulsars à grand $\dot E$. En particulier, on peut s'attendre à ce que d'autres paramètres entrent en jeu, l'âge ou l'intensité du champ magnétique en surface ou au cylindre de lumière. Le scénario d'évolution peut également avoir son importance : existe-t-il des pulsars milliseconde émetteurs de rayons $\gamma$ ? Si c'est le cas, le processus d'émission est-il commun aux populations « jeunes » et « recyclées » ? Ces questions ont leur importance pour les synthèses de population \citep{Gonthier2002,Story2007}, qui cherchent à déterminer le nombre de pulsars dans la Galaxie et leurs propriétés. Enfin, la non-détection de pulsars à grandes valeurs de $\dot E$ est tout aussi importante dans la mesure où elle renseigne sur la géométrie des faisceaux aux différentes longueurs d'onde.

Enfin, si l'étude des pulsars en rayons $\gamma$ touche aux modèles d'émission théoriques et à l'évolution des pulsars dans un contexte plus général, il est bien évident que l'identification de pulsars $\gamma$ peut permettre la découverte et l'étude de sources avoisinantes. De même, on s'attend à ce que les pulsars éloignés, non détectables mais tout de même émetteurs, contribuent à l'émission diffuse qui est source de bruit pour beaucoup d'études. La contribution des pulsars à l'émission diffuse est encore mal connue, mais des modèles prédisent une contribution importante, en particulier par les pulsars milliseconde (voir par exemple \citet{Faucher2009,Wang2005}). Les raisons mentionnées ici sont autant de motivations pour la recherche de pulsars $\gamma$ avec le Large Area Telescope à bord de Fermi.

%% file: part2/chap3.tex
\chapter[Chronométrie des pulsars]{Chronométrie des pulsars}

\minitoc

\chaptabst{L}{a}{spécificité des pulsars est leur émission cadencée, par conséquent l'analyse temporelle est centrale dans toute étude de ces objets. Pour construire les courbes de lumière des pulsars, c'est-à-dire le signal émis en fonction de la phase rotationnelle, il faut connaître la fréquence de rotation (ou de façon équivalente, la période) en fonction du temps. Connaissant cette fréquence, on peut calculer les phases pour un lot de dates donné, et ainsi accéder à la morphologie de l'émission. La mesure des paramètres de rotation, comme par exemple la fréquence $f_0$ et ses premières dérivées, constitue la « chronométrie » des pulsars, que nous détaillons dans la deuxième partie de ce chapitre. Elle consiste en plusieurs étapes, parmi lesquelles l'enregistrement de « temps d'arrivée » de pulsations, le transfert de ces temps vers le barycentre du système solaire, et le calcul des résidus de datation. L'aboutissement de la chronométrie est la construction d'éphémérides, qui permettent de calculer la phase rotationnelle dans un intervalle de temps donné. Dans la troisième partie nous revenons sur la campagne de datation des pulsars candidats à la détection en rayons $\gamma$ par le LAT. En particulier, nous détaillons l'algorithme de construction et de mise à jour des éphémérides construites à partir des données du radiotélescope de Nançay. Les éphémérides ont été centralisées dans une base de données accessibles par les pulsaristes de la collaboration Fermi-LAT. Outre la chronométrie, les instruments impliqués dans la campagne de datation ont également permis la vérification des outils d'analyse temporelle mis en place pour les données du LAT, ce que nous présentons dans la quatrième partie du chapitre.}

\section{Courbes de lumière de pulsar}
\label{courbelumiere}

Pour un observateur fixe, les pulsars émettent un signal modulé par leur fréquence de rotation. Construire la courbe de lumière (ou phasogramme) d'un pulsar consiste à découper le signal accumulé pendant un certain temps d'observation en morceaux correspondant à une rotation de l'objet sur lui-même, puis les empiler. On a vu précédemment que les pulsars ralentissent, que leur taux de ralentissement varie, et ainsi de suite. Une façon d'exprimer la fréquence de rotation $f$ d'un pulsar en fonction du temps est alors d'utiliser un développement en série de Taylor, que l'on tronque à l'ordre approprié :

\begin{eqnarray}
f(t) = f_0 + \dot f_0 \times (t - T_0) + \frac{\ddot f_0}{2} \times (t - T_0)^2 + \dots
\label{eqfrequence}
\end{eqnarray}

où $f_0$, $\dot f_0$ et $\ddot f_0$ sont la fréquence de rotation et ses premières dérivées à une date de référence $T_0$. Pour un signal continu puis échantillonné, tel qu'en radio, ou un signal discret reçu sous la forme de photons espacés dans le temps, on dispose de dates pour lesquelles on souhaite connaître la phase rotationnelle. D'après l'équation \ref{eqfrequence}, cette phase est donnée par :

\begin{eqnarray}
\Phi(t) = \Phi_0 + f_0 \times (t - T_0) + \frac{\dot f_0}{2} \times (t - T_0)^2 + \frac{\ddot f_0}{6} \times (t - T_0)^3 + \dots
\label{eqphase}
\end{eqnarray}

Dans cette expression, $\Phi_0$ est la phase (arbitraire) du pulsar à la date $T_0$. La partie entière de $\Phi(t)$ donne le nombre de tours du pulsar effectués entre les temps $T_0$ et $t$. La partie fractionnaire indique le stade de la rotation, au cours d'un tour. 

Pour construire la courbe de lumière il faut donc connaître $f_0$ et un certain nombre de ses dérivées avec précision. Le besoin de précision est d'autant plus important que le signal est faible : plusieurs semaines d'observations étaient nécessaires à EGRET pour détecter certains pulsars en rayons $\gamma$, ce qui pouvait représenter plusieurs millions à milliards de rotations des étoiles sur elles-mêmes. Comme nous le verrons dans la suite de ce chapitre, une connaissance approximative des paramètres de rotation a pour conséquence le décalage progressif du profil ou son étalement, empêchant \emph{in fine} la détection du pulsar.

Pour cette thèse, consacrée à la recherche de pulsars milliseconde émetteurs $\gamma$ dans les données de Fermi, le besoin de précision sur $f(t)$ est donc crucial. Deux stratégies sont envisageables pour rechercher les $f_0$, $\dot f_0$, $\ddot f_0$, \emph{etc}. qui permettront éventuellement la détection en rayons $\gamma$ :

\begin{itemize}
\item Une technique consiste à rechercher les paramètres optimaux dans le signal lui-même, en essayant un certain nombre de valeurs de $f_0$ et de ses dérivées (et éventuellement d'autres paramètres, par exemple orbitaux si l'on recherche des systèmes binaires), à la recherche de la configuration qui optimise la significativité du signal. Cette technique « à l'aveugle » a été mise en oeuvre pour la recherche de pulsars dans les données de Fermi, avec succès \citep{FermiCTA1,Ziegler2008}. Plus généralement, elle est surtout employée dans le domaine radio, domaine dans lequel presque tous les pulsars connus ont été découverts.
\item L'autre méthode, mise à l'oeuvre dans le cadre de cette thèse, est la « chronométrie » des pulsars, dans les domaines d'énergie où le signal est plus abondant (essentiellement en radio). Cette technique est présentée dans la suite de ce chapitre.
\end{itemize}

\section{Principe de la chronométrie}

Le but de la chronométrie (ou datation) d'un pulsar est la construction d'un modèle décrivant la rotation de l'étoile en fonction du temps avec précision et dans un intervalle de temps donné. Pour cela, on commence par observer le pulsar à plusieurs reprises, et on détermine pour chaque observation un « temps d'arrivée » (\emph{time of arrival}, TOA), à savoir, une date à laquelle l'objet était à un stade donné de sa rotation (par exemple, le point où l'émission est maximale). Disposant de $N$ dates auxquelles la phase du pulsar était identique, on recherche alors une fonction\footnote{Il n'y a pas nécessairement unicité de la solution.} qui reproduit au mieux la rotation du pulsar.

\subsection{Temps d'arrivée de pulsations}
Si la morphologie de l'émission peut varier fortement d'une rotation à l'autre \citep{Xilouris1999}, l'expérience montre que des profils accumulés sur une certaine durée sont des empreintes stables. Ces profils intégrés $f_i$ peuvent être décrits par la relation :

\begin{eqnarray}
f_i(\phi) = a + b \times F (\phi^\prime) + c(\phi)
\end{eqnarray}

Dans cette équation, $a$ et $b$ sont des constantes, $c$ est une fonction représentant la contribution du bruit, et $F$ est un « profil idéal » de l'émission du pulsar, décalé de $\Delta \phi = \phi^\prime - \phi$. L'enregistrement de TOAs revient à déterminer le déphasage $\Delta \phi_i$ pour chaque observation $f_i$, puis connaissant les dates des $N$ observations, en déduire les dates $t_i$ pour lesquelles le pulsar était au même point de sa rotation. 

Plusieurs choix sont possibles pour la définition du profil de référence $F$. Une solution simple est un pic de Dirac, auquel on aligne un point arbitraire des observations individuelles $f_i$, par exemple le maximum de l'émission. Avec cette solution on omet cependant la morphologie de l'émission du pulsar. On peut créer le profil $F$ en modélisant une observation par une somme de gaussiennes d'amplitudes et de phases centrales différentes. On peut également lisser une observation individuelle, c'est-à-dire réduire la contribution du bruit. Enfin, on peut créer un profil idéal de manière itérative, en alignant $N$ observations entre elles grâce à un modèle de datation obtenu au préalable, et en les additionnant pour améliorer le rapport signal-sur-bruit.

\subsection{Modèle de datation}

Les TOAs déterminés précédemment sont généralement topocentriques, c'est-à-dire enregistrés par un observatoire en mouvement par rapport à la source de l'émission, et dans une échelle de temps variant de façon irrégulière par rapport au temps propre du pulsar. Le barycentre du système solaire est, en première approximation, en mouvement inertiel par rapport aux pulsars de la Galaxie ; on choisit donc de convertir les dates  « topocentriques » ($t_i$) en temps « barycentriques » ($T_i$). L'équation de transfert reliant les ($t_i$) et ($T_i$) est la suivante :

\begin{eqnarray}
T = t + \Delta_C - \Delta_D + \Delta_{R,\odot} + \Delta_{E,\odot} - \Delta_{S,\odot} + \Delta_B
\label{bary}
\end{eqnarray}

Dans cette équation, $\Delta_C$ représente les corrections d'horloges à apporter entre l'appareil de mesure, qui enregistre des temps dans une certaine échelle, UTC (\emph{Temps Universel Coordonné}), TAI (\emph{Temps Atomique International}) ou TT (\emph{Temps Terrestre}), et une échelle standard proche du temps propre du pulsar, corrigée des irrégularités dues à la rotation de la Terre. L'utilisation de l'échelle TT, liée au temps atomique et dont l'unité est la seconde SI, est préconisée pour cette dernière. Plus de détails sur les définitions de systèmes temps-coordonnées sont disponibles dans \citet{Andersen1999,Rickman2001,McCarthy2004}

Le membre $\Delta_D$ est un terme dispersif. Dans le milieu interstellaire, les rayons lumineux sont dispersés par les électrons présents sur leur parcours. Le terme dominant est :

\begin{eqnarray}
\Delta_D = \dfrac{DM}{k \times f_\mathrm{SSB}^2}
\label{EqDispersion}
\end{eqnarray}

où $DM$ est la « mesure de dispersion », correspondant à la densité de colonne d'électrons pour une direction donnée, $f_\mathrm{SSB}$ est la fréquence d'observation et $k = 2,410 \times 10^{-4}$ MHz$^{-2}$ cm$^{-3}$ pc s$^{-1}$ est la constante de dispersion \citep{ManchesterTaylor1977}. Notons que pour un observateur situé sur Terre, la fréquence d'observation effective varie au cours de l'année, principalement à cause de l'effet Doppler dû au mouvement de la Terre. La fréquence $f_\mathrm{SSB}$ de l'expression précédente est par conséquent celle d'un observateur situé au centre de masse du système solaire. 

\`{A} cause de cet effet dispersif, des photons de fréquences différentes mais émis simultanément arrivent en décalage (une illustration de l'effet du terme $\Delta_D$ est montré en Figure \ref{disp}). \`{A} l'inverse, il est crucial de maîtriser cette correction si l'on veut mesurer l'écart temporel entre des pics émis à des énergies différentes, ces pics pouvant être réellement décalés dans le temps et donc signifier des origines différentes dans la magnétosphère du pulsar. Corriger de cet effet revient à passer à une observation faite à fréquence infinie. Remarquons enfin que la mesure de dispersion $DM$ peut varier au cours du temps, pour les cas où la densité de colonne d'électrons sur la ligne de visée évolue \citep{IsaacmanRankin1977,Ramachandran2006}. Dans ce cas on peut ajouter à \ref{EqDispersion} des dérivées temporelles $\dot{DM}$, $\ddot{DM}$, \emph{etc}. La Figure \ref{DM1824} illustre l'effet d'une variation de la mesure de dispersion.

\fig[Illustration du décalage temporel induit par la dispersion du signal radio]{width=12cm}{chap3/dispersion.pdf}{Effet de la dispersion interstellaire, sur une bande de fréquence de 288 MHz centrée à 1380 MHz. Les pulsations arrivent à l'observatoire avec un décalage en temps, variant comme l'inverse du carré de la fréquence. Figure extraite de \citet{Handbook}.}{disp}

\fig[Effet d'une variation de la mesure de dispersion $DM$]{width=12cm}{chap3/0285fig2.pdf}{Résidus de chronométrie pour le pulsar milliseconde B1821$-$24 observé par le radiotélescope de Nan\c{c}ay. Les croix indiquent des TOAs enregistrés à 1,4 GHz, les carrés sont des observations à 2 GHz. Pour ce graphe une mesure de dispersion constante a été utilisée. Les TOAs enregistrés aux deux fréquences d'observations s'écartent. Dans ce cas il est nécessaire de tenir compte d'une variation de la mesure de dispersion, $\dot{DM}$, dans le modèle de datation. Figure issue de \citet{PapierTiming}.}{DM1824}

Le membre $\Delta_R$ est appelé terme de \emph{Roemer}, et contient le temps de trajet classique de la lumière, entre l'instrument de détection et le centre de masse du système solaire. On peut le décomposer de la manière suivante :

\begin{eqnarray}
\Delta_R = - \dfrac{\left( \vec{r_{BT}} + \vec{r_{Tt}} \right) \vec{r_{BP}}}{c}
\label{EqRoemer}
\end{eqnarray}

où $\vec{r_{BT}}$ est un vecteur pointant du barycentre du système solaire vers le centre de la Terre, $\vec{r_{Tt}}$ relie le géocentre et le télescope, et $\vec{r_{BP}}$ est un vecteur unitaire donnant la direction du pulsar, depuis le centre de masse du système solaire. Pour calculer ces vecteurs, on a recours aux « éphémérides planétaires » telles que DE200 \citep{Standish1990} et DE405 \citep{Standish1998} publiées par le \emph{Jet Propulsion Laboratory}\footnote{http://www.jpl.nasa.gov}. Ces éphémérides donnent la position des principaux astres du système solaire en fonction du temps. Le terme de Roemer est largement dominant dans le transfert de dates vers le barycentre. Sa magnitude est approximativement $(\mathrm{1\ U.A.}) \times \cos(\beta) / c$, où $\beta$ est la latitude écliptique du pulsar. Pour un pulsar situé dans le plan de l'écliptique (c'est-à-dire $\beta = 0$), l'effet est maximal et son amplitude vaut environ 500 s. 

Certains pulsars sont animés d'un mouvement apparent dans le plan du ciel, le mouvement propre. L'ascension droite et la déclinaison varient au cours du temps, modifiant ainsi le vecteur $\vec{r_{BP}}$. Si le mouvement du pulsar n'est pas pris en compte, le TOA au barycentre est erroné ; l'écart avec le temps d'arrivée « vrai » correspond à l'erreur positionnelle, modulée par l'orbite terrestre.

Enfin, l'effet de parallaxe de chronométrie\footnote{Différent de la parallaxe classique.}, induit par la variation de la courbure du front d'onde de l'émission du pulsar au cours de l'orbite terrestre, est visible pour certains pulsars proches (l'équation \ref{EqRoemer} fait en effet l'approximation d'une onde plane). Son amplitude est donnée par $(\mathrm{1\ U.A.})^2 \times \cos(\beta) / (2cd)$, et donne donc accès à la distance $d$ du pulsar, avec une grande précision dans certains cas \citep{Lommen2006,Hotan2006}.

Les termes $\Delta_{E,\odot}$ et $\Delta_{S,\odot}$ sont des corrections relativistes. Ce dernier, l'effet \emph{Shapiro}, correspond aux retards induits par la déformation de l'espace-temps à proximité des masses du système solaire \citep{Shapiro1964}. Il est le plus intense pour les photons passant près du Soleil (l'effet est typiquement de 100 $\mu$s). La contribution des planètes du système solaire varie de quelques ns à quelques centaines de ns. L'effet \emph{Einstein}, représenté par le terme $\Delta_{E,\odot}$, correspond à la dilatation de l'espace-temps au cours du mouvement de la Terre dans le potentiel gravitationnel du système solaire \citep{Backer1986}. 

Pour finir, le terme $\Delta_B$ correspond aux corrections à prendre en compte pour les pulsars appartenant à des systèmes multiples. Ceux-ci sont animés d'un mouvement orbital, ce qui se traduit en des effets Roemer, Shapiro et Einstein locaux, analogues à ce qui concerne le système solaire. Plusieurs modèles permettent la description de ces orbites, selon leur excentricité ou leur complexité, selon le nombre de compagnons ou leur masse. Par exemple, le modèle BT (Blandford-Teukolsky) qui est adapté aux orbites lentes obéissant à la mécanique newtonienne, DD (Damour-Deruelle) qui est une extension du modèle BT convenant aux systèmes en champ fort, ou encore ELL1, adapté aux orbites quasi-circulaires \citep{TaylorWeisberg1989}. Ils mettent en jeu les paramètres Képleriens classiques : période orbitale $P_b$, demi grand-axe projeté $x$, excentricité $e$, longitude du périastre $\omega$ et temps de passage au périastre $T_0$, ainsi que des paramètres « post-képleriens », pour les théories le permettant.

\subsection{\'Ephéméride de pulsar et signatures de résidus}

On a vu dans le paragraphe \ref{courbelumiere} que l'on peut écrire la fréquence de rotation $f$ sous la forme d'un développement en série de Taylor tronqué, autour d'une date de référence $T_0$, avec ($f_0$, $\dot{f}_0$, $\ddot{f}_0$, \dots) la fréquence de rotation et ses dérivées à cette date (\emph{cf.} équation \ref{eqfrequence}). Construire une éphéméride pour un pulsar consiste à rechercher des paramètres de rotation ($f_0$, $\dot{f}_0$, $\ddot{f}_0$, \dots), des paramètres astrométriques (ascension droite $\alpha$, déclinaison $\delta$, mouvement propre $\mu_\alpha = \dot{\alpha}\ \cos(\delta)$ et $\mu_\delta = \dot{\delta}$, parallaxe $\pi = 1/d_\mathrm{kpc}$), et des paramètres orbitaux le cas échéant, permettant de reproduire au mieux la rotation du pulsar observée au télescope. 

Pour y parvenir, on se donne un lot de TOAs enregistrés à l'observatoire, $t_i$, que l'on convertit en temps barycentriques, $T_i$, d'après \ref{bary}. Pour chaque $T_i$, on calcule la phase $\Phi(T_i)$ prédite par le modèle que l'on souhaite améliorer, à l'aide de l'équation \ref{eqphase}. Une éphéméride optimale minimise les « résidus » de chronométrie, donnés par :

\begin{eqnarray}
R = P \times \sum_{i=1}^{N}{\frac{\Phi(T_i) - n_i}{\sigma(T_i)}}
\end{eqnarray}

Dans cette expression, $\sigma(T_i)$ désigne le poids attribué à $T_i$ (en effet, l'incertitude sur la mesure des TOAs étant variable, on peut choisir des pondérations différentes) et $n_i$ est l'entier le plus proche de $\Phi(T_i)$.

En pratique, on débute l'analyse des temps d'arrivée avec une éphéméride minimale, donnant une connaissance approximative de la phase en fonction du temps. De manière itérative, on optimise les paramètres du modèle un à un, en fonction de la forme des résidus de datation. Plus précisément, un paramètre mal adapté donne lieu à une signature caractéristique dans le graphe des résidus. La Figure \ref{0030_POS_PM} illustre quelques exemples de signatures, pour le pulsar J0030+0451, observé avec le radiotélescope de Nan\c{c}ay sur quatre ans. Dans l'exemple (A), la fréquence de rotation est faussée de $10^{-8}$ Hz. \`{A} cause de cette erreur sur la fréquence du pulsar, l'écart entre phases prédites et observées grandit linéairement dans le temps. Dans ce graphe on constate un saut d'une rotation ($\sim$ 4,9 ms) vers MJD 53900, signifiant que le modèle de chronométrie diffère d'un tour complet par rapport à la rotation réelle. La Figure (B) correspond à une erreur sur $\dot f_0$ de $5 \times 10^{-17}$ Hz$^2$, induisant un écartement en parabole. Les cas (C) et  (D) correspondent à des erreurs sur la position du pulsar. Pour le premier cas, la position du pulsar dans le modèle de chronométrie est erronée de 1 arcseconde. La signature est une sinuso\"ide annuelle, due à l'orbite terrestre. Pour les pulsars ayant un mouvement propre, comme PSR J0030+0451, on peut observer des résidus tels que ceux montrés dans l'exemple (D) : ici on n'a utilisé aucun mouvement propre dans l'éphéméride. La signature correspondante est une sinuso\"ide annuelle dont l'amplitude augmente linéairement avec le temps. 

\sfig[Exemples de signatures de résidus caractéristiques]{chap3/TimingCorrections.tex}{Exemples de signatures de résidus caractéristiques, pour le pulsar J0030+0451 observé à Nan\c{c}ay.}{0030_POS_PM}

Ayant amélioré le modèle de fa\c{c}on itérative, on parvient à une éphéméride optimale, et un graphe de résidus semblable à celui montré en Figure \ref{0030Residus}. Ici les résidus sont répartis aléatoirement autour de 0, et la déviation standard, ici de 4,3 $\mu$s, est du même ordre de grandeur que la précision moyenne sur la mesure des TOAs individuels. L'éphéméride correspondante est donnée dans la Table \ref{Ephem0030}.

\fig[Exemple de résidus de chronométrie optimaux]{width=12cm}{chap3/0030_TOAres.pdf}{Résidus de chronométrie optimaux, pour le pulsar J0030+0451 observé au radiotélescope de Nan\c{c}ay. La déviation standard de ces résidus est de 4,3 $\mu$s, l'incertitude moyenne sur les TOAs individuels étant de 3,5 $\mu$s. L'éphéméride correspondante est donnée en Table \ref{Ephem0030}.}{0030Residus}

\begin{table}[bt]
\begin{center}
\begin{minipage}[b][1.05\totalheight][c]{16cm}
\begin{center} 
\input{tableaux/Ephem0030.tex}
\parbox{15cm}{\caption[Exemple d'éphéméride de pulsar]{Exemple d'éphéméride de pulsar, ici pour PSR J0030+0451. Les nombres donnés entre parenthèses sont les incertitudes à 1 $\sigma$ sur les dernières décimales citées.}\label{Ephem0030}}\end{center}
\end{minipage}
\end{center}
\end{table}

Les résidus de datation accessibles pour les pulsars milliseconde sont généralement dépourvus de bruit, à l'instar de ce qui est présenté en Figure \ref{0030Residus} pour PSR J0030+0451. La plupart des pulsars milliseconde sont extrêmement stables et ne produisent pas de \emph{glitches} ou de \emph{timing noise}, au contraire des pulsars jeunes, beaucoup plus bruités. Les \emph{glitches} sont des augmentations soudaines de la fréquence de rotation, probablement provoqués par une modification de la structure de l'étoile à neutrons. Un quart environ des pulsars suivis dans le cadre de la campagne de chronométrie (\emph{cf.} \ref{campagne}) sont connus pour avoir produit des \emph{glitches} \citep{Melatos2008}. Le \emph{timing noise} est la variation aléatoire de la fréquence de rotation ou d'une de ses dérivées, avec une pseudo-périodicité comprise entre quelques jours à quelques mois voire années. Des illustrations de \emph{timing noise} sont données dans \citet{Hobbs2006}. Ces deux comportements perturbent la précision des modèles de datation. Pour les \emph{glitches}, il faut incrémenter la fréquence pré-\emph{glitch} d'une certaine quantité pour pouvoir suivre la rotation à nouveau. Dans le cas du \emph{timing noise}, on peut lisser les résidus grâce à un grand nombre de dérivées de la fréquence. Une autre possibilité, offerte par le logiciel TEMPO2, est d'ajuster la marche aléatoire due au \emph{timing noise} par un développement en sinuso\"ides\footnote{\emph{cf.} fonctionnalité \emph{FITWAVES} de TEMPO2}.\label{labeltimingnoise}

\section{Campagne de datation}

\subsection{Couverture des pulsars}

Comme on l'a déjà évoqué, une campagne de chronométrie de pulsars prometteurs a été mise en place dans les mois précédant le lancement de Fermi \citep{PapierTiming}. Des radiotélescopes parmi les principaux dans le monde ont participé : le télescope de 64 m de Parkes, en Australie \citep{Manchester2001}, le télescope Lovell de 76 m à Jodrell Bank, en Angleterre \citep{Hobbs2004}, le radiotélescope de 110 m de Green Bank aux \'Etats-Unis \citep{Kaplan2005}, et la station de Nançay en France \citep{Theureau2005}, télescope méridien équivalent à une antenne de 94 m (\emph{cf.} Figure \ref{Nancay}). En outre, des contributions ponctuelles ont été apportées par les télescopes de Westerbork aux Pays-Bas \citep{Voute2002} et Arecibo, à Porto Rico \citep{Dowd2000}. 

\fig[Le radiotélescope de Nançay]{width=12cm}{chap3/RadioTelescope_Nancay.jpg}{Le radiotélescope de Nançay (région Centre) vu du ciel. Le miroir visible en arrière-plan (200 m $\times$ 40 m) joue le rôle de premier réflecteur. C'est un miroir plan et orientable, au contraire du second réflecteur au premier plan (300 m $\times$ 35 m), fixe et de forme sphérique. Les miroirs sont séparés de 460 m. Ce second réflecteur concentre les ondes incidentes vers le point focal, au centre. Le point visé dans le ciel se déplace au cours du temps à cause de la rotation de la Terre, et par conséquent le point focal se déplace. Pour cette raison, les antennes sont disposées dans un « chariot focal » (on peut le distinguer au centre de l'image) pouvant se déplacer sur des rails. Grâce à ce mouvement, le radiotélescope de Nançay peut couvrir à tout moment $\pm \frac{1}{2}^h$ en ascension droite. Le chariot focal embarque deux récepteurs observant dans les gammes de fréquence 1,1 à 1,8 GHz et 1,7 à 3,5 GHz, permettant de couvrir la raie de l'hydrogène HI et au-delà. L'ensemble forme un télescope méridien, équivalent à une antenne simple de 94 m de diamètre.}{Nancay}

Les radiotélescopes cités précédemment permettent de couvrir l'essentiel des pulsars connus. Pour certains pulsars ne produisant pas ou peu d'émission détectable dans le domaine radio, mais détectables en rayons X, la chronométrie a été réalisée à l'aide des satellites RXTE\footnote{http://heasarc.gsfc.nasa.gov/docs/xte/XTE.html} et XMM\footnote{http://xmm.esac.esa.int/}. 

Ces instruments ont couvert prioritairement les 218 pulsars considérés comme prometteurs. Ce suivi des pulsars candidats à la détection par Fermi est en fait similaire aux travaux réalisés auparavant, dans le contexte de la mission CGRO \citep{Arzoumanian1994,Johnston1995,DAmico1996} à la différence que Fermi fonctionne principalement en mode de balayage, et non de pointé. Cela signifie que si pour CGRO, la durée et la direction des pointés définissaient la liste des pulsars à observer et la fréquence des observations nécessaire dans un temps limité, un suivi continu et homogène est requis pour Fermi qui balaye le ciel en permanence. Une autre différence est bien évidemment la différence de performance entre EGRET et le LAT. Le LAT, plus sensible, a accès à davantage de sources.

Les périodes d'observation prolongées favorisent l'apparition de \emph{timing noise} dans les résidus de chronométrie des pulsars jeunes et bruités, généralement à grande valeur de $\dot E$ et donc bons candidats à la détection en rayons $\gamma$. La conséquence est que la bonne connaissance de la phase en fonction du temps requiert plus de complexité dans le modèle de chronométrie. Une illustration de ce besoin de complexité est le cas du pulsar du Crabe, quotidiennement observé à l'aide d'un des instruments de la station de Jodrell Bank près de Manchester. Avec ces observations des éphémérides sont construites chaque mois, de telle sorte que leur période de validité ne couvre que la durée du mois en question\footnote{\emph{cf.} http://www.jb.man.ac.uk/~pulsar/crab.html}. Sauf exceptions, un simple développement de la fréquence de rotation en fonction du temps en $f_0$, $\dot f_0$ et $\ddot f_0$ suffit à assurer une précision de l'éphéméride de quelques dizaines de $\mu$s. Lorsque la plage d'observation du pulsar du Crabe se prolonge, le \emph{timing noise} peut devenir prépondérant et imposer l'utilisation de termes d'ordre supérieur sur la fréquence de rotation, voire un développement en sinusoïdes. La Figure \ref{ResCrabeBad} (A) montre les résidus obtenus pour 697 TOAs enregistrés conjointement à l'aide des radiotélescopes de Jodrell Bank et de Nançay, entre le 20 juin 2008 et le 9 avril 2009, avec un modèle de rotation restreint à $f_0$, $\dot f_0$ et $\ddot f_0$. Le \emph{timing noise} domine, l'excursion en résidus s'étendant sur 2 ms environ, soit plus de 5\% de la période de rotation du pulsar. Dans la Figure \ref{ResCrabeBad} (B) on a lissé ce bruit oscillant par un développement en harmoniques. Cette fois l'amplitude typique des résidus de chronométrie est de 50 $\mu$s, de l'ordre du millième de rotation du pulsar.

\sfig[Le phénomène de \emph{timing noise} pour le pulsar du Crabe]{chap3/TimingNoiseCrabe.tex}{Résidus de chronométrie pour le pulsar du Crabe, observé à 697 reprises entre le 20 juin 2008 et le 9 avril 2009 par les radiotélescopes de Nançay et de Jodrell Bank. (A) : seule la fréquence de rotation du pulsar et ses deux premières dérivées ont été ajustées. (B) : le modèle de datation contient en plus un développement en harmoniques de la phase rotationnelle.}{ResCrabeBad}

La Figure \ref{Campagne} dresse un bilan de la couverture des pulsars dans le cadre de la mission Fermi, jusqu'à mai 2009. Les pulsars définis \emph{a priori} comme prioritaires sont indiqués par des points épais. Rappelons que le seuil à $\dot E = 10^{34}$ erg/s qui avait été fixé ne tenait pas compte de l'effet Shklovskii, ce qui explique que certains d'entre eux se trouvent en deçà de la ligne pointillée. Les carrés vides indiquent les pulsars qui n'étaient pas considérés comme prioritaires, mais pour lesquels des paramètres de rotation ont été fournis au moins ponctuellement. Le nombre de ces pulsars est de 500 environ. Les objets non couverts pendant la première année de campagne sont indiqués par des points fins. 

La centralisation des éphémérides issues de cette campagne de chronométrie dans une base de données constitue l'un des travaux effectués au cours de cette thèse. Afin de rendre accessible les paramètres de rotation aux pulsaristes $\gamma$ de la collaboration Fermi-LAT, nous avons construit une base de données php/MySQL accessible par internet\footnote{http ://www.cenbg.in2p3.fr/ephem/}. Le mode de fonctionnement est le suivant : en amont, le site internet absorbe et conserve les données provenant des radiotélescopes et télescopes X, à savoir les éphémérides et éventuellement les graphes de résidus et les profils de référence. En aval, la base de données permet à l'utilisateur côté LAT d'effectuer des requêtes : pulsar étudié, gamme de validité de l'éphéméride souhaitée, observatoire souhaitée, \emph{etc}. Puis, une fois la requête passée, le site propose à l'utilisateur de télécharger l'éphéméride voulue au format initial (propre aux télescopes radio/X) ou dans un format spécial adapté aux outils de la collaboration Fermi. Sur une base mensuelle à bimestrielle, cette base de données a été mise à jour de façon à disposer d'un suivi continu des pulsars. \emph{In fine}, les éphémérides utilisées dans des publications de Fermi sont mises à disposition libre sur un serveur du Fermi Science Support Center (FSSC)\footnote{http://fermi.gsfc.nasa.gov/ssc/data/access/lat/ephems/}.

\fig[Couverture des pulsars par la campagne de chronométrie]{width=14cm}{chap3/TimingCampaign.pdf}{Couverture des pulsars dans le cadre de la campagne de datation. Les points épais sur ce diagramme $P$ -- $\dot P$ sont les pulsars à grand $\dot E$, définis avant le lancement comme bons candidats à l'émission $\gamma$. Les carrés vides indiquent les pulsars à valeur de $\dot E$ moindre, pour lesquels des éphémérides ont tout de même été fournies au cours de la première année. Les points fins sont les pulsars non couverts.}{Campagne}

Enfin, soulignons que par le biais de la campagne de chronométrie, presque tous les pulsars milliseconde du champ galactique ont été suivis. Des éphémérides contemporaines avec les observations de Fermi n'ont pas été construites pour six des 72 objets de la liste \ref{tableauMSPsGAL} seulement ; à savoir PSR J1125$-$6014, J1216$-$6410, J1618$-$39, J1629$-$6902, J1757$-$5322 et J1933$-$6211. Cependant, ces six pulsars ont des valeurs de $\dot E$ intermédiaires et sont situés à des distances relativement grandes parmi les MSPs connus. Disposant de chronométrie pour l'essentiel des 72 pulsars milliseconde galactiques, nous avons eu l'opportunité de rechercher des pulsations parmi un échantillon de pulsars limité non pas par une hypothèse formulée \emph{a priori} sur les meilleurs candidats à l'émission, mais par la complétude des recherches de MSPs conduites dans le domaine radio.

\subsection{Analyse des données de Nançay}

En plus de la centralisation des éphémérides dans une base de données accessible par internet, la construction d'éphémérides à partir des données radio du télescope de Nançay a été un autre volet de la participation à la campagne de chronométrie au cours de cette thèse. 

Le radiotélescope de Nançay a servi à la construction d'éphémérides pour environ 150 pulsars. La procédure de mise à jour des éphémérides pour ce grand nombre de pulsars a été rendue automatique. La Figure \ref{organigramme} montre les différentes étapes de la procédure automatique d'optimisation des paramètres de rotation. On dispose initialement de l'éphéméride valide au moment de la mise à jour précédente. Si de nouvelles observations ont été réalisées pour le pulsar considéré, les profils sont alignés et additionnés. \`{A} partir des observations, nouvelles et anciennes, et utilisant comme référence le profil intégré, on calcule les temps d'arrivée. Pour ces TOAs, on calcule les résidus de chronométrie « pré-fit », correspondant à l'éphéméride initiale, ainsi que les résidus « post-fit » pour lesquels un ou plusieurs paramètres de l'éphéméride ont été ajustés. Si les résidus « pré-fit » sont meilleurs que les résidus « post-fit », il n'est pas nécessaire de mettre à jour l'éphéméride : les paramètres initiaux ajustent toujours les données de façon satisfaisante. Dans le cas contraire, la procédure est répétée jusqu'à la convergence des résidus de chronométrie. 

\fig[Algorithme de mise à jour des éphémérides à partir des données de Nançay]{width=12cm}{chap3/organigramme.pdf}{Processus de construction des éphémérides à partir des observations de pulsars faites à Nançay.}{organigramme}

\section{Vérification des outils}

Pour l'analyse des données LAT, une distribution d'outils dédiés a été mise en place, incluant des codes consacrés au calcul des phases des pulsars, à partir des dates enregistrées au télescope. Cette distribution, appelée \emph{Science Tools}, est développée, maintenue et distribuée par le FSSC\footnote{\emph{cf.} http://fermi.gsfc.nasa.gov/ssc/data/analysis/software/}, avec l'aide des membres de la collaboration Fermi. Outre la construction d'éphémérides couvrant les observations $\gamma$ par Fermi, les télescopes radio et X ont également offert la possibilité de tester les outils d'analyse de Fermi à l'aide de données extérieures, en particulier \emph{gtbary}, procédure de barycentrisation des dates des photons $\gamma$, et \emph{gtpphase}, outils de calcul des phases des pulsars à partir des dates barycentrées. Ce travail de vérification est détaillé dans \citet{PapierTiming}. Ici nous avons retenu les deux tests les plus contraignants, réalisés avec des pulsations géantes du pulsar milliseconde B1937+21 et des TOAs du pulsar milliseconde en orbite binaire J0437$-$4715.

\subsection{Pulsations géantes de B1937+21}

Le phénomène de pulsations radio géantes (\emph{Giant Radio Pulses}, GRPs) se manifeste pour quelques pulsars, jeunes ou anciens, tels que le pulsar du Crabe ou certains pulsars milliseconde \citep{Johnston2004,Knight2006}, et consiste en de brèves bouffées d'émission radio, beaucoup plus intenses que le niveau d'émission moyen. Le pulsar milliseconde isolé B1937+21 ($P = 1,56$ ms) est un émetteur de GRPs \citep{Cognard1996}. L'émission de GRPs pour ce pulsar consiste en des composantes principale et secondaire, décalées de 55 à 70 $\mu$s par rapport aux pics standard et intervenant dans des fenêtres de temps inférieures à 10 $\mu$s. 

Entre le 13 mars et le 7 avril 2007, 251 GRPs pour lesquels le rapport signal-sur-bruit était supérieur à 30 déviations standard ont été enregistrés au radiotélescope de Nançay. Une éphéméride a été produite à l'aide des mêmes observations. Pour vérifier les outils de Fermi, les GRPs ainsi enregistrés ont été traités de la manière suivante :

\begin{itemize}
\item Conversion des TOAs, donnés dans un format de temps spécifique à l'observatoire, vers des dates MET (\emph{Mission Elapsed Time}, format des dates de Fermi ainsi que de certaines missions spatiales), en l'occurence le nombre de secondes depuis le 1$^{er}$ janvier 2001 à minuit (UTC), dans l'échelle de Temps Terrestre (TT).
\item Calcul de la position de l'observatoire dans le système solaire en fonction du temps et conversion vers le format spécifique des données de Fermi.
\item Transfert des temps à l'observatoire, vers le barycentre du système solaire, grâce à \emph{gtbary}. Il faut cependant corriger des effets dispersifs au préalable si besoin est : \emph{gtbary} ne permet pas cette correction puisqu'elle est négligeable pour les photons $\gamma$ dont la fréquence est quasi-infinie.
\item Enfin, calcul de la phase du pulsar pour chaque TOA, et ce grâce à l'outil \emph{gtpphase}.
\end{itemize}
 
En réalisant les étapes précédentes, on obtient les phasogrammes montrés en Figure \ref{GRP1937} A et B. Les composantes principale et secondaire des GRPs retardent de respectivement 60,1 et 67,3 $\mu$s par rapport aux pics d'émission classique, avec des déviations standard de 1,9 et 2,4 $\mu$s. Ce résultat, en accord avec ceux de \citet{Kinkhabwala2000}, démontre que les outils d'analyse de Fermi, de la procédure de barycentrisation au calcul des phases, sont précis à l'échelle de quelques $\mu$s. Cependant, le pulsar B1937+21 est isolé, par conséquent la procédure de correction des orbites binaires n'est pas testée ici. De plus, le faible intervalle de temps considéré (trois semaines environ) ne permet pas de déceler des erreurs de positionnement de l'observatoire, éventuellement visibles à plus long terme.

\sfig[Pulsations géantes de PSR B1937+21 analysées avec les outils de Fermi]{chap3/gp1937.tex}{Pulsations radio géantes du pulsar milliseconde B1937+21 enregistrées à Nançay et analysées avec les outils de Fermi (ligne continue, échelle de gauche). Un profil radio standard du pulsar à 1,4 GHz est montré en ligne pointillée (échelle de droite). Le pic de GRPs a une largeur d'environ 2 $\mu$s, démontrant la précision des outils d'analyse. La figure (A) est issue de \citet{PapierTiming}. (B) : zoom sur la composante principale des pulsations géantes.}{GRP1937}

\subsection{Orbites complexes : cas de J0437$-$4715}

Le pulsar milliseconde J0437$-$4715 ($P = 5,76$ ms) appartient à un système binaire. La période de l'orbite est de 5,7 jours environ, et cette orbite permet la mesure de paramètres post-képleriens \citep{vanStraten2001,Verbiest2008}. \`{A} partir de l'éphéméride pour ce pulsar fournie dans \citet{Verbiest2008}, de 300 ns de précision, 37 TOAs au télescope de Parkes ont été simulés sur un intervalle de temps de 500 jours (de décembre 2003 à mai 2005), grâce à la fonctionnalité \emph{FAKE} de TEMPO2. La procédure de test est la même que celle décrite précédemment, à la différence que cette fois la procédure permettant à \emph{gtpphase} de corriger de l'orbite des pulsars binaires est mise à l'épreuve. Les 37 phases calculées à l'aide des outils de Fermi sont montrées dans la Figure \ref{Parkes0437}. En moyenne, les TOAs traités dans ce test ont un retard moyen de 320 ns par rapport aux mêmes TOAs analysés avec TEMPO2. On en déduit ici que les outils de barycentrisation et de calcul des phases \emph{gtbary} et \emph{gtpphase} donnent des résultats compatibles avec ceux de l'outil standard TEMPO2, pour un pulsar milliseconde dans un système binaire complexe et pour des dates couvrant un intervalle de temps de plus d'un an d'observation.

\fig[Simulation de temps d'arrivée radio du pulsar J0437$-$4715, observé à Parkes]{scale=0.5}{chap3/parkes_0437.pdf}{Temps d'arrivée simulés pour le pulsar binaire J0437$-$4715, observé à Parkes, et analysés avec les outils de Fermi. La phase absolue est définie de telle sorte que le pic radio principal soit centré à 0,5. Les TOAs ont un retard moyen de 0,32 $\mu$s, indiquant que les outils d'analyse ont une précision de l'ordre de la $\mu$s. Figure issue de \citet{PapierTiming}.}{Parkes0437}

D'autres tests ont été réalisés au cours de cette thèse, utilisant notamment des GRPs du pulsar du Crabe enregistrés à Nançay, et des observations en rayons X de PSR J0218+4232 avec XMM. Ces autres vérifications ont amené à des conclusions identiques. Par conséquent, les outils d'analyse temporelle distribués par le FSSC fournissent des résultats compatibles avec ceux des outils déjà existant, à l'échelle de la $\mu$s.

%% file: tableaux/Ephem0030.tex
\begin{scriptsize}
\begin{tabular}{ll}
\hline\hline
Nom\dotfill & J0030+0451 \\ 
Intervalle de temps (MJD)\dotfill & 53285,9---54829,7 \\ 
Nombre de TOAs\dotfill & 413 \\
RMS des résidus ($\mu s$)\dotfill & 4,26 \\
Ascension droite, $\alpha$\dotfill &  00$:$30$:$27,4301(11) \\ 
Déclinaison, $\delta$\dotfill & +04$:$51$:$39,78(4) \\ 
Fréquence, $f_0$ (s$^{-1}$)\dotfill & 205,530699274958(19) \\ 
Première dérivée de la fréquence, $\dot{f_0}$ (s$^{-2}$)\dotfill & $-$4,2988(7)$\times 10^{-16}$ \\ 
Mouvement propre en ascension droite, $\mu_{\alpha}$ (mas\,yr$^{-1}$)\dotfill & $-$3,5(18) \\ 
Mouvement propre en déclinaison, $\mu_{\delta}$ (mas\,yr$^{-1}$)\dotfill & $-$7(5) \\ 
Parallaxe, $\pi$ (mas)\dotfill & 3,4(5) \\ 
Date de référence de la fréquence (MJD)\dotfill & 50984,4 \\ 
Date de référence de la position (MJD)\dotfill & 50984,4 \\ 
Mesure de dispersion, $DM$ (cm$^{-3}$pc)\dotfill & 4,33822 \\ 
\'Ephéméride planétaire\dotfill & DE200 \\
Unités\dotfill & TDB \\
\hline
\end{tabular}
\end{scriptsize}

%% file: part2/chap4.tex
\chapter[Le Large Area Telescope à bord de Fermi]{Le Large Area Telescope à bord de Fermi}

\minitoc

\chaptabst{D}{ans}{ce chapitre nous présentons l'instrument utilisé en majorité au cours de cette thèse, le Large Area Telescope (LAT) à bord du satellite Fermi, lancé en juin 2008. Dans un premier temps nous faisons un rappel des propriétés de Fermi. Celui-ci embarque deux instruments : le GBM dédié à la détection de sursauts $\gamma$ à des énergies comprises entre 8 keV et 40 MeV, et le LAT, instrument principal embarqué par l'observatoire, sensible à des photons d'énergie entre 20 MeV et 300 GeV. Le LAT est lui-même composé de différents éléments dont nous rappelons le fonctionnement dans la deuxième partie de ce chapitre. Ceux-ci sont le bouclier anti-coïncidences, élément participant au rejet des rayons cosmiques, le trajectographe, déterminant la direction des photons détectés, et le calorimètre qui mesure l'énergie qu'ils déposent. Ces différents sous-systèmes interagissent pour rejeter les particules chargées indésirables, et pour reconstruire la direction et l'énergie des photons tombant dans le télescope. Nous nous penchons enfin sur les performances du LAT, résolution angulaire, résolution en énergie ; mais également la précision en temps, vitale pour l'étude des pulsars milliseconde.}

\section{L'observatoire Fermi}

Le \emph{Fermi Gamma-ray Space Telescope} (FGST), à l'origine nommé GLAST (pour \emph{Gamma-Ray Large Area Telescope}), est un observatoire dédié à la détection de sources de rayonnement $\gamma$, mis en orbite autour de la Terre le 11 juin 2008. La mission Fermi succède à CGRO, en activité pendant les années 90. \`{A} l'image de CGRO qui embarquait quatre instruments, l'observatoire spatial Fermi dispose à son bord du GBM (\emph{Glast Burst Monitor}) \citep{Meegan2007}, instrument consacré à la détection de sursauts $\gamma$ d'énergie comprise entre 8 keV et 40 MeV, et du LAT (\emph{Large Area Telescope}, \emph{cf.} Figure \ref{Fermi}), instrument principal de Fermi dont nous allons rappeler le fonctionnement dans la suite de ce chapitre. Le lecteur intéressé pourra trouver davantage de détails sur le LAT dans \citet{FermiLAT}. Fermi et ses instruments devraient fonctionner de cinq à dix ans.

\fig[Le Fermi Gamma-Ray Space Telescope]{width=8cm}{chap4/213572main_Observatory_lg.jpg}{Le Fermi Gamma-Ray Space Telescope, surmonté du LAT, visible dans la partie supérieure de l'image.}{Fermi}

Le satellite est placé sur une orbite basse, quasi-circulaire, à environ 565 km d'altitude et 25,6$^\circ$ d'inclinaison, pour une période orbitale de $\sqrt{\frac{4 \pi^2 a^3}{G M_T}} =$ 95 minutes, avec $a = R_T +$ 565 km. Il fonctionne essentiellement en mode de balayage :  l'axe du satellite pointe dans une direction à 35$^\circ$ du zénith vers le pôle de l'orbite, pendant une orbite. Au cours de la rotation suivante l'axe du satellite pointe à -35$^\circ$ du zénith, et ainsi de suite. Le ciel est balayé de façon quasi-uniforme toutes les deux orbites (soit environ trois heures) grâce au grand champ de vue du LAT (2,4 sr). \`{A} plusieurs reprises au cours des 60 jours qui ont suivi le lancement Fermi a opéré en pointé, vers le pulsar de Vela notamment. Dans ce mode le satellite pointe son axe vers la cible désirée et bascule vers une autre direction lorsque la Terre s'approche de la ligne de visée.

\section{Le Large Area Telescope}

\subsection{Principe de fonctionnement}

Le Large Area Telescope (LAT) est un télescope sensible au rayonnement $\gamma$ d'énergie comprise entre 20 MeV et 300 GeV. Ses trois éléments principaux sont le trajectographe, le calorimètre et le bouclier anti-coïncidences, auxquels il faut ajouter l'électronique associée, enregistrant les informations. La technique de détection du rayonnement $\gamma$ est la suivante : un photon $\gamma$ tombe dans le télescope et interagit avec la matière pour se convertir en une paire électron et positron. Ces particules chargées sont détectables par les éléments du télescope, qui, en déterminant leur trajectoire et leur énergie, permettent de remonter à la direction et l'énergie totale du photon $\gamma$ initial. Ce principe de fonctionnement est illustré par la Figure \ref{LAT2}. 

Les photons $\gamma$ se convertissent en paires en présence de matière, par conséquent l'atmosphère terrestre empêche l'observation au sol dans cette gamme d'énergie, d'où la nécessité de satelliser le LAT. Placé en orbite autour de la Terre, Fermi est cependant soumis à un intense flux de particules chargées qui interagissent également avec les éléments du LAT. Ces interactions peuvent être interprétées à tort comme des détections de photons $\gamma$. Pour cette raison, le détecteur est entouré du bouclier anti-coïncidences chargé de rejeter le plus possible de ces particules chargées indésirables. Nous reviendrons par la suite sur le rejet du fond.

\fig[Schéma en coupe du LAT]{width=8cm}{chap4/LAT.pdf}{Schéma en coupe du LAT, montrant un photon $\gamma$ tombant dans le trajectographe et interagissant avec la matière pour former une paire électron / positron. Le calorimètre, dans la partie inférieure, absorbe les particules chargées et mesure leur énergie. L'ensemble est protégé par le bouclier anti-coïncidences.}{LAT2}

\subsection{Le bouclier anti-coïncidences}

Le but principal du bouclier anti-coïncidences (ACD) est le rejet des rayons cosmiques, particules chargées provenant de toutes les directions de l'espace. Comme on l'a évoqué précédemment elles peuvent interagir avec le détecteur, et engendrer de fausses détections de photons. L'objectif fixé pour le rejet des particules non $\gamma$ est de 99,97\%. L'ACD est composé de 89 scintillateurs plastiques d'un cm d'épaisseur, dits tuiles, et de huit rubans couvrant les interstices. L'ACD recouvre le trajectographe et le calorimètre. 

Le principe de fonctionnement est le suivant : les rayons $\gamma$ interagissent peu avec l'ACD. On s'attend alors à ce que la détection d'un rayon $\gamma$ par le trajectographe coïncide avec une absence de détection de signal par l'ACD (c'est-à-dire, une « anti-coïncidence »). Au contraire, une particule chargée tombant sur le télescope dépose une partie de son énergie dans l'ACD, puis interagit avec le trajectographe. Ainsi on s'attend à ce que pour les particules chargées, la détection de signal par les deux instruments coïncide. 

La nouveauté du système ACD embarqué par le LAT est qu'il ne s'agit pas d'une structure monobloc mais de nombreuses tuiles et rubans, permettant cette fois de connaître les positions où des particules ont été détectées. Pour les rayons $\gamma$ de haute énergie, typiquement au-delà de quelques GeV, la gerbe engendrée par l'interaction dans les instruments peut contenir des photons sortants, dirigés vers le bouclier anti-coïncidences. Ce phénomène est appelé \emph{back-splash}. La probabilité que ces photons d'énergie moindre (typiquement 100 keV à 1 MeV) interagissent avec l'ACD est faible. S'il y a néanmoins interaction, un ACD monobloc sans informations spatiales n'a aucun moyen de distinguer ce back-splash de rayonnement cosmique traditionnel. En conséquence l'ACD produit systématiquement un signal de \emph{self-veto} qui empêche la détection du photon initial. Cet effet a rendu EGRET très peu sensible aux photons au-dessus de 10 GeV. L'ACD du LAT est segmenté, permettant l'accès à l'information spatiale en cas de signal, et réduisant ainsi l'effet de self-veto. Il a été conçu pour que seuls 20\% des photons de 300 GeV tombant sur le LAT ne soient rejetés en raison de coïncidences fortuites. 

\subsection{Le trajectographe}

Cet élément (également appelé \emph{Tracker}, TKR) est conçu pour déterminer la direction du rayon $\gamma$ tombant dans le télescope. Le principe est ici d'avoir un matériau riche en protons, pour que la probabilité que des photons $\gamma$ incidents s'y convertissent en paire électron / positron soit maximale. Puis, en cas de conversion, de mesurer les points de passage de ces particules chargées dans le détecteur, et en joignant les points, en déduire la position où la conversion a eu lieu (le « vertex ») et l'angle d'incidence de ce photon.

Le TKR est un assemblage de 4 $\times$ 4 tours, chacune consistant en un empilement de 19 plateaux. Chaque plateau contient deux plans de détecteurs à pistes de silicium (\emph{Silicon-Strip Detectors}, SSDs). disposés aux deux faces et dont les pistes sont orientées dans la même direction. Les plateaux aux extrémités ne comportent qu'un seul plan de SSDs. Chaque plateau est orienté perpendiculairement à celui qui le précède. De cette manière, l'ensemble se répartit en deux orientations distinctes, $x$ et $y$. Deux plans successifs d'orientations perpendiculaires forment des doublets $xy$ séparés de seulement 3 mm. Ainsi, on peut localiser en trois dimensions les points de passage des particules chargées en repérant les doublets $xy$ qui ont détecté un signal, et leur position sur l'axe $z$, orthogonal au plan $xy$.

Afin de maximiser la conversion de photons $\gamma$ en paires électron / positron, des couches de feuilles de tungstène ($Z = 74$) d'épaisseur plus ou moins importante ont été intercalées dans les différents plateaux. Le schéma de répartition du tungstène dans le détecteur répond au besoin de disposer à la fois d'une bonne reconstruction spatiale et d'une surface effective de collection des photons maximale. La précision de la reconstruction spatiale se heurte principalement à la diffusion multiple des particules de basse énergie. \`{A} cause de cet effet, les faibles épaisseurs de matériau sont privilégiées. Cependant, une grande épaisseur de tungstène apporte plus de longueurs de radiation\footnote{La définition de la longueur de radiation $X_0$ est la suivante : un électron traversant une épaisseur $X_0$ d'un matériau donné voit son énergie diminuer d'un facteur $e$ par pertes radiatives (par Bremsstrahlung essentiellement).}, accroissant ainsi la capacité de conversion du matériau pour une surface donnée. La conception adoptée pour le trajectographe est la suivante : dans les 12 premiers plateaux, les couches de tungstène sont fines représentent 0,03 longueurs de radiation. Pour les quatre plateaux qui suivent, elles sont épaisses et équivalent à 0,18 longueurs de radiation. Enfin, les trois derniers plateaux ne contiennent pas de tungstène.

\subsection{Le calorimètre}

Le calorimètre (CAL) a pour but de mesurer l'énergie des particules chargées nées de la conversion du photon $\gamma$ initial dans le trajectographe, pour remonter à l'énergie de ce photon. Chacune des 16 tours du TKR possède un module de calorimètre dans sa partie inférieure. Un module comprend huit couches de 12 barreaux d'iodure de césium (CsI) enrichi au thallium (Tl), chaque barreau ayant deux paires de photodiodes situées à ses extrémités, mesurant la lumière produite par scintillation du cristal. Grâce à sa densité, le calorimètre a une épaisseur équivalente de 8,6 longueurs de radiation. \`{A} nouveau, les barreaux sont arrangés perpendiculairement d'une couche à l'autre, de façon à ce que l'on puisse mesurer la position ($x$,$y$) du dépôt d'énergie dans la couche considérée. La précision sur la mesure de la position est de quelques mm pour les faibles dépôts d'énergie (typiquement 10 MeV), à moins d'un mm pour $E > 1$ GeV. L'information spatiale apportée par le CAL participe à l'algorithme de calcul des trajectoires des particules chargées dans le télescope, en particulier pour les photons $\gamma$ de haute énergie, dont les cascades ont un développement plus grand dans le détecteur. La résolution en énergie ($\Delta E / E$) du calorimètre est inférieure à 20\%, voire à 10\% dans certaines bandes en énergie.

\subsection{Reconstruction des événements}

Les paragraphes qui précèdent présentent les différents éléments du LAT comme relativement indépendants. En réalité les informations délivrées par les sous-systèmes doivent le plus souvent être mises en relation pour déterminer la direction d'incidence et l'énergie de la particule initiale, créant une gerbe électromagnétique dans le détecteur. 

La détermination de la trace est centrale dans la reconstruction des événements : elle fournit des \emph{a priori} sur ce que devraient observer le CAL et l'ACD. \`{A} partir des positions $xy$ du TKR où un signal a été repéré, des algorithmes génèrent des trajectoires de particules, et il s'agit alors de retenir la plus plausible. Le critère utilisé le plus souvent fait intervenir le calorimètre : on fait l'hypothèse que le barycentre du dépôt d'énergie dans le CAL doit être sur la trajectoire de la particule primaire. \`{A} haute énergie où la diffusion multiple est la moins importante, la gerbe est étroite : ainsi ce barycentre de dépôt dans le CAL fournit une contrainte de plus en plus forte sur la trajectoire. Il se peut cependant que peu ou pas d'énergie soit déposée dans le calorimètre. Dans ce cas l'algorithme de recherche de trajectoire optimale ne peut tenir compte que de la position des doublets $xy$ ayant été activés dans le TKR. 

La trajectoire ainsi déterminée permet alors de raffiner la mesure d'énergie par le calorimètre. L'estimation la plus rudimentaire se fait en effet en sommant les énergies mesurées par tous ses cristaux. Selon le profil de la gerbe, on peut estimer l'énergie perdue par échappement des particules sur les côtés, à l'arrière, ou encore dans les interstices séparant les composants du calorimètre. \`{A} basse énergie, une part importante de l'énergie est laissée dans le trajectographe, et par conséquent il est important de mesurer ce dépôt et de l'ajouter à la quantité mesurée par le calorimètre. Le nombre de pistes de silicium ayant détecté un signal fournit une estimation de l'énergie ainsi déposée dans le TKR. Dans tous les cas, la forme de la cascade se développant dans le télescope est ajustée de façon à bien estimer l'énergie. 

\subsection{Déclenchement et rejet du fond}

Le LAT est essentiellement traversé par des rayons cosmiques, les photons $\gamma$ ne représentant qu'une part marginale des particules incidentes. Le rejet des particules chargées indésirables est donc indispensable. Cependant, on ne peut pas simplement rejeter les événements pour lesquels du signal est détecté dans l'ACD : le \emph{backsplash} engendré par les photons $\gamma$ est souvent récupéré par des tuiles de l'ACD. Les tuiles où la détection de signal est autorisée sont voisines de la trajectoire hypothétique de la particule primaire, et l'énergie reconstruite permet de fixer la taille de la région de l'ACD autorisée. Plus la particule primaire est énergétique, plus la gerbe est étroite, moins grande est la surface de l'ACD où du signal doit être détecté. Si des tuiles en dehors de la région permise sont activées, l'événement est rejeté. Ceci est un exemple des vérifications faites pour le rejet du fond, faisant intervenir là encore les différents sous-systèmes pour déclencher l'acquisition des données. Sur les 3 kHz d'événements considérés par le LAT en moyenne, seuls 400 Hz environ sont enregistrés et transmis au sol. 

Les événements transmis sont ensuite analysés, afin de déterminer la qualité de la reconstruction : précision de la direction d'incidence, de la mesure de l'énergie de la particule primaire, et probabilité que la particule corresponde à un photon. \`{A} ce stade, les algorithmes mettent en oeuvre des arbres décisionnels (\emph{Classification Trees}, CT), entraînés par des simulations de la réponse attendue du LAT. Les événements attribués à des photons sont distribués en classes, selon la qualité de leur reconstruction : \emph{Transient}, \emph{Source} et \emph{Diffuse}, constituant une nouvelle couche de rejet du fond et optimisées pour différents objectifs scientifiques\footnote{Avec l'amélioration des algorithmes décisionnels, de nouvelles classes pourraient être définies à l'avenir pour l'analyse des données du LAT.}. 

La classe \emph{Transient} forme le jeu de coupure le plus lâche, donnant le plus de statistique, au prix d'une qualité de reconstruction moindre et d'une contamination par des événements de fond importante. Le taux d'événements \emph{Transient} est d'environ 5 Hz dont 2 Hz de contamination par du fond. Les événements de type \emph{Source} forment une sous-classe des événements \emph{Transient}, de meilleure qualité et correspondant à des photons $\gamma$ avec une plus grande probabilité. Pour cette classe le taux d'événements total est d'environ 1 Hz, avec un fond estimé à environ 0,4 Hz. Enfin, les événements de classe \emph{Diffuse}, une sous-catégorie de la classe \emph{Source}, sont les mieux reconstruits et ont la plus grande probabilité de correspondre à de véritables photons, avec un taux de contamination de 0,1 Hz environ sur 0,5 Hz au total. En pratique, c'est cette dernière classe qui est utilisée pour l'étude des pulsars avec le LAT, le rapport signal-sur-bruit y étant le meilleur. Les événements n'appartenant à aucune de ces trois classes se répartissent en électrons, positrons, ions lourds, et déchets.

\subsection{Performances}

Les performances du LAT et certaines de ses caractéristiques sont données dans la Table \ref{Performances}. Les propriétés du LAT sont généralement déterminées par les performances individuelles des différents instruments ainsi que par les algorithmes d'identification des rayons $\gamma$, à la fois dans la reconstruction de leur énergie et de leur direction d'incidence, mais également dans le rejet des détections fortuites. 

\tab[Caractéristiques et performances du LAT]{LATPerformances.tex}{Caractéristiques et performances du LAT.}{Performances}

L'observation de pulsars en rayons $\gamma$ avec le LAT regroupe trois aspects : spatial, spectral et temporel. Les performances de datation du LAT sont mentionnées dans la suite de ce chapitre. Le bon fonctionnement des horloges embarquées par Fermi est primordial pour l'analyse des pulsars dans la mesure où la dimension temporelle est ce qui fait leur spécificité.

\`{A} chaque classe d'événement, \emph{Transient}, \emph{Source} ou \emph{Diffuse}, correspond une « fonction de réponse instrumentale » (\emph{Instrument Response Function}, IRF), que l'on peut formuler de la manière suivante\footnote{Notons que cette formulation est élémentaire et ne rend pas compte de la topologie de l'instrument. Par exemple, l'énergie et la trace des particules interagissant au niveau des interstices séparant les tours du TKR sont moins bien reconstruites, ce qui introduit une dépendance en ($x$,$y$) de la qualité de reconstruction.} :

\begin{eqnarray}
\mathrm{IRF}(\theta,E) = \mathrm{A_{eff}}(\theta,E) \times \mathrm{PSF}(\theta,E) \times \Delta E (\theta,E)
\label{irf}
\end{eqnarray}

Dans cette expression, $\theta$ et $E$ désignent l'angle d'incidence par rapport à l'axe du télescope et l'énergie reconstruits. $\mathrm{A_{eff}}(\theta,E)$ est la surface effective de collection. Elle est la plus importante dans la classe \emph{Transient}, la plus faible pour la classe \emph{Diffuse}. Elle est illustrée en Figure \ref{aeff}, pour la classe \emph{Source}. 

\fig[Surface effective de collection du LAT en fonction de l'énergie]{width=10cm}{chap4/aeff.pdf}{Surface effective de collection en fonction de l'énergie, pour des photons en incidence normale (trait plein), et à 60$^\circ$ d'inclinaison (ligne pointillée), dans la classe d'événements \emph{Source}. Figure extraite de \citet{FermiLAT}.}{aeff}

La \emph{Point Spread Function}, décrite par le terme $\mathrm{PSF}(\theta,E)$ dans l'équation \ref{irf}, représente la réponse spatiale du système pour une source ponctuelle. La résolution angulaire concerne tous les aspects de l'analyse des données du LAT. Elle résulte essentiellement des performances du trajectographe. Son comportement en fonction de l'énergie des photons incidents est montré en Figure \ref{AngRes} pour la classe \emph{Source}. Comme déjà mentionné, la reconstruction des directions est meilleure à haute énergie ; en effet la diffusion multiple est importante à basse énergie. La Figure \ref{AngRes} indique que 68\% des photons de 100 MeV réellement émis par la source ponctuelle considérée sont détectés avec une séparation angulaire inférieure ou égale à 3,5$^\circ$ (en incidence normale). \`{A} 1 GeV, la distance angulaire correspondante est de 0,6$^\circ$, et continue de décroître lorsque l'énergie augmente. 

\fig[Résolution angulaire du LAT en fonction de l'énergie]{width=10cm}{chap4/AngRes.pdf}{Résolution angulaire du LAT en fonction de l'énergie dans la classe d'événements \emph{Source}, pour des photons en incidence normale (ligne pleine), et à 60$^\circ$ d'inclinaison (ligne pointillée), pour des conversions ayant lieu dans la région du trajectographe où les couches de tungstène sont fines. Figure issue de \citet{FermiLAT}.}{AngRes}

La \emph{Point Spread Function}, décrite par le terme $\mathrm{PSF}(\theta,E)$ dans l'équation \ref{irf}, représente la réponse spatiale du système pour une source ponctuelle. La résolution angulaire concerne tous les aspects de l'analyse des données du LAT. Elle résulte essentiellement des performances du trajectographe. Son comportement en fonction de l'énergie des photons incidents est montré en Figure \ref{AngRes} pour la classe \emph{Source}. Comme déjà mentionné, la reconstruction des directions est meilleure à haute énergie ; en effet la diffusion multiple est importante à basse énergie. La Figure \ref{AngRes} indique que 68\% des photons de 100 MeV réellement émis par la source ponctuelle considérée sont détectés avec une séparation angulaire inférieure ou égale à 3,5$^\circ$ (en incidence normale). \`{A} 1 GeV, la distance angulaire correspondante est de 0,6$^\circ$, et continue de décroître lorsque l'énergie augmente.

\fig[Résolution en énergie du LAT en fonction de l'énergie]{width=10cm}{chap4/EnergyRes.pdf}{Résolution en énergie du LAT en fonction de l'énergie, pour des photons en incidence normale (ligne pleine) et à 60$^\circ$ d'inclinaison (ligne pointillée), dans la classe d'événements \emph{Source}. Figure extraite de \citet{FermiLAT}.}{EnergyRes}

En résumé, les performances du LAT sont très supérieures à celles de son prédécesseur, EGRET. Il est moins assujetti au back-splash, ainsi sa gamme d'énergie s'étend de 20 MeV à 300 GeV, là où EGRET n'observait que jusqu'à quelques dizaines de GeV. Le champ de vue est de 2,4 sr contre 0,4 pour EGRET, et la surface effective maximale, qui est de 9500 cm$^2$, est six fois supérieure environ. Entre les deux télescopes le temps mort de l'électronique a été réduit de façon spectaculaire : 100 ms pour EGRET, contre 26,5 $\mu$s pour le LAT. Les résolutions angulaire et en énergie ont également été améliorées. Globalement, la stratégie d'observation (balayage plutôt que pointé), la surface effective et la précision de localisation rendent le LAT 30 fois plus sensible que son prédécesseur à des sources ponctuelles. 

\subsection{Précision de datation}

La précision de datation des événements mesurés par le télescope est importante pour l'étude des sursauts $\gamma$, nécessaire pour l'observation de pulsars jeunes et cruciale pour les pulsars milliseconde. C'est pour cette raison qu'avant le lancement de Fermi des tests de la précision de mesure des dates ont été conduits. Une paire de scintillateurs a été disposée à proximité du satellite de façon à ce que des muons atmosphériques puissent traverser à la fois le TKR et les scintillateurs. Pour ces derniers, les dates étaient enregistrées à l'aide d'un système GPS utilisé auparavant par le télescope CELESTE et dont la précision avait été démontrée grâce à des observations du pulsar du Crabe dans le domaine optique \citep{deNaurois2002}. 

\fig[Résultats des premiers tests de la précision de datation du LAT]{width=16cm}{chap4/SASS14193.pdf}{Tests de la précision de datation du LAT, par comparaison avec un système GPS de référence (ici, test 77014193 réalisé le 22 février 2007). \emph{\`{A} gauche} : histogramme des différences de temps entre le LAT et les scintillateurs. Le pic légèrement en deçà de 0 représente le signal de muons. \emph{\`{A} droite} : écarts entre les dates enregistrés par les deux télescopes, en fonction du temps. On observe ici un comportement en dents de scie, dont la pente est de -3,4 $\mu$s/s.}{SASS14193}

Les premiers essais de la datation du LAT, monté sur le satellite et utilisant son système GPS ont été effectués les 22 et 23 février 2007\footnote{Ces tests ont été réalisés par D.~A. Smith, J.~E. Grove, D. Dumora, D.~P. Sandora et E.~J. Siskind.}. Il s'agissait de tester deux configurations : 1) le LAT reçoit et interprète le signal GPS pour calculer les dates des muons, 2) on empêche le LAT d'utiliser le signal GPS pour calculer les dates, celles-ci sont dérivées d'horloges internes. Pour le premier cas, les temps enregistrés par les deux instruments diffèrent, et les écarts forment un motif de dents de scie au cours du temps. Les écarts s'échelonnent de 0 à $-1$ ms, avec une pente de quelques $\mu$s/s (un exemple est présenté en Figure \ref{SASS14193}). Le deuxième cas est plus critique encore : au lieu de revenir à 0 lorsque $\Delta T = -1$ ms, les écarts continue de croître, et atteignent quelques dizaines de ms.

Une erreur de programmation dans les algorithmes de calcul des dates à été diagnostiquée. Après correction, les tests finaux ont révélé que les deux systèmes GPS étaient en accord à 300 ns près (\emph{cf.} Figure \ref{absolutetime}). Le contrôle des données de télémétrie indique que la datation des événements par le LAT sur orbite se comporte comme lorsque le télescope était au sol \citep{FermiCalibration}. La précision de datation est donc inférieure à 1 $\mu$s, garantissant la possibilité d'étudier les pulsars, \emph{a fortiori} milliseconde, avec le LAT.

\fig[Tests définitifs de la précision temporelle du LAT]{width=16cm}{chap4/absolutetime.pdf}{Résultats des tests de datation finaux. \emph{\`{A} gauche} : écarts entre les dates mesurées par les deux dispositifs en fonction du temps. \emph{\`{A} droite} : histogramme des différences de temps. L'écart moyen est de 0,3 $\mu$s. Figure extraite de \citet{FermiCalibration}.}{absolutetime}

%% file: part3/chap5.tex
\chapter[Pulsars milliseconde du champ galactique]{Pulsars milliseconde du champ galactique}

\minitoc

\chaptabst{D}{isposant}{d'éphémérides contemporaines aux observations du LAT pour la plupart des pulsars milliseconde situés en dehors d'amas globulaires, nous avons recherché des pulsations en rayons $\gamma$, et détecté huit MSPs avec une significativité supérieure à 5 $\sigma$. Les huit objets détectés sont J0030+0451, J0218+4232, J0437$-$4715, J0613$-$0200, J0751+1807, J1614$-$2230, J1744$-$1134 et J2124$-$3358. En plus de ces pulsars, deux MSPs montrent des signaux intéressants, J0034$-$0534 et J1713+0747, et la progression de leur significativité semble indiquer une détection probable dans les mois à venir. Nous présentons dans ce chapitre les courbes de lumière $\gamma$ et les propriétés spectrales des huit MSPs détectés au cours de cette thèse. Nous montrons aussi une analyse de données XMM pour J0030+0451, où contrairement à des résultats antérieurs, nous déterminons la position des pics X par rapport aux pics radio. Les composantes radio et X sont en alignement proche, ce qui conforte l'idée que l'émission X de J0030+0451 est produite aux pôles, par des processus thermiques. De manière générale, les phasogrammes $\gamma$ des huit MSPs ressemblent à ce qui est observé par ailleurs pour les pulsars normaux, de même que leurs spectres. En outre, on note que les MSPs détectés dominent cette classe d'objets en terme de $\dot E / d^2$, tout comme chez les pulsars ordinaires. Ces points communs laissent à penser que la production de rayonnement $\gamma$ dans la magnétosphère des pulsars normaux et des MSPs a sensiblement les mêmes causes et origines. En l'occurence, les modèles d'émission dans la magnétosphère externe sont privilégiés chez les MSPs, comme chez les pulsars normaux. Ce chapitre constitue une mise à jour des résultats présentés dans \citet{FermiJ0030, FermiMSPs}, avec ici un lot de données plus large.}

\section{Détection de huit pulsars milliseconde}

Parmi les 72 pulsars milliseconde du champ galactique (\emph{cf.} Annexe \ref{annexeA}), 66 ont été suivis dans le cadre de la campagne de datation. Nous avons utilisé les éphémérides ainsi obtenues et recherché des pulsations dans les données du LAT, de manière analogue à la recherche effectuée une décennie auparavant dans les données d'EGRET \citep{Fierro1995}. 

\subsection{Analyse des données}

Pour la recherche de pulsations des pulsars milliseconde du champ galactique, nous avons utilisé les données enregistrées par le LAT depuis sa mise en service, le 30 juin 2008, jusqu'au 2 juin 2009, essentiellement en mode de balayage. Nous avons retenu les événements d'énergie reconstruite supérieure à 100 MeV, et appartenant à la classe \emph{Diffuse}. Cette classe regroupe les événements qui ont la plus grande probabilité d'être de véritables photons $\gamma$, comme nous l'avons évoqué dans le chapitre précédant. Pour s'affranchir du maximum de photons $\gamma$ d'albédo terrestre (photons émanant des interactions des rayons cosmiques dans la  haute atmosphère terrestre), nous avons exclu les événements dont la direction reconstruite forme un angle de plus de 105$^\circ$ avec le zénith. \`{A} cause de l'émission diffuse galactique, causée par l'interaction des rayons cosmiques dans le gaz interstellaire, les sources situées dans le plan galactique (donc à faible latitude galactique $b$) baignent dans un fond de photons $\gamma$ intense. Les sources à haute latitude sont contaminées par un fond extragalactique isotrope, moins intense. Pour cette raison, le rayon $\rho$ des régions d'intérêt retenues autour de chaque pulsar a été adapté à leur position en coordonnées galactiques. En l'occurence :

\begin{equation}
\left\langle
	\begin{array}{ll}
		\rho = 0,5^\circ & |b| < 10^\circ \\
		\rho = 1^\circ & \mbox{sinon} \\
	\end{array}
\right.
\end{equation}

Remarquons néanmoins que nous avons fait une exception pour deux pulsars, PSR J0218+4232 et J1614$-$2230, dont la latitude galactique est supérieure à 10$^\circ$ mais pour lesquels nous avons réduit $\rho$ à 0,5$^\circ$. J0218+4232 est en effet situé à 1$^\circ$ du blazar 3C 66A, très brillant en rayons $\gamma$. Quant à J1614$-$2230, celui-ci est situé dans une région riche en sources $\gamma$. La conséquence de ce changement de coupure angulaire vis-à-vis de la significativité des signaux est discutée dans la suite. 

Nous avons ensuite transféré les dates des événements sélectionnés au barycentre du système solaire, selon l'équation \ref{bary}, puis calculé les phases rotationnelles à partir des éphémérides fournies par les radiotélescopes. Certains pulsars n'ont pas été couverts, à savoir J1125$-$6014, J1216$-$6410, J1618$-$39, J1629$-$6902, J1757$-$5322 et J1933$-$6211. Pour ces objets nous avons utilisé les paramètres de rotation fournis en ligne par le catalogue de pulsars de l'ATNF. Ces éphémérides étant anciennes et/ou approximatives, nous prendrons soin de considérer les résultats les concernant avec précaution. Pour la réalisation des étapes décrites ici, nous avons utilisé le logiciel TEMPO2. Les outils standard d'analyse temporelle \emph{gtbary} et \emph{gtpphase} présentent en effet un certain nombre de limitations : ils ne permettent pas de corriger d'un éventuel mouvement propre ou effet de parallaxe de chronométrie, d'employer plus de deux dérivées temporelles dans le développement de Taylor de la fréquence de rotation du pulsar (\emph{cf.} équation \ref{eqfrequence}), ou encore de s'affranchir d'éventuel \emph{timing noise} par un développement en harmoniques. En pratique, un module de TEMPO2 a été développé au cours de cette thèse, permettant l'analyse des données du LAT à l'aide des fonctionnalités de TEMPO2. Le module \emph{fermi\_plug.C} est livré dans la distribution de TEMPO2\footnote{http://tempo2.sourceforge.net/}.

\subsection{Recherche de pulsations}

Un certain nombre de méthodes statistiques sont disponibles pour déterminer si un phasogramme donné est compatible avec un signal pulsé, l'absence de pulsation correspondant à une distribution uniforme en phase. Parmi ces tests de périodicité, citons par exemple le test de $\chi^2$ de \emph{Pearson}, le test de \emph{Rayleigh}, $Z_m^2$ ou encore le \emph{H-test} \citep{deJager1989}, ces méthodes étant adaptées à différentes situations. Nous avons opté pour le \emph{H-test}, qui présente l'avantage de ne dépendre d'aucun paramètre arbitraire introduisant des nombres d'essais, comme par exemple le nombre de subdivisions du phasogramme (\emph{binning}) nécessaire au test de $\chi^2$, ou le nombre d'harmoniques du test $Z_m^2$. 

Le \emph{H-test} est construit de la manière suivante : soit $f$ la fonction normalisée décrivant l'échantillon des $N$ phases mesurées. On peut représenter $f$ par une somme de fonctions $\delta$ :

\begin{eqnarray}
f(\phi) = \frac{1}{N} \sum_{i=1}^N \delta\left( \phi_i \right)
\label{dirac}
\end{eqnarray}

On peut également écrire $f$ comme un développement en série de Fourier d'ordre $m$ :

\begin{eqnarray}
f_m (\phi) = \alpha_0 + \frac{1}{\pi} \sum_{k=1}^m \left[ \alpha_k \cos(k \phi) + \beta_k \sin(k \phi) \right]
\end{eqnarray}

Les coefficients $\alpha_0$, $\alpha_k$ et $\beta_k$ sont donnés par :

\begin{eqnarray}
\alpha_0 & = & \frac{1}{2 \pi} \int_0^{2 \pi} f(\phi) d \phi = \frac{1}{2 \pi}\\
\alpha_k & = & \int_0^{2 \pi} f(\phi) \cos(k \phi) d \phi \\
\beta_k & = & \int_0^{2 \pi} f(\phi) \sin(k \phi) d \phi \\
\end{eqnarray}

Le développement en fonctions $\delta$ (\emph{cf.} équation \ref{dirac}) conduit à :

\begin{eqnarray}
\alpha_k & = & \frac{1}{N} \sum_{i=1}^N \cos(k \phi_i) \\
\beta_k & = & \frac{1}{N} \sum_{i=1}^N \sin(k \phi_i) \\
\end{eqnarray}

Soit $Z_m^2$ l'écart entre la distribution de phase observée, représentée par la fonction $f_m$, et le phasogramme uniforme (qui prend pour valeur $\frac{1}{2 \pi}$ sur l'ensemble de la rotation) :

\begin{eqnarray}
Z_m^2 = 2 \pi N \int_0^{2 \pi} \left[ f_m(\phi) - \frac{1}{2 \pi} \right]^2 d\phi = 2 N \sum_{k=1}^m \left( \alpha_k^2 + \beta_k^2 \right)
\end{eqnarray}

Comme le montre l'expression précédente, $Z_m^2$ est un critère de périodicité, qui dépend cependant du paramètre $m$, à choisir en fonction de la forme de profil attendue. Le \emph{H-test} remédie à ce problème en recherchant l'harmonique $m$ optimale :

\begin{eqnarray}
H = \max_{1 \leq m \leq 20} \left( Z_m^2 - 4 m + 4 \right)
\end{eqnarray}

La probabilité d'obtenir une valeur supérieure à $H$ de manière fortuite a été évaluée par simulations Monte Carlo. Elle est donnée par :

\begin{equation}
P (> H) \simeq 
\left\langle
	\begin{array}{ll}
		0,9999755\ e^{-0,39802\ H} & H \leq 23 \\
		1,210597\ e^{-0,45901\ H + 0,00229\ H^2} & 23 < H \leq 50 \\
		< 4 \times 10^{-8} & \mbox{sinon} \\
	\end{array}
\right.
\end{equation}

Le \emph{H-test} s'avère être parmi les plus efficaces\footnote{L'efficacité d'un test de périodicité est définie comme étant la probabilité qu'une source périodique soit identifiée par ce test en tant que telle au-dessus d'un certain seuil de vraisemblance. Cette efficacité dépend des propriétés de la source étudiée : nombre et largeur des pics, intensité, \emph{etc}. \emph{Cf.} \citet{deJager1989} pour une comparaison de l'efficacité du \emph{H-test} avec celle d'autres tests de périodicité.} pour des profils d'émission à un pic raisonnablement large, ou des profils à deux pics de type Vela, Crabe ou Geminga. Néanmoins, notons que le test de Rayleigh est plus efficace pour des profils sinuso\"idaux (type de profil d'émission jusqu'à présent non rencontré en rayons $\gamma$) et que les tests $Z_m^2$ et de $\chi^2$ sont plus performants pour des profils à plus de deux composantes, ou lorsque les pics sont très étroits (typiquement, lorsque le cycle utile de la courbe de lumière est inférieur à 5\%). Ce choix de test de périodicité introduit par conséquent un biais de sélection : le \emph{H-test} favorise des détections de profils tels que ceux observés auparavant en rayons $\gamma$, par exemple par EGRET. Cependant, en l'absence de détections fermes de MSPs avant cette étude, il était raisonnable d'utiliser le type de profil des émetteurs $\gamma$ connus comme hypothèse \emph{a priori}.

La recherche de pulsations dans le lot de données décrit précédemment conduit à la distribution de significativités montrée en Figure \ref{histoHtest}. Rappelons que pour deux des pulsars, J0218+4232 et J1614$-$2230, un essai supplémentaire a été introduit du fait du changement de coupure angulaire. Pour ces deux pulsars, la probabilité de détection fortuite est inférieure à $4 \times 10^{-8}$ d'après le \emph{H-test} pour des rayons d'intégration $\rho =$ 0,5$^\circ$. Un essai supplémentaire amène donc à $P < 8 \times 10^{-8}$. Par conséquent, la significativité est toujours supérieure à 5 $\sigma$, et le changement de jeu de coupure reste sans conséquence. Huit pulsars milliseconde sont détectés avec certitude au moment de la rédaction de ce manuscrit (soit un lot de données d'environ un an) : PSR J0030+0451, J0218+4232, J0437$-$4715, J0613$-$0200, J0751+1807, J1614$-$2230, J1744$-$1134 et J2124$-$3358. En outre, deux pulsars sont détectés marginalement, avec des niveaux de significativité de 3,7 et 4,2 $\sigma$ : PSR J1713+0747 et J0034$-$0534.

\fig[Résultats de la recherche de pulsations pour les 72 MSPs du champ galactique]{width=12cm}{chap5/histoTS.pdf}{Résultats de la recherche de pulsations pour les 72 MSPs du champ galactique. Les noms des pulsars pour lesquels la significativité est d'au moins 3, 4, ou 5 $\sigma$ sont indiqués.}{histoHtest}

Avant de présenter les courbes de lumière et les propriétés spectrales des huit MSPs détectés avec certitude, ainsi que de discuter des candidats à la détection au cours des mois à venir, remarquons que la majorité des 72 MSPs traités ici présentent un signal dont la significativité est inférieure à 3 $\sigma$. Pour les huit pulsars détectés, on s'aperçoit que la valeur de $H$ augmente de façon linéaire avec le temps, en première approximation (la Figure \ref{prog_lin} illustre quelques exemples). En supposant qu'un signal actuellement compris entre 0 et 1 $\sigma$ (soit $H \leq 2,9$) voie sa valeur de $H$ augmenter linéairement, on aboutit à $H \leq 14,5$ au bout de cinq ans. Or une détection à 3 $\sigma$ correspond à une valeur de $H$ d'environ 15. On peut par conséquent avancer que les pulsars dont la significativité n'a pas atteint 1 $\sigma$ au cours de cette première année ne seront pas détectés dans les cinq premières années de fonctionnement du LAT, avec les jeux de coupures utilisés ici. En revanche, pour les pulsars dont le signal est compris entre 1 et 2 $\sigma$ (soit $H \leq 7,76$) le même raisonnement conduit à des valeurs de $H$ inférieures à 38,8 après cinq ans. Cette valeur correspond approximativement à une détection à 5 $\sigma$, par conséquent le LAT pourrait détecter certains de ces objets dans les quatre années à venir. 

\fig[Progression de la valeur de $H$ en fonction du temps pour des MSPs détectés]{width=12cm}{chap5/prog_lin.pdf}{Exemples de progression de la valeur de $H$ en fonction du temps, pour quatre des huit MSPs détectés. La progression est linéaire, en première approximation.}{prog_lin}

Cependant, les 62 MSPs pour lesquels la significativité est inférieure à 3 $\sigma$ ne présentent pas de telles croissances linéaires de $H$ en fonction du temps, indicatrices de détections à venir. Pour les futures recherches de pulsations $\gamma$ dans les données LAT, il faudra donc envisager des jeux de coupures différents, en angle ou en énergie, ou opter pour un test de périodicité différent : rappelons en effet que le \emph{H-test} est efficace pour des profils d'émission à un ou deux pics relativement larges, mais son efficacité chute pour les pics étroits ou nombreux.

\subsection{Observations X et $\gamma$ de PSR J0030+0451}

Le premier MSP pour lequel des pulsations en rayons $\gamma$ ont été détectées au cours de cette thèse est PSR J0030+0451. Ce pulsar, de période $P = 4,865$ ms, a été découvert indépendamment par deux recherches de pulsars dans le domaine radio, conduites respectivement à Arecibo \citep{Somer2000} et à Bologne \citep{DAmico2000}. La chronométrie a conduit à une mesure de parallaxe de $3,3 \pm 0,9$ mas, correspondant à une distance de $300^{+110}_{-65}$ pc \citep{Lommen2006}. Le mouvement apparent du pulsar est relativement faible. Les valeurs de $\dot P$ apparentes et intrinsèques diffèrent de moins d'un pourcent. Les quantités dérivées $\dot E$, $\tau$, $B_S$ et $B_{LC}$ sont données en Annexe \ref{annexeA}.

Ce pulsar n'était pas encore connu lors de la recherche de MSPs dans les données d'EGRET \citep{Fierro1995}. Cependant il est intéressant de noter que EGR J0028+0457, appartenant au catalogue de sources d'EGRET révisé, co\"incide avec la position du pulsar \citep{Casandjian2008}. Les perspectives de détection de J0030+0451 avec le LAT étaient donc importantes. 

Par ailleurs, PSR J0030+0451 a également été détecté en rayons X en tant que source pulsée, d'abord par ROSAT \citep{Becker2000}, puis par XMM--Newton \citep{Becker2002}. Les deux télescopes X ont observé un profil d'émission comprenant deux pics larges, séparés d'environ 180$^\circ$, mais n'ont pas permis de déterminer le décalage en phase des composantes radio et X, les horloges de ces télescopes ne le permettant pas à l'époque des observations. Depuis, des corrections ont été apportées à la calibration des horloges embarquées par XMM--Newton, de sorte que celles-ci délivrent désormais des dates dont la précision absolue est 300 $\mu$s, soit 0,06 rotations de J0030+0451\footnote{\emph{cf.} http://xmm2.esac.esa.int/docs/documents/CAL-TN-0045-1-0.pdf}. Parallèlement à l'analyse des photons $\gamma$ du LAT, nous avons ré-analysé les observations de J0030+0451 par XMM, afin de mesurer l'alignement relatif entre les émissions radio, X et $\gamma$.

PSR J0030+0451 a été observé par XMM les 19 et 20 juin 2001, pour une durée totale de 29 ks. Pour construire une éphéméride radio couvrant à la fois les observations récentes de Fermi ainsi que celles de XMM, nous avons utilisé 700 TOAs radio enregistrés au radiotélescope de Nançay entre juillet 1999 et aujourd'hui. L'essentiel des observations a été réalisé à 1,4 GHz (1360 $\pm$ 72 MHz avant 2002, 1398 $\pm$ 32 MHz après 2004 et 1398 $\pm$ 64 MHz depuis juillet 2008), en complément d'observations à 2 GHz afin de contraindre au mieux la mesure de dispersion. Cette dernière induit en effet un retard entre la composante radio et l'émission de haute énergie, qu'il faut corriger pour pouvoir mesurer le décalage intrinsèque (\emph{cf.} équation \ref{EqDispersion}). L'incertitude moyenne sur les TOAs est de 3,6 $\mu$s pour les données enregistrées après 2004, et 8,6 $\mu$s pour l'ensemble du lot de données. Les résidus de chronométrie sont montrés en Figure \ref{residusJ0030_NBPP_BON}. La rms des résidus est de 3,7 $\mu$s. Nous mesurons ainsi une parallaxe chronométrique de 4,1 $\pm$ 0,7 mas, comparable à celle de \citet{Lommen2006}. De même, la valeur de $DM$ obtenue dans cette étude, de 4,333 $\pm$ 0,001 cm$^{-3}$ pc, est en accord\footnote{La construction de TOAs à partir des observations NBPP et BON de J0030+0451 a été réalisée par I. Cognard.}.

\fig[Résidus pour l'éphéméride de J0030+0451 couvrant les observations XMM et LAT]{width=12cm}{chap5/residusJ0030_NBPP_BON.pdf}{Résidus en fonction du temps pour le pulsar milliseconde J0030+0451, observé à Nançay. Ces TOAs radio couvrent à la fois les observations du MSP par XMM, en juin 2001, et les récentes observations du Fermi-LAT.}{residusJ0030_NBPP_BON}

Pour l'analyse des données X, nous avons utilisé le signal enregistré par l'instrument \emph{EPIC-pn} en mode de datation. Nous avons sélectionné les événements d'énergie comprise entre 300 eV et 2,5 keV. Une précision temporelle optimale est obtenue en enregistrant le signal des CCD du \emph{pn} en continu, ce qui a pour effet d'étaler l'information dans la direction $y$. Par conséquent, les événements sont conservés en écrasant les données dans une seule dimension, où chaque pixel de la direction $x$ contient la somme des pixels de la direction $y$. Nous avons sélectionné une région centrée sur le pulsar, large de sept pixels dans la direction $x$. Le niveau de fond a été estimé en utilisant une région voisine libre de sources X et de même dimension. Nous l'avons ensuite moyenné sur le nombre de subdivisions du phasogramme X et soustrait au phasogramme. Enfin, les dates ont été transférées au barycentre du système solaire grâce à l'outil dédié \emph{barycen}\footnote{La sélection et la barycentrisation des données XMM ont été réalisées par N. Webb et B. Pancrazi.}. 

\fig[Courbes de lumière radio, X et $\gamma$ de PSR J0030+0451]{width=12cm}{chap5/0030p0451_X_2.pdf}{Courbes de lumière radio, X et $\gamma$ de PSR J0030+0451. Deux rotations du pulsar sont montrées.}{J0030_radioXgamma}

Les courbes de lumière radio, X et $\gamma$ de J0030+0451 sont présentées en Figure \ref{J0030_radioXgamma}. Le profil d'émission $\gamma$ comprend deux pics, séparés de $\Delta =$ 0,45 $\pm$ 0,01. Comme on l'a vu dans le deuxième chapitre, ce type de profil à deux pics séparés de 0,4 à 0,5 rotations est relativement commun parmi les pulsars émetteurs $\gamma$ comme Vela, le Crabe, Geminga ou encore PSR J2021+3651 \citep{Thompson2004,FermiJ2021}. Le premier pic $\gamma$ est décalé de $\delta =$ 0,16 du maximum de l'émission radio, ce qui est également une caractéristique commune : $\delta$ est compris entre 0,11 et 0,16 pour Vela, PSR B1951+32 ou J2021+3651. La séparation des pics et le décalage radio / $\gamma$ indique que le rayonnement $\gamma$ est vraisemblablement produit dans la magnétosphère externe, au contraire de l'émission radio.

La Figure \ref{J0030_radioXgamma} montre que l'émission en rayons X est en alignement proche avec la composante radio (rappelons que la précision temporelle est de 300 $\mu$s, soit 0,06 fractions de tour), et donc décalée de l'émission $\gamma$. Ce résultat indique que les émissions X et radio ont des origines communes dans la magnétosphère, et que le rayonnement $\gamma$ est produit dans une région différente. L'alignement radio/X conforte l'idée que PSR J0030+0451 fait partie des émetteurs de rayonnement X thermique, à l'instar de J0437$-$4715, pour lesquels l'émission est produite près des calottes polaires, là où la température est la plus importante \citep{Zavlin2007}. Par ailleurs, l'analyse spectrale des données XMM a montré que l'émission est compatible avec un modèle purement thermique \citep{Becker2002}.

\subsection{Confirmation de PSR J0218+4232, détection de six autres MSPs}

En plus de la détection de PSR J0030+0451, sept autres MSPs ont été détectés avec le LAT au cours de cette thèse. Leurs courbes de lumière $\gamma$ sont présentées en Figures \ref{phasos1} et \ref{phasos2}. Remarquons néanmoins que ces courbes de lumière sont basées sur des coupures angulaires fixes, ce qui a tendance à biaiser les profils $\gamma$ vers les hautes énergies, compte tenu de la résolution angulaire du LAT. Des courbes de lumière obtenues avec des coupures angulaires dépendantes de l'énergie sont présentées en Annexe \ref{annexeC}. Les phasogrammes globaux au-dessus de 100 MeV et de 1 GeV sont semblables qualitativement avec ceux présentés en Figures \ref{phasos1} et \ref{phasos2}, avec le lot de données actuel. 

Les sept pulsars sont les suivants :

\begin{itemize}
\item \textbf{J0218+4232} : ce MSP appartenant à un système binaire a été découvert à Westerbork puis confirmé au radiotélescope de Jodrell Bank \citep{Navarro1995}. Comme nous l'avons évoqué dans le second chapitre, des pulsations marginales ont été découvertes dans les données d'EGRET \citep{Kuiper2000}. La Figure \ref{phasos1} montre que le profil $\gamma$ de J0218+4232 est très large, une caractéristique que l'on retrouve dans le domaine radio. Le pic principal radio, ici centré autour de 0, possède deux composantes symétriques et la structure globale est large ($\simeq$ 0,4). On constate que les profils radio et $\gamma$ sont similaires : l'émission $\gamma$ forme un faisceau large, et à haute énergie (E $>$ 1 GeV) deux pics semblent émerger, bien que le nombre de photons ne permet pas pour le moment de l'affirmer. Les profils radio et $\gamma$ sont décalés d'environ 0,5. En raison du faible nombre de photons et de la largeur du profil $\gamma$, il est difficile pour le moment de déterminer le décalage radio / $\gamma$ avec plus de précision. Nous avons donc confirmé la détection de J0218+4232 par EGRET, bien que le phasogramme $\gamma$ obtenu dans cette étude diffère qualitativement de celui d'EGRET (\emph{cf.} Figure \ref{Kuiper0218}). 
\item \textbf{J0437$-$4715} : découvert avec le radiotélescope de Parkes \citep{Johnston1993}, ce MSP est le plus proche connu à présent. Malgré sa faible distance, la recherche de pulsations $\gamma$ dans les données d'EGRET n'a pas abouti \citep{Fierro1995}. La contribution de l'effet Shklovskii pour ce pulsar est importante, si bien que la valeur de $\dot E$ intrinsèque est environ quatre fois inférieure à la valeur apparente. Le pulsar présente un seul pic $\gamma$, séparé de $\delta =$ 0,45 avec le pic radio. Comme chez PSR J0030+0451, l'émission X est alignée avec la composante radio, et le spectre est compatible avec un rayonnement X thermique \citep{Zavlin2007}, tandis que l'émission $\gamma$ est décalée, implicant des régions de production séparées. 
\item \textbf{J0613$-$0200} : ce pulsar situé dans un système binaire a été découvert au radiotélescope de Parkes \citep{Lorimer1995}. Le profil d'émission $\gamma$ est analogue à celui de J0437$-$4715. Le pic $\gamma$ est ici séparé du maximum de l'émission radio de $\delta =$ 0,42. En supposant que la similitude avec J0437$-$4715 persiste aux autres longueurs d'onde, on pourrait s'attendre à ce que des observations en X révèlent un pic en phase avec l'émission radio et un spectre thermique.

\sfig[Courbes de lumière radio et $\gamma$ pour quatre MSPs détectés]{chap5/phasos1.tex}{Courbes de lumière $\gamma$ au-dessus de 1 GeV et de 100 MeV, et profil radio pour quatre des huit MSPs détectés par le LAT. Deux rotations sont montrées.}{phasos1}

\item \textbf{J0751+1807} : parmi les objets présentés ici, ce pulsar est à part : il a en effet été découvert en radio dans la boîte d'erreur d'une source $\gamma$ non identifiée du premier catalogue d'EGRET, GRO J0749+17 \citep{Lundgren1995}. La luminosité de la source d'EGRET était cependant trop importante pour être due à la seule contribution de PSR J0751+1807. Par la suite, la recherche de pulsations dans les données d'EGRET n'a pas permis la détection du pulsar (Fierro et al 1995). L'analyse de données XMM a révélé des pulsations marginales \citep{Webb2004}. Le décalage des composantes radio et X n'a cependant pas pu être mesuré, les horloges ne le permettant pas. Le phasogramme $\gamma$ de J0751+1807 vu par le LAT possède des caractéristiques communes avec celui de J0218+4232 (\emph{cf.} Figure \ref{phasos2}). En effet pour ce pulsar le faisceau radio présente deux sous-pics formant une largeur totale d'environ 0,2, largeur que l'on retrouve dans le pic d'émission $\gamma$ ici centré autour de 0,5. La structure du faisceau $\gamma$ semble présenter deux pics, en particulier à haute énergie, la séparation n'étant pas significative avec le lot de données actuel. Comme pour J0218+4232, le décalage $\delta$ est d'environ 0,5. Avec plus de photons, il sera possible de mesurer le décalage plus précisément et de conclure sur la présence d'un éventuel deuxième pic.
\item \textbf{J1614$-$2230} : ce MSP a également été découvert au cours d'une recherche de pulsars dans les boîtes d'erreur de sources EGRET non identifiées \citep{Crawford2006}. Cette fois il s'agissait de la source du troisième catalogue 3EG J1616$-$2221. La courbe de lumière $\gamma$ présente deux pics, séparés de $\Delta =$ 0,5 et dont le premier succède au pic radio de $\delta =$ 0,2. Comme pour J0030+0451, le décalage radio / $\gamma$ et la séparation des pics sont semblables à ceux des pulsars jeunes. En revanche, si un bridge est clairement observé chez J0030+0451, la statistique est trop faible à présent pour déterminer si J1614$-$2230 présente une émission inter-pics. 
\item \textbf{J1744$-$1134} : ce pulsar milliseconde isolé a été découvert par le radiotélescope de Parkes \citep{Bailes1997}. ROSAT a montré que la source RX J1744.4$-$1134 est vraisemblablement associée au pulsar milliseconde, mais le nombre de photons enregistrés était insuffisant pour la recherche de pulsations \citep{Becker1999}. La complexité de l'émission $\gamma$ de J1744$-$1134 à haute énergie est frappante. Un pic principal se dégage, séparé de $\delta =$ 0,85 du pic radio. Il est pour l'instant difficile de dénombrer les pics. Cependant le phasogramme au-dessus de 1 GeV suggère qu'un certain nombre de composantes secondaires existent.
\item \textbf{J2124$-$3358} : tout comme J1744$-$1134, ce pulsar est isolé et a été détecté en radio à Parkes \citep{Bailes1997}. Des pulsations ont été découvertes en rayons X grâce aux télescopes ROSAT \citep{Becker1999} et XMM \citep{Zavlin2006}. Ces instruments ont mis en évidence un profil d'émission X comportant un pic principal asymétrique, mais n'ont malheureusement pas pu mesurer le décalage radio/X. La complexité de l'émission radio de PSR J2124$-$3358 est remarquable, suggérant que le pulsar émet tout au long de sa rotation. Un pic principal est néanmoins observable en radio, que l'on trouve en rayons $\gamma$ décalé de $\delta =$ 0,85 comme chez J1744$-$1134. Un seul pic $\gamma$ se dégage dans les données actuelles. \'Etant donnée la structure complexe de l'émission radio, la courbe de lumière $\gamma$ pourrait également exhiber de multiples composantes. 
\end{itemize}

\sfig[Courbes de lumière radio et $\gamma$ pour quatre MSPs détectés]{chap5/phasos2.tex}{Courbes de lumière $\gamma$ au-dessus de 1 GeV et de 100 MeV, et profil radio pour les quatre autres MSPs détectés par le LAT. Deux rotations sont montrées.}{phasos2}

La Table \ref{proprietes_msps} liste les valeurs de $\delta$ et $\Delta$. En résumé, J0030+0451 et J1614$-$2230 possèdent des profils $\gamma$ à deux pics séparés de $\Delta \simeq$ 0,45 et retardant de $\delta \simeq$ 0,15 par rapport à l'émission radio, des caractéristiques identiques à celles de beaucoup des pulsars $\gamma$ ordinaires. J0437$-$4715 et J0613$-$0200 ne présentent qu'un pic, décalé d'environ 0,4 par rapport à la radio, ce qui les rend qualitativement similaires à J2229+6114, pulsar $\gamma$ jeune \citep{FermiJ2229}. Pour les quatre MSPs restants, le nombre de photons est encore insuffisant pour conclure sur la multiplicité des pics et leur position. 

\subsection{Détections à venir ?}

Outre les huit détections de MSPs mentionnées précédemment, la Figure \ref{histoHtest} indique que PSR J0034$-$0534 et J1713+0747 sont détectés marginalement, avec des significativités de 4,2 $\sigma$ et 3,7 $\sigma$, respectivement. On a vu que pour les détections certaines, la valeur de $H$ croît de façon linéaire (\emph{cf.} Figure \ref{prog_lin}). Le même graphe de progression de $H$ en fonction du temps constitue par conséquent un indicateur des détections à venir : on s'attend en effet à ce qu'un pulsar détectable voie sa significativité augmenter régulièrement avec le temps, et que la significativité varie de façon erratique pour un pulsar non émetteur.

La Figure \ref{prog_lin2} montre, pour les deux pulsars marginalement détectés, la progression de $H$ en fonction du temps. Pour les deux pulsars, $H$ est globalement croissant (en particulier pour PSR J0034$-$0534, où la croissance est quasi-monotone). L'allure des courbes amène à penser que ces pulsars sont bel et bien émetteurs $\gamma$. Des ajustements linéaires sont montrés par des lignes pointillées obliques. Ils atteignent $H =$ 39,5 (équivalent à 5 $\sigma$) entre 2,85 et 2,95 $\times 10^8$ MET s. Les pulsars devraient donc être détectés par le LAT au cours des mois à venir en utilisant le même jeu de coupure, en supposant que ces régimes de progression se maintiennent. Nous verrons par la suite que ces deux pulsars sont associés à des sources d'émission continue détectées par le LAT, renforçant l'hypothèse selon laquelle les deux MSPs sont émetteurs de rayons $\gamma$. 

\sfig[\emph{H-test} en fonction du temps pour J0034$-$0534 et J1713+0747]{chap5/prog_lin2.tex}{Progression de la valeur de $H$ en fonction du temps, pour les pulsars milliseconde J0034$-$0534 et J1713+0747. La valeur de $H$ équivalant à une détection à 5 $\sigma$ est indiquée par une ligne pointillée horizontale, et un ajustement linéaire est montré par une ligne pointillée oblique.}{prog_lin2}

Les courbes de lumière actuelles pour PSR J0034$-$0534 et J1713+0747 au-dessus de 100 MeV et de 1 GeV sont présentées en Figure \ref{phasos_0034_1713}, ainsi que leur profil d'émission radio à 1,4 GHz. Pour J0034$-$0534, il semble que le profil d'émission $\gamma$ comprenne deux pics, séparés d'environ 0,3, et en alignement avec les pics radio. Dans la mesure où les huit MSPs détectés à présent ont leurs pics $\gamma$ et radio en décalage, J0034$-$0534 pourrait être le premier MSP à avoir ses pics alignés dans les deux gammes d'énergie, tout comme le Crabe. Il pourrait également être intéressant d'observer le pulsar en rayons X. Jusqu'à présent les recherches en X se sont avérées infructueuses \citep{Zavlin2006}. 

En ce qui concerne PSR J1713+0747, la description est moins aisée, bien qu'une structure semble se dégager vers 0,4 en phase. Si le MSP est confirmé en tant qu'émetteur $\gamma$ pulsé, il semble à présent qu'il entrerait dans la catégorie des pulsars à un seul pic radio et $\gamma$, séparés d'environ 0,4 en phase, comme J0437$-$4715, J0613$-$0200 ou encore le pulsar jeune J2229+6114 \citep{FermiJ2229}.

\sfig[Phasogrammes $\gamma$ de PSR J0034$-$0534 et J1713+0747]{chap5/phasos_0034_1713.tex}{Phasogrammes $\gamma$ de PSR J0034$-$0534 et J1713+0747, probables futures détections de MSPs par le LAT. Les cadrans supérieurs montrent les courbes de lumière au-dessus de 100 MeV et 1 GeV. Les parties inférieures montrent les profils radio à 1,4 GHz, acquis avec le radiotélescope de Nançay.}{phasos_0034_1713}

\subsection{Analyse spectrale}

L'analyse spectrale des huit MSPs détectés dans cette étude a été réalisée à l'aide des méthodes décrites dans \citet{FermiVela}\footnote{Dans cette partie, les résultats spectraux présentés sont ceux présentés dans \citet{FermiMSPs}, correspondant à un lot de données s'étendant jusqu'au 15 mars 2009.}. De façon schématique, l'émission $\gamma$ est ajustée par un modèle comprenant des fonds galactique, extragalactique et instrumental (ces deux dernières sources de fond étant isotropes), ainsi que les sources avoisinantes et le pulsar lui-même, dont on souhaite mesurer les propriétés spectrales. La fonctionnelle utilisée pour l'ajustement du spectre des pulsars milliseconde est une loi de puissance à coupure exponentielle, conformément à l'équation \ref{eqEmission}, avec le paramètre de forme $b$ égal à 1 : 

\begin{equation}
\frac{dN}{dE} = N_0 \left( \frac{E}{\mathrm{1\ GeV}}  \right)^{-\Gamma} e^{- \frac{E}{E_c}}
\end{equation}

Rappelons que dans cette expression $E$ représente l'énergie, $\Gamma$ est l'indice spectral et $E_c$ est l'énergie de coupure du spectre. Le flux $dN/dE$ correspond au nombre de photons provenant de la source et détecté par unité de temps, d'énergie, et de surface du détecteur. La réponse de l'instrument (résolution angulaire, surface effective, résolution en énergie) est variable en fonction de l'énergie, comme on l'a évoqué dans le chapitre précédent ; elle doit donc être prise en compte par l'estimateur lors du calcul de $N_0$, $\Gamma$ et $E_c$. La fonction de réponse de l'instrument utilisée pour cette analyse spectrale est P6\_V1\_DIFFUSE. Ces estimations de la réponse du LAT ont été établies avant lancement. Il semble à présent que ces IRFs sous-estiment les flux mesurés, comme indiqué par les erreurs systématiques citées en légende de la Table \ref{proprietes_msps}. Par intégration de $dN/dE$ sur l'énergie, on peut calculer le flux de photons $f$ au-dessus de 100 MeV, ainsi que le flux en énergie $h$, par :

\begin{eqnarray}
f & = & \int_{\mathrm{100\ MeV}}^\infty \left( \frac{dN}{dE} \right) dE\\
h & = & \int_{\mathrm{100\ MeV}}^\infty E \left( \frac{dN}{dE} \right) dE
\end{eqnarray}

La Table \ref{proprietes_msps} donne les valeurs de $\Gamma$, $E_c$, $f$ et $h$ pour les 8 MSPs. Deux exemples de spectres en énergie sont présentés en Figure \ref{spectres_J0030_J1614}. PSR J0030+0451 est le pulsar milliseconde le plus brillant vu de la Terre ; son analyse spectrale est relativement aisée. PSR J1614$-$2230 est un exemple de pulsar moins lumineux pour lequel les paramètres spectraux sont plus difficilement contraints, la conséquence étant ici une erreur statistique importante sur le flux de photons $f$ (\emph{cf.} Table \ref{proprietes_msps}), en particulier à cause du mauvais ajustement à basse énergie, où la majorité des photons sont émis.

\tab[Propriétés spectrales des huit pulsars milliseconde détectés]{spectres.tex}{Propriétés des MSPs détectés avec le Fermi-LAT. Le paramètre $\delta$ correspond au décalage en phase entre le maximum de l'émission radio et le pic $\gamma$ le plus proche, $\Delta$ est la séparation des pics $\gamma$ pour les profils à deux pics. Les paramètres $f$ et $h$ donnent les flux en photons et en énergie, au-dessus de 100 MeV. Enfin, $\Gamma$, $E_c$ et $L_\gamma = 4 \pi h d^2$ sont l'indice spectral, l'énergie de coupure et la luminosité $\gamma$ au-dessus de 100 MeV. Les incertitudes citées ici sont statistiques. Les erreurs systématiques, liées à l'incertitude sur la réponse de l'instrument et sur le fond diffus sont (-0,1 ; +0,3) pour $\Gamma$, (-10\% ; +20\%) pour $E_c$, (-10\% ; +30\%) pour $f$ et (-10\% ; +20\%) pour $h$.}{proprietes_msps}

\sfig[Distributions spectrales en énergie pour PSR J0030+0451 et J1614$-$2230]{chap5/spectres.tex}{Distributions spectrales en énergie pour les pulsars J0030+0451 et J1614$-$2230, avec leurs ajustements par des lois de puissance à coupure exponentielle. Les paramètres de ces ajustements sont donnés en Table \ref{proprietes_msps}. Crédit : T.~J. Johnson, M. Kerr.}{spectres_J0030_J1614}


Les valeurs de luminosité $\gamma$ $L_\gamma = 4 \pi f_\Omega h d^2$ (\emph{cf.} équation \ref{Lgamma}) sont également indiquées dans la Table \ref{proprietes_msps}. En l'absence d'hypothèse sur la géométrie de l'émission $\gamma$ des MSPs, nous avons utilisé $f_\Omega = 1$. Notons néanmoins que $f_\Omega$ est certainement variable d'un pulsar à l'autre, aussi une incertitude importante est introduite dans le calcul de $L_\gamma$. Pour six des huit MSPs, la distance est fiable, provenant de mesures de parallaxes trigonométriques ou chronométriques \citep{Lommen2006,Hotan2006,Deller2008}. Les distances de PSR J0218+4232 et J1614$-$2230 sont basées sur le modèle NE2001 de densité d'électrons dans la Galaxie \citep{NE2001}. Si les distances dérivées du modèle NE2001 sont fiables en moyenne, des comparaisons particulières avec des distances issues de mesures de parallaxes ont montré que NE2001 produit des résultats erronés dans de nombreux cas (\emph{cf.} Figure \ref{ne2001error}). L'efficacité d'émission $\gamma$ au-dessus de 100 MeV, $\eta = L_\gamma / \dot E$, varie de 2\% pour J0437$-$4715 à environ 30\% pour J1744$-$1134, à l'exception de PSR J1614$-$2230 pour lequel $\eta$ est d'environ 100\%. Ceci indique que la distance du pulsar est vraisemblablement surestimée par le modèle NE2001. Une distance plus faible réduirait l'effet Shklovskii dont la contribution à $\dot P$ est $\dot P_{Shk} = \mu^2 d / (c P)$, augmentant ainsi $\dot E$ ; tout en diminuant $L_\gamma$. La valeur de l'efficacité s'en trouverait ainsi réduite. Si l'on ne tient pas compte de la variation de $\dot E$ due à l'effet Shklovskii, une efficacité de 10\% requiert une distance inférieure d'un facteur $\simeq 3$, amenant J1614$-$2230 à $d \simeq$ 0,4 kpc.

\fig[Histogramme des erreurs sur les distances dérivées du modèle NE2001]{width=12cm}{chap5/ne2001error.pdf}{Histogramme des erreurs sur les distances dérivées du modèle NE2001, pour les pulsars ayant une mesure de parallaxe. Les erreurs sont nulles en moyenne, mais ont une déviation standard de 2 dB et s'étendent jusqu'à 6 dB. Figure extraite de \citet{Deller2009}.}{ne2001error}

Bien que le cas de figure ne soit pas prévu par les différents modèles théoriques, remarquons qu'il sera judicieux à l'avenir d'ajuster les spectres des MSPs en introduisant le paramètre de forme $b$ (\emph{cf.} équation \ref{eqEmission}), et ainsi s'assurer que celui-ci est compatible avec 1. Les données actuelles ne permettent pas à présent d'effectuer cette vérification et d'en tirer des conclusions significatives.

\subsection{\'Emission continue}

La recherche de sources d'émission continue parmi les 72 MSPs galactiques est complémentaire de la recherche de pulsations. En effet elle fournit un \emph{a priori} fort sur les pulsars pour lesquels la détection de pulsations est envisageable ou ne l'est pas.

On peut résumer la méthode mise en place dans le cadre de la construction des catalogues de sources continues de Fermi de la façon suivante : dans un premier temps on recherche des régions du ciel où la densité de photons détectés dévie d'un modèle de fond galactique et extragalactique (un seuil est fixé sur la déviation pour limiter le nombre de sources -- par exemple 4 $\sigma$). La seconde étape consiste à affiner la position de chaque source candidate et déterminer l'incertitude sur celle-ci. L'étape finale est l'estimation de la significativité de l'émission continue observée, c'est-à-dire la probabilité que cette émission provienne de fluctuations statistiques. Pour cela un algorithme de vraisemblance compare les probabilités des situations \{fond uniquement\} et \{fond + source hypothétique\} où la source est représentée par une loi de puissance (soit deux paramètres : flux et indice spectral). La quantité $TS = 2 \Delta$ln(vraisemblance) est une mesure de la significativité de la source. La probabilité d'obtenir au moins $TS$ est environ égale à 0,5 fois l'espérance associée à une distribution de $\chi^2$ à deux degrés de liberté, de sorte qu'une significativité de 5 $\sigma$ correspond à $TS \simeq$ 30. Pour plus de détails, le lecteur pourra se reporter à \citet{FermiBSL}. La Table \ref{TS_MSPs} donne les valeurs de $TS$ pour les MSPs associés à des sources d'émission continue dont la significativité est supérieure à 5 $\sigma$.

\tab[Pulsars milliseconde associés à des sources $\gamma$ continues]{TS_MSPs.tex}{Table des quinze MSPs galactiques associés à des sources continues, dont la significativité est supérieure à 5 $\sigma$. Les objets pour lesquels un signal pulsé est détecté sont indiqués par les lettres $p$ (détection certaine) ou $m$ (détection marginale).}{TS_MSPs}

On note que les huit détections pulsées de MSPs sont également observés comme sources continues. De même, J0034$-$0534 et J1713+0747 sont associés à des sources continues, ce qui renforce la perspective de détection pulsée pour ces deux pulsars, comme présenté précédemment. 

La situation est moins claire pour ce qui est des cinq pulsars restants, J0610$-$2100, J0621+1002, J1600$-$3053, J1939+2134 et J1959+2048. Aucun de ces cinq MSPs, pour lesquels des éphémérides contemporaines étaient disponibles lors de cette étude, n'est détecté comme source $\gamma$ pulsée, y compris marginalement. Davantage de photons et un jeu de coupures différent pourraient permettre la détection de ces pulsars. En outre, la nébuleuse à vent de pulsar (\emph{Pulsar Wind Nebula}, PWN) G59.2$-$4.7, émettrice de rayons X, est associée à J1959+2048 \citep{Stappers2003}. Cette nébuleuse pourrait être responsable de l'émission continue en $\gamma$. Une analyse détaillée des propriétés spectrales et spatiale de l'émission pourrait révéler la nature de la source.

\section{Discussion}

\subsection{Propriétés des MSPs détectés}

Les courbes de lumière $\gamma$ des pulsars milliseconde et l'alignement relatif entre les composantes radio et $\gamma$ sont similaires à ce qui est observé par ailleurs pour les pulsars normaux. Par exemple, les profils à deux pics de J0030+0451 et de J1614$-$2230, séparés de $\Delta \simeq$ 0,45 et dont le premier pic retarde de $\delta \simeq$ 0,15 par rapport à la radio sont relativement communs parmi les pulsars ordinaires \citep{FermiVela,FermiJ1028,FermiJ0205,FermiJ2021}. Un plus grand nombre de MSPs détectés présente un seul pic, séparé de $\delta \simeq$ 0,4 -- 0,5 de l'émission radio, à l'image de J2229+6114 \citep{FermiJ2229}. On en conclut que la géométrie de l'émission $\gamma$ dans la magnétosphère des MSPs d'une part et des pulsars jeunes d'autre part est relativement similaire.

L'analyse spectrale, dont les résultats sont donnés dans la Table \ref{proprietes_msps}, a montré que les indices sont généralement durs, les valeurs de $\Gamma$ étant typiquement inférieures à 2. Les indices spectraux sont donc comparables à ceux des pulsars normaux. Le spectre le plus dur est celui de PSR J1614$-$2230, dont l'indice a été mesuré à 1,0 $\pm$ 0,3. Le durcissement des spectres en fonction de l'âge caractéristique avait été observé pour les pulsars d'EGRET \citep{Fierro1993}. Il est intéressant de remarquer que les valeurs de $\Gamma$ mesurées pour les MSPs semblent décorrélées de $B_S$, $B_{LC}$ et $\dot E$. La Figure \ref{tendances} (A) montre l'âge caractéristique $\tau = P / (2 \dot P)$ en fonction de l'indice spectral $\Gamma$ pour les MSPs détectés. Aucune corrélation ne semble se dégager entre ces deux grandeurs. Cependant, pour ces pulsars dont la rotation a été accélérée par un processus de recyclage, l'âge caractéristique n'est pas représentatif de l'âge véritable de ces objets, comme on l'a évoqué précédemment. Dans ces conditions, une corrélation éventuelle entre l'âge et l'indice spectral est probablement indécelable sans une connaissance de l'histoire d'accrétion. 

\sfig[Propriétés spectrales des MSPs détectés]{chap5/tendances.tex}{Indice spectral en fonction de l'âge caractéristique $\tau$ et énergie de coupure $E_c$ en fonction du champ magnétique surfacique $B_S$ pour les huit MSPs détectés, où $\tau$ et $B_S$ ont été corrigés de l'effet Shklovskii.}{tendances}

Les énergies de coupure exponentielle $E_c$ sont comprises entre 1 et 4 GeV, si l'on fait exception de J0218+4232 pour lequel l'incertitude sur $E_c$ est importante ($E_c = 7 \pm 4$ GeV). \`{A} nouveau, ces caractéristiques sont comparables à celles des pulsars émetteurs $\gamma$ normaux. Ces énergies de coupure ne présentent pas de corrélation particulière avec $B_{LC}$ et $\dot E$. Comme pour l'indice spectral des pulsars vus par EGRET, il est apparu que l'énergie de coupure des pulsars de CGRO semble diminuer lorsque $B_S$ augmente (\emph{cf.} Figure 4 de \citet{Thompson2008_2}). Il faut cependant mettre un bémol, car la tendance repose encore une fois sur les valeurs minimale et maximale, celles de B1509$-$59 et B1951+32. Pour ce dernier, l'analyse des données du LAT révèle que $E_c$ a été surestimée par EGRET\footnote{Information apportée par T. Reposeur. L'analyse de B1951+32 par le LAT sera publiée prochainement.}. La Figure \ref{tendances} (B) montre les valeurs de $E_c$ en fonction du champ magnétique surfacique, $B_S$. La dispersion est grande, principalement à cause des incertitudes sur $E_c$. Néanmoins, bien qu'aucune corrélation claire ne se dégage d'après le graphe, il semble que la tendance vue pour les pulsars de CGRO ne soit pas confirmée, à savoir, l'énergie de coupure exponentielle $E_c$ ne diminue pas en fonction du champ magnétique surfacique $B_S$.

La luminosité $\gamma$ au-dessus de 100 MeV, $L_\gamma = 4 \pi f_\Omega h d^2$ est représentée en fonction de $\dot E$ pour les MSPs et les pulsars normaux détectés jusqu'à présent dans la Figure \ref{LgammaEdot}. Pour les MSPs, une valeur de 1 a été utilisée pour le facteur de correction du flux $f_\Omega$. Les valeurs de $f_\Omega$ prédites par le modèle \emph{Outer Gap} ont été adoptées ici pour les pulsars normaux (l'utilisation des facteurs de correction issus du modèle \emph{Slot Gap} ne modifierait pas qualitativement la Figure \ref{LgammaEdot}), d'après \citet{Watters2009}. Une ligne pointillée, représentant un ajustement de la luminosité $\gamma$ des pulsars normaux par une loi en $\sqrt{\dot E}$ est également tracée. La luminosité de J0218+4232, un MSP très énergétique, est en bon accord avec cette loi. Les sept autres MSPs sont distincts de J0218+4232 de par leur faible $\dot E$. Pour ces objets (à l'exception de J1614$-$2230 pour lequel la luminosité $\gamma$ est certainement surestimée comme on l'a déjà évoqué), $L_\gamma$ est inférieur à la valeur attendue selon $\sqrt{\dot E}$. En particulier, la luminosité $\gamma$ de J0437$-$4715 n'est pas conciliable avec la loi empirique, à moins d'un facteur de correction $f_\Omega$ irréaliste. On en conclut qu'il doit exister une valeur de $\dot E$ seuil, en deçà de laquelle la relation entre $L_\gamma$ et $\dot E$ s'infléchit, afin d'expliquer les efficacités $\gamma$ de ces MSPs. D'autres détections de pulsars, milliseconde ou normaux, dans ce régime de faible $\dot E$ pourraient aider à comprendre la relation entre luminosité et perte d'énergie par freinage électromagnétique. 

\fig[Luminosité $\gamma$ des MSPs détectés en fonction de $\dot E$]{angle=270,width=14cm}{chap5/LgammaEdot.pdf}{$L_\gamma$ en fonction de $\dot E$ pour les huit pulsars milliseconde détectés (points et traits pleins) et les pulsars normaux (carrés vides et traits pointillés). Les valeurs de $L_\gamma$ pour les MSPs sont données dans la Table \ref{proprietes_msps}. Elles sont calculées en utilisant $f_\Omega = 1$. Les barres d'erreur sur $L_\gamma$ pour les MSPs ne tiennent pas compte de la contribution de l'incertitude sur $f_\Omega$. Pour les pulsars normaux, les valeurs de $f_\Omega$ et les incertitudes associées proviennent du modèle \emph{Outer Gap} (Table 1 de \citet{Watters2009} et K. Watters, communication privée), à savoir $f_\Omega = 1$ sauf pour CTA1 (0,6 $\pm$ 0,3), Geminga (0,125 $\pm$ 0,025), J0205+6449 (0,95 $\pm$ 0,05), J1028$-$5819 (1,1), J1057$-$5226 (0,55), J1709$-$4429 (0,85 $\pm$ 0,15), J1952+3252 (0,925 $\pm$ 0,175) et J2021+3651 (1,05).}{LgammaEdot}

Dans cette étude nous avons recherché des pulsations à partir d'éphémérides contemporaines pour tous les MSPs, à six exceptions près. Par conséquent, il est possible de distinguer les objets pour lesquels une émission pulsée a été détectée, parmi la population de MSPs connus ou plus généralement parmi la population des pulsars. La Figure \ref{DistEdot} montre les 72 MSPs galactiques connus dans un diagramme $\dot E$ -- distance. On constate que les pulsars pour lesquels des pulsations sont détectées sont énergétiques et généralement proches. J0218+4232 est lui plus éloigné, mais a un $\dot E$ plus fort. PSR J0034$-$0534, dont les perspectives de détection sont importantes, est également un pulsar à grand $\dot E$ et à distance faible. En résumé, les MSPs détectés se distinguent par un grand $\dot E / d^2$. 

\fig[$\dot E$ en fonction de la distance pour les 72 MSPs du champ galactique connus]{width=14cm}{chap5/DistEdot.pdf}{$\dot E$ en fonction de la distance pour les 72 MSPs du champ galactique connus. Les huit pulsars détectés comme sources pulsées sont indiquées par des points pleins. Les triangles indiquent les pulsars associés à des sources $\gamma$ continues. Les carrés vides sont les MSPs pour lesquels des éphémérides contemporaines n'étaient pas disponibles. Les autres pulsars sont indiqués par des cercles.}{DistEdot}

La Figure \ref{Edotd2_P0} montre $\dot E / d^2$ en fonction de $P$ pour la population de pulsars milliseconde ainsi que pour les pulsars normaux du champ galactique. Conformément à ce qui est observé dans la Figure \ref{DistEdot}, les MSPs détectés dominent en terme de $\dot E / d^2$, de même que les pulsars normaux émetteurs $\gamma$ sont distribués dans la partie haute. On remarque que tous les pulsars détectés sont au-dessus d'un seuil situé à environ $5 \times 10^{33}$ erg/s/kpc$^2$, un autre point commun entre les deux populations de pulsars. Un certain nombre d'objets non détectés sont néanmoins situés au-dessus de cette valeur (\emph{cf.} Table \ref{nondetectes}). Différentes causes peuvent l'expliquer : d'abord les distances peuvent être erronées, notamment lorsqu'elles sont calculées à partir d'un modèle de distribution galactique d'électrons, ce qui est le cas pour PSR J1012+5307, J1843$-$1113, J1911$-$1114, J1933$-$6211 et J2129$-$5721. Rappelons cependant que les paramètres de rotation utilisés pour PSR J1933$-$6211 sont anciens, aussi le résultat de la recherche de pulsation pour cet objet est à considérer avec précaution. Néanmoins ce pulsar n'est pas détecté en tant que source d'émission continue, contrairement aux huit MSPs pour lesquels des pulsations sont observées, ce qui signifie que le pulsar n'est vraisemblablement pas détectable à présent. En revanche, on dispose pour PSR J1024$-$0719 et J1909$-$3744 de distances fiables basées sur la mesure de leur parallaxe chronométrique \citep{Hotan2006}. Une autre raison possible de ces non-détections est l'orientation de l'émission : le faisceau $\gamma$, s'il existe, peut être orienté de manière défavorable par rapport à la ligne de visée, empêchant ainsi la détection. Enfin, nous avons vu dans le premier chapitre que l'incertitude sur le moment d'inertie $I$ des étoiles à neutrons est importante : celui-ci pourrait donc prendre une valeur plus faible que les $10^{45}$ g cm$^2$ généralement employés ; de sorte qu'à un ralentissement $\dot P$ donné corresponde un taux de perte d'énergie $\dot E$ plus faible.

\tab[Table des MSPs à $\dot E / d^2 > 5 \times 10^{33}$ erg/s/kpc$^2$ non détectés]{nondetections.tex}{Table des MSPs non-détectés en tant que sources d'émission pulsée ou continue, et dont le flux $\dot E / d^2$ est supérieur à $5 \times 10^{33}$ erg/s/kpc$^2$.}{nondetectes}

\fig[$\dot E$ normalisé à la distance au carré en fonction de la période]{width=14cm}{chap5/Edotd2_P0.pdf}{$\dot E$ divisé par la distance au carré en fonction de la période de rotation pour les pulsars situés en dehors des amas globulaires. Les MSPs et pulsars normaux émetteurs $\gamma$ pulsés sont indiqués par des points pleins (pulsars normaux : détections de COMPTEL, EGRET et AGILE, et détections récentes du LAT \citep{FermiJ0205,FermiBSL}). Les MSPs probablement associés à des sources $\gamma$ continues sont montrés par des triangles. Les MSPs pour lesquels des éphémérides contemporaines n'étaient pas disponibles pour cette étude sont représentés par des carrés vides. Les pulsars non détectés jusqu'à présent sont indiqués par des cercles pour les MSPs, et par des points fins pour les pulsars normaux. Figure issue de \citep{FermiMSPs}.}{Edotd2_P0}

\subsection{Conclusion}

Les similitudes des courbes de lumière, des propriétés spectrales et du rôle de $\dot E$ dans l'émission $\gamma$ entre les MSPs et les pulsars normaux détectés suggèrent fortement que les mêmes mécanismes de production de rayonnement $\gamma$ opèrent chez ces deux classes d'objets, et ce malgré les écarts de période de rotation et de taux de ralentissement. Les champs magnétiques surfaciques $B_S$ diffèrent de plusieurs ordres de grandeur entre les MSPs et les pulsars ordinaires. En revanche, les valeurs du champ magnétique au cylindre de lumière $B_{LC}$ pour les pulsars émetteurs $\gamma$, milliseconde ou ordinaires, sont tout à fait comparables. De plus ces pulsars sont généralement distribués aux grandes valeurs de $B_{LC}$, comme le montre la Figure \ref{Blc_P0}. Ceci suggère que les propriétés électromagnétiques de la magnétosphère externe ont un rôle important dans le processus de rayonnement $\gamma$.

\fig[Champ magnétique au cylindre de lumière en fonction de la période]{width=14cm}{chap5/Blc_P0.pdf}{Champ magnétique au cylindre de lumière $B_{LC}$ en fonction de la période de rotation pour les pulsars situés en dehors des amas globulaires (\emph{cf.} Figure \ref{Edotd2_P0} pour la signification des différents symboles).}{Blc_P0}

Pour les modèles \emph{Polar Cap}, le rayonnement prend son origine au-dessus des calottes polaires. Par conséquent ces modèles prédisent des faisceaux de rayons $\gamma$ en alignement proche avec l'émission radio. Au contraire, les modèles \emph{Slot Gap} et \emph{Outer Gap} prévoient une production de photons $\gamma$ en altitude dans la magnétosphère, formant des faisceaux larges généralement non alignés avec l'émission radio. Les phasogrammes présentés dans les Figures \ref{phasos1} et \ref{phasos2} montrent que les pics $\gamma$ sont toujours en décalage par rapport à la radio, ce qui suggère à nouveau une origine du rayonnement dans la magnétosphère externe. Enfin, les énergies de coupure mesurées sont de quelques GeV, ce qui indique que l'accélération de particules ne se produit pas au-dessus des pôles où le champ électrique est plus intense, et où les énergies de coupure atteindraient 10 GeV ou plus \citep{Bulik2000}.

En résumé, les modèles théoriques pour lesquels l'accélération de particules se produit dans des régions situées en altitude, comme \emph{Slot Gap} ou \emph{Outer Gap}, sont favorisés pour les MSPs comme ils le sont pour les pulsars normaux \citep{FermiJ1028,FermiJ2021,FermiJ0205,FermiVela}. Cependant, les observations ne permettent pas à présent de discriminer parmi ces deux types de modèle. Plus de détections de MSPs, plus de détails dans les courbes de lumière et davantage de précision sur les propriétés spectrales pourraient permettre des comparaisons particulières entre modèles théoriques.

%% file: part3/chap6.tex
\chapter[Pulsars milliseconde des amas globulaires]{Pulsars milliseconde des amas globulaires}

\minitoc

\chaptabst{P}{lus}{de la moitié des MSPs recensés appartiennent à des amas globulaires, tels que 47 Tucanae ou Terzan 5. Nous avons vu précédemment que certains MSPs émettent un rayonnement $\gamma$ détectable par le LAT. De plus, des amas globulaires sont détectés en rayons $\gamma$ en tant que sources d'émission continue. Les perspectives de détection de MSPs au sein de ces amas sont donc importantes, motivant la recherche de pulsations pour ces pulsars. Dans ce dernier chapitre nous présentons les résultats de cette recherche de pulsations. Aucun pulsar n'est fermement détecté en tant que source de rayonnement $\gamma$ pulsé, bien que pour la plupart d'entre eux, ces pulsars n'ont pas été couverts lors de la campagne de chronométrie. Un autre facteur rédhibitoire est certainement leur grande distance, par comparaison aux huit MSPs galactiques détectés. Nous nous penchons enfin sur le cas de J1824$-$2452A dans l'amas M28, marginalement détecté par AGILE et de prime abord non détecté dans notre étude. Bien que la courbe de lumière vue par le LAT ne corresponde pas à une détection ferme, elle est qualitativement différente de celle observée par AGILE.}

\section{Intérêt}

Les amas globulaires sont des groupes d'étoiles très denses, situés dans le halo galactique. Ce sont des formations anciennes (typiquement, quelques milliards d'années), intéressantes du point de vue de l'évolution stellaire. En raison de la grande densité des amas globulaires, les collisions entre étoiles y sont plus fréquentes que dans le champ galactique \citep{Verbunt1987}. Une conséquence des interactions fréquentes entre étoiles est le grand nombre de systèmes multiples \citep{Clark1975}, ce qui est propice au recyclage des étoiles à neutrons en pulsars milliseconde. Ainsi, les amas globulaires sont riches en MSPs. \`{A} ce jour, près de 150 pulsars ont été détectés dans 26 amas globulaires (\emph{cf.} Annexe \ref{annexeB}, et voir \citet{Camilo2005} pour une revue récente).

Avant l'arrivée du LAT, les amas globulaires avaient été observés et détectés dans toutes les longueurs d'onde, à l'exception du domaine $\gamma$. Depuis, le LAT a détecté l'amas 47 Tucanae en tant que source d'émission continue \citep{Fermi47Tuc}. Le spectre d'émission de 47 Tucanae est bien représenté par une loi de puissance à coupure exponentielle, avec un indice spectral $\Gamma =$ 1,3 $\pm$ 0,3 et une énergie de coupure $E_c = $2,5$_{-0,8}^{+1,6}$ GeV. Par ailleurs, le catalogue de sources continues de Fermi après six mois d'observation comprend trois autres amas globulaires dans lesquels des MSPs sont connus (\emph{cf.} Table \ref{DCTS_GC}). Pour expliquer l'émission $\gamma$ de 47 Tucanae, un scénario proposé est la production de vents de particules relativistes par les MSPs, interagissant avec le vent des autres étoiles ou d'autres MSPs de l'amas, créant ainsi des chocs capables d'accélérer des particules chargées jusqu'au TeV \citep{Bednarek2007}. Les auteurs prédisent un flux en énergie comparable à celui observé par le LAT, mais en se basant sur une valeur de $\dot E$ moyenne pour les MSPs de 47 Tucanae d'un facteur 30 trop importante \citep{Fermi47Tuc}. Ce scénario semble \emph{a priori} ne pas pouvoir expliquer l'émission $\gamma$ observée.

\begin{table}[bt]
\begin{center}
\begin{minipage}[b][1.05\totalheight][c]{16cm}
\begin{center} 
\input{tableaux/DCTS_GC.tex}
\parbox{15cm}{\caption[Amas globulaires détectés par le LAT et comprenant des MSPs]{Table des amas globulaires coïncidant avec une source continue détectée par le LAT avec une significativité supérieure à 5 $\sigma$, et comprenant au moins un pulsar milliseconde.}\label{DCTS_GC}}\end{center}
\end{minipage}
\end{center}
\end{table}

Comme le montre la Table \ref{proprietes_msps}, les propriétés spectrales des huit MSPs galactiques détectés par le LAT sont tout à fait comparables à ce qui est observé pour 47 Tucanae. Compte tenu du fait que les pulsars milliseconde peuvent émettre du rayonnement $\gamma$ pulsé, comme nous l'avons vu dans le chapitre précédent, l'émission continue détectée par le LAT dans la direction de l'amas globulaire pourrait être due à la contribution ajoutée des pulsars milliseconde de l'amas, dont 23 sont connus mais dont le nombre total pourrait s'élever à 60 objets \citep{Camilo2005}. Pour cette raison, nous avons également recherché des pulsations pour les MSPs des amas globulaires. 

\section{Recherche de pulsations}

\subsection{Analyse des données}

Le lot de données considéré ici est similaire à celui du chapitre précédent (données du LAT enregistrées entre le 30 juin 2008 et le 2 juin 2009). De même, nous avons retenu les événements de classe \emph{Diffuse}, d'énergie supérieure à 100 MeV, et compris dans un rayon de 0,5$^\circ$ autour du pulsar si $|b| < 10^\circ$, ou dans un rayon de 1$^\circ$ sinon. 

Contrairement aux MSPs du champ galactique, l'essentiel des pulsars des amas globulaires n'a pas été suivi par la campagne de datation (\emph{cf.} chapitre II). La Table \ref{ephemGC} indique les pulsars pour lesquels des éphémérides contemporaines ou récentes ont été obtenues.

\tab[Pulsars d'amas globulaires disposant de chronométrie récente ou contemporaine]{ephemGC.tex}{Pulsars des amas globulaires pour lesquels des éphémérides récentes ou contemporaines aux observations du LAT ont été obtenues.}{ephemGC}

Pour les autres objets, les éphémérides utilisées pour la recherche de pulsations sont issues de la base de données de l'ATNF, à l'exception des paramètres de datation des pulsars de 47 Tucanae, mises à jour des résultats de \citet{Freire2003} mais non contemporains. En tout, des éphémérides ont été collectées pour 106 pulsars. Pour la majorité d'entre eux, les éphémérides sont anciennes ou minimales. Par conséquent les résultats de la recherche de pulsations sont à considérer avec prudence. 

La Figure \ref{histoTS_GC} montre l'histogramme des significativités des courbes de lumière obtenues pour les pulsars des amas globulaires, d'après le \emph{H-test}. Aucun MSP n'est détecté à présent avec une significativité de 3 $\sigma$ au moins. La grande distance de ces objets est certainement une cause importante des non-détections : les MSPs étudiés ici ont des distances comprises entre 4 et 10 kpc, tandis que les MSPs détectés par le LAT sont proches, avec des distances inférieures à 1 kpc pour la majorité d'entre eux. Il faut cependant considérer séparément les cas de PSR J0218+4232 et des sept autres pulsars milliseconde émetteurs $\gamma$, nettement en deçà en terme de $\dot E$. Un MSP de luminosité $\gamma$ égale à celle de J0218+4232 et placé à $d \geq$ 4 kpc produirait un flux en énergie $h = L_\mathrm{J0218}/ \left( 4 \pi f_\Omega d^2 \right)$, soit $h \leq$ 1,6 $\times 10^{-11}$ erg/cm$^2$/s, en prenant $f_\Omega = 1$. Cette limite supérieure est tout à fait comparable aux flux en énergie vus par le LAT, pour les huit MSPs émetteurs $\gamma$ galactiques (voir Table \ref{proprietes_msps}). Un pulsar tel que PSR J0218+4232 semble donc détectable, à faible distance ($d \simeq 4$ kpc). Néanmoins le flux attendu décroît rapidement vers des niveaux pour le moment non détectables par le LAT au-delà de 4 kpc. Les sept autres MSPs galactiques émetteurs $\gamma$ sont moins lumineux et plus proches. Leur luminosité $\gamma$ étant de $\overline{L_\gamma} \simeq$ 1,2 $\times 10^{33}$ erg/s en moyenne, le flux en énergie attendu au-delà de 4 kpc est $h \leq 6 \times 10^{-13}$ erg/cm$^2$/s, ce qui très inférieur aux valeurs mesurées actuellement. 


D'autre part, les éphémérides utilisées ici sont généralement approximatives, et empêchent certainement le calcul précis des phases pour certains pulsars. Néanmoins, remarquons que les pulsars correctement couverts, dont les noms sont donnés en Table \ref{ephemGC}, ne sont pas détectés à présent. En particulier, le cas de PSR J1824$-$2452A est intéressant, dans la mesure où une détection marginale de ce MSP a été rapportée par le télescope AGILE au-dessus de 100 MeV. 

\fig[Résultats de la recherche de pulsations pour 106 pulsars d'amas globulaires]{width=12cm}{chap6/histoGC.pdf}{Résultats de la recherche de pulsations pour 106 pulsars d'amas globulaires. Les noms des pulsars pour lesquels la significativité est d'au moins 2 $\sigma$ sont indiqués.}{histoTS_GC}

\subsection{Le cas de PSR J1824$-$2452A dans M28}

Le télescope $\gamma$ AGILE a rapporté la détection du pulsar milliseconde J1824$-$2452A, dans l'amas globulaire M28, avec un niveau de confiance de 4,2 $\sigma$ \citep{Pellizzoni2009}. Cependant, cette valeur de significativité n'est atteinte que pour une partie seulement des observations d'AGILE, alors que le lot de données complet ne mène à aucune détection. Pour l'intervalle de temps favorable, la courbe de lumière vue par AGILE est semblable à une sinusoïde. Le pic principal radio à 1,4 GHz coïncide avec le pic $\gamma$. 

La recherche de pulsations dans les données du LAT avec le jeu de coupures décrit dans le paragraphe précédent conduit, pour J1824$-$2452A, à une significativité inférieure à 1 $\sigma$. Bien que l'article d'AGILE ne détaille pas les coupures angulaires utilisées pour ce pulsar en particulier, il est indiqué que la région d'intégration utilisée respecte la résolution angulaire de l'instrument, avec un rayon d'ouverture maximal de 2$^\circ$, pour les régions bruitées. Nous avons utilisé ce schéma pour produire la courbe de lumière de J1824$-$2452A, à savoir : $\rho = \min(68\%\ \mathrm{PSF}, 2^\circ)$ (rappelons que la PSF du LAT est donnée en Figure \ref{AngRes}). Le phasogramme est montré en Figure \ref{phasoJ1824}. La significativité correspondante est de 3,1 $\sigma$, si l'on ne tient pas compte de l'essai supplémentaire. 

Au-dessus de 100 MeV, le profil est en désaccord avec les résultats d'AGILE \citep{Pellizzoni2009} : on n'observe pas de modulation sinusoïdale, et le pic principal radio à 1,4 GHz n'est aligné avec aucune structure en $\gamma$. En revanche, il est intéressant de constater que les pics radio secondaire et tertiaire (ici à 0,75 et 0,25 en phase, respectivement) sont alignés avec des structures émergeant du niveau de fond dans la courbe de lumière $\gamma$ au-dessus de 100 MeV. Au contraire, le pic radio principal semble ne pas être accompagné d'une émission $\gamma$. Or, ce pulsar milliseconde est émetteur de GRPs, principalement en phase avec le pic radio secondaire, les GRPs restants s'alignant avec le pic tertiaire \citep{Knight2006_2}. Ce parallèle entre émission de GRPs et rayonnement $\gamma$ est intéressant, mais il faut néanmoins attendre davantage de photons pour confirmer la détection de J1824$-$2452A. 

\fig[Courbes de lumière du pulsar milliseconde J1824$-$2452A]{scale=0.37}{chap6/1824m2452_2.pdf}{Courbe de lumière $\gamma$ pour le pulsar milliseconde J1824$-$2452A, dans l'amas globulaire M28. Un profil radio de référence, enregistré au radiotélescope de Nançay à 1,4 GHz est également montré.}{phasoJ1824}

\section{Conclusion}

Parmi les pulsars des amas globulaires, aucun n'est donc détecté à présent. Outre le fait que les éphémérides utilisées dans cette étude n'étaient pas contemporaines des observations $\gamma$ pour la plupart d'entre elles, la non-détection de pulsars s'explique vraisemblablement par leur éloignement. De plus, rappelons qu'à cause de la résolution angulaire du LAT, qui est de quelques degrés à basse énergie, plusieurs sources $\gamma$ sont potentiellement présentes dans le champ de vue, de sorte que chacune d'entre elles participe au bruit de fond des autres objets. Dans l'hypothèse où un pulsar est détecté dans un amas, une stratégie à envisager pour la recherche des autres pulsars est donc d'isoler les photons appartenant à ses pics, puis de les éliminer. Ceci diminuerait le niveau de fond pour les autres sources présentes dans le champ de vue, ce qui pourrait favoriser la détection de pulsations.

La détection de 47 Tucanae par le LAT laisse néanmoins à penser que celui-ci pourrait détecter des pulsations pour des MSPs des amas globulaires au cours des années à venir. Il serait donc intéressant de disposer d'observations radio de ces pulsars afin de construire des éphémérides couvrant les prochains mois d'observation $\gamma$. Bien que la plupart des pulsars milliseconde soient stables, certains présentent des orbites complexes, et leur chronométrie requiert un soin particulier. La détection d'au moins un MSP émetteur $\gamma$ dans un amas globulaire permettrait vraisemblablement de lever l'ambigu\"ité sur le mécanisme d'émission $\gamma$ dans ces formations stellaires.

Enfin, dans l'hypothèse où des pulsars des amas globulaires soient détectés, l'étude du spectre d'émission de ces objets permettrait la comparaison de leurs propriétés spectrales avec celles des pulsars galactiques. La mesure des luminosités $\gamma$ fournirait des contraintes sur les valeurs intrinsèques de $\dot E$ pour ces pulsars, en supposant une efficacité d'émission typique des MSPs connus par ailleurs ($\eta \simeq$ 10\%). \`{A} cause des mouvements locaux, les taux de perte d'énergie rotationnelle intrinsèques sont généralement inaccessibles par chronométrie. Bien que les incertitudes soient importantes, en particulier sur $\eta$ ou sur la distance, des estimations de $\dot E$ apporteraient des informations sur le budget énergétique des pulsars des amas globulaires, dont le scénario d'évolution et l'environnement sont très différents de ceux des pulsars galactiques. 

%% file: tableaux/DCTS_GC.tex
\begin{scriptsize}
\begin{center}
\begin{tabular}{ccc}
\hline
\hline
Amas globulaire & $TS$ & N$_{\mathrm{MSPs}}$ \\
\hline
47 Tucanae & 395,2 & 23 \\
Terzan 5 & 310,6 & 33 \\
M62 & 112,6 & 6 \\
M28 & 106,5 & 12 \\
\hline
\end{tabular}
\end{center}
\end{scriptsize}

%% file: conclusion/conclusion.tex
\partie*{Conclusion} 


Le satellite Fermi est en orbite autour de la Terre depuis le 11 juin 2008. \`{A} son bord, le \emph{Large Area Telescope} (LAT) observe depuis le ciel $\gamma$ entre 20 MeV et plus de 300 GeV, avec une sensibilité bien supérieure à celle de son prédécesseur EGRET, ou d'AGILE, autre télescope $\gamma$ opérant à l'heure actuelle. Avant les observations du ciel par le LAT, seuls neuf pulsars avaient été détectés dans le domaine des rayons $\gamma$, grâce notamment au télescope EGRET. De façon remarquable, le LAT a augmenté le nombre de pulsars $\gamma$ connus d'un facteur cinq : des pulsations ont en effet été observées pour une cinquantaine de pulsars à l'heure actuelle. Ce nombre continuera de croître, la mission Fermi n'achevant que sa première année d'activité. \`{A} l'origine de ces découvertes, le gain en sensibilité du LAT par rapport aux instruments antérieurs a naturellement occupé une place importante. Pour ce qui est des pulsars, deux autres facteurs ont joué des rôles clés : \newline

\begin{itemize}
\item La précision temporelle, testée avant le lancement de Fermi. Ces tests ont justement mis au jour des problèmes importants dans la datation des événements, si bien que moins de pulsars $\gamma$ auraient vraisemblablement été détectés sans ces corrections. En particulier, on peut penser qu'aucun pulsar milliseconde n'aurait été découvert sans cela. On sait désormais que les photons enregistrés par le LAT sont datés avec une précision inférieure à la $\mu$s, ce qui conduira à un niveau de détails sans précédent dans les courbes de lumière, y compris chez les pulsars les plus rapides.\newline
\item la campagne de chronométrie des pulsars émetteurs radio et/ou X, dont le succès a eu différentes conséquences. D'une part la collaboration avant lancement a permis, \emph{via} l'utilisation de données réelles, la vérification et la validation des outils d'analyse temporelle de Fermi. D'autre part, le bon fonctionnement de la coordination des efforts de chronométrie pour le LAT a permis la couverture de presque tous les pulsars définis comme prioritaires (ceux à $\dot E > 10^{34}$ erg/s). L'enthousiasme créé autour des résultats de Fermi permet désormais de chercher des pulsations pour de nombreux pulsars initialement non prioritaires, donc à caractéristiques différentes, ce qui pourrait conduire à des résultats inattendus. Cette thèse a directement bénéficié du bon déroulement de cette campagne de datation, puisque l'essentiel des 72 MSPs galactiques connus a été suivi, de sorte que cette population d'objets a quasiment été étudiée dans son ensemble.\newline
\end{itemize}

La recherche de pulsations en rayons $\gamma$ pour les pulsars milliseconde, présentée dans cette thèse, a conduit à huit détections fermes. Ainsi, le débat de l'émission des MSPs dans le domaine $\gamma$ est clos : les pulsars milliseconde sont de puissants accélérateurs de particules, créant une émission $\gamma$ détectable à grande distance. L'analyse temporelle, basée sur des éphémérides très précises construites à partir d'observations des radiotélescopes de Nançay, Parkes et Green Bank, a révélé que les huit MSPs présentent un ou deux pics $\gamma$, décalés de l'émission radio. Alors que les MSPs sont différents des pulsars normaux \emph{a priori}, de par leur cylindre de lumière beaucoup plus étroit et leur champ magnétique plus faible, les phasogrammes sont tout à fait similaires à ceux des pulsars ordinaires. Les spectres d'émission des MSPs ont des indices relativement durs ($\Gamma < 2$) et des énergies de coupure de l'ordre du GeV. \`{A} nouveau, ces propriétés différent peu de ce qu'on observe pour les autres pulsars. Enfin, on constate que pour les deux classes d'objets, normaux et milliseconde, le LAT détecte les pulsars dont le flux de freinage électromagnétique ($\dot E / d^2$) est important. Le seuil de détection actuel paraît commun entre les deux populations. Ces différents éléments amènent à la conclusion que le mécanisme de production de rayonnement $\gamma$ est commun entre les pulsars normaux et milliseconde. Les modèles d'émission $\gamma$ de type \emph{Polar Cap} semblent exclus par les observations, au profit des modèles \emph{Slot Gap} ou \emph{Outer Gap}. \newline

La prochaine étape dans la compréhension de l'émission $\gamma$ des pulsars est donc la discrimination des modèles situant l'accélération des particules chargées dans la magnétosphère externe. Pour la première fois nous disposons de données concrètes sur l'émission $\gamma$ des MSPs, tant du point de vue temporel que spectral. Les différentes observables sont maintenant à confronter aux prédictions théoriques, afin d'affiner ou d'infirmer les modèles. Les prochaines années d'observation du ciel $\gamma$ par le LAT amèneront plus de photons, donc plus de détails dans les courbes de lumière, et en particulier la possibilité d'étudier l'émission dans des bandes d'énergie étroites. Au bout de cinq à dix ans d'activité, le seuil de détection devrait diminuer d'un facteur deux à trois, augmentant ainsi le nombre de pulsars observables en $\gamma$ et donc la variété des caractéristiques. En particulier, le LAT pourrait détecter des MSPs appartenant à des amas globulaires, ce qui est pour le moment prématuré. Pour les MSPs les plus brillants, comme PSR J0030+0451, une statistique accrue permettra l'analyse spectrale résolue en phase, donnée importante pour la comparaison des modèles théoriques. Par ailleurs, on ne dispose d'observations en rayons X avec datation absolue par rapport à l'émission radio que pour un petit nombre d'objets parmi les huit MSPs détectés dans cette thèse. Ces huit détections motivent l'observation dans le domaine des rayons X, afin de cartographier l'émission au sein de leur magnétosphère dans les différentes gammes d'énergie. Une fois améliorés, les modèles théoriques vont permettre des synthèses de population de MSPs plus réalistes, et ainsi fournir des prédictions plus précises au sujet des MSPs de la Galaxie, leur nombre, leurs propriétés et leur contribution à l'émission $\gamma$ diffuse.\newline

Finalement, la détection de MSPs émetteurs $\gamma$ en bon nombre amène à penser que certaines sources non identifiées détectées par le LAT, notamment à haute latitude galactique, sont des MSPs pour le moment inconnus. Le faisceau d'émission des MSPs dans le domaine radio est plus large que celui des pulsars normaux, à cause du cylindre du lumière, plus étroit chez les MSPs. On attend donc moins de MSPs muets dans le domaine radio que chez les pulsars normaux. Par conséquent la recherche de pulsations pour ces sources non identifiées pourrait être efficace en radio comme en $\gamma$. Les perspectives de détection de MSPs émetteurs radio ou non par le LAT au cours des années à venir sont donc grandes. Le LAT a déjà permis une avancée sans précédent dans l'étude de l'émission des pulsars à haute énergie. Tout porte à croire que lorsque la mission Fermi viendra à son terme, le LAT aura également constitué un allié de poids pour les radiotélescopes, dans la compréhension globale de ces objets extrêmes et de leur environnement. 

%% file: annexes/annexeA.tex
\chapter{Table des pulsars milliseconde galactiques}
\label{annexeA}

Les valeurs de la d\'{e}riv\'{e}e de la p\'{e}riode $\dot P$, de la luminosit\'{e} de ralentissement $\dot E$, des champs magn\'{e}tiques surfacique $B_S$ et au cylindre de lumiere $B_{LC}$ et de l'\^{a}ge caract\'{e}ristique $\tau$ donn\'{e}es dans les tableaux, ainsi que les incertitudes associ\'{e}es, tiennent compte de l'effet Shklovskii \citep{Shklovskii1970} lorsque possible. Les différentes grandeurs données dans les tableaux qui suivent sont obtenues à partir des formules ci-dessous.

\begin{itemize}

\item Derivée de la periode :

\begin{eqnarray}
\dot P = \dot{P}_\mathrm{obs} - \dot{P}_\mathrm{Shk}
\end{eqnarray}
\begin{eqnarray}
\Delta \dot P = \sqrt{ \Delta \dot{P}^2 + \Delta \dot{P}_\mathrm{Shk}^2 }
\end{eqnarray}

où $ \dot{P}_\mathrm{Shk} \simeq 2.43 \times 10^{-21} P\ D\ \mu^2$ est la correction de l'effet Shklovskii, avec $P$ en s, $D$ en kpc et $\mu$ en mas/yr. L'incertitude associée est donc :

\begin{eqnarray}
\Delta \dot{P}_\mathrm{Shk} = \dot{P}_\mathrm{Shk} \times \sqrt{ \left( \frac{\Delta D}{D} \right)^2 + 2 \left( \frac{\Delta \mu}{\mu} \right)^2 + \left( \frac{\Delta P}{P} \right)^2 }
\end{eqnarray}

La composante due à l'incertitude sur la période est négligeable, les contributions de la distance $D$ et du mouvement propre transverse $\mu$ sont typiquement de l'ordre de 10\%.

\emph{Les valeurs de $\dot P$ fournies dans les tableaux sont corrigées de l'effet Shklovskii si la distance $D$ et le mouvement propre transverse $\mu$ sont connus. De plus on interdit $\dot P < 0$, ce qui peut se produire si la distance $D$ est surestimée. Si les conditions précédentes ne sont pas réalisées, on utilise la valeur de $\dot P$ apparente.} 

\item Luminosité de ralentissement :

\begin{eqnarray}
\dot E = 4 \pi^2 I \frac{\dot P}{P^3}
\end{eqnarray}
\begin{eqnarray}
\Delta \dot E = \dot E \times \sqrt{ 3 \left( \frac{\Delta P}{P} \right)^2 + \left( \frac{\Delta \dot P}{\dot P} \right)^2  }
\label{uncEdot}
\end{eqnarray}

\emph{On utilise si possible la valeur de $\dot P$ corrigée de l'effet Shklovskii. Sinon, les valeurs de $\dot E$ et de son incertitude sont celles du catalogue ATNF \citep{CatalogueATNF}, qui n'emploient que la valeur de $\dot P$ apparente.}

\item Champ magnétique surfacique :

\begin{eqnarray}
B_S = \sqrt{  \frac{3 I c^3}{8 \pi^2 R^6}   P \dot P}
\label{BS}
\end{eqnarray}
\begin{eqnarray}
\Delta B_S = B_S \times \sqrt{ \frac{1}{2} \left( \frac{\Delta P}{P} \right)^2 + \frac{1}{2} \left( \frac{\Delta \dot P}{\dot P} \right)^2     }
\label{uncBS}
\end{eqnarray}

\emph{Détail du calcul: \emph{cf.} $\dot E$. }

\item Champ magnétique au cylindre de lumière :

\begin{eqnarray}
B_{LC} = 4 \pi^2 \sqrt{ \frac{3 I}{2 c^3} } P^{-5/2} \dot P^{1/2}
\label{BLC}
\end{eqnarray}
\begin{eqnarray}
\Delta B_{LC} = B_{LC} \times  \sqrt{ \frac{5}{2} \left( \frac{\Delta P}{P} \right)^2 + \frac{1}{2} \left( \frac{\Delta \dot P}{\dot P} \right)^2 }
\label{uncBLC}
\end{eqnarray}

\emph{Détail du calcul: \emph{cf.} $\dot E$. }

\item Age caractéristique :

\begin{eqnarray}
\tau = \frac{\dot P}{2 P}
\end{eqnarray}
\begin{eqnarray}
\Delta \tau = \tau \times \sqrt{ \left( \frac{\Delta P}{P} \right)^2 + \left( \frac{\Delta \dot P}{\dot P} \right)^2 }
\end{eqnarray}

\emph{Détail du calcul: \emph{cf.} $\dot E$. }

\item Remarques : 
\begin{itemize}
\item Les valeurs de $\dot E$, $B_S$ et $B_{LC}$ dépendent du moment d'inertie de l'étoile à neutrons $I$, et les champs magnétiques $B_S$ et $B_{LC}$ dépendent également de son rayon $R$. Les incertitudes sur $I$ et $R$ devraient donc intervenir dans les équations \ref{uncEdot}, \ref{uncBS} et \ref{uncBLC}. De plus, on a fait l'hypothèse de rotateurs orthogonaux dans les équations \ref{BS} et \ref{BLC}. Il faudrait en principe tenir compte de l'inclinaison magnétique $\alpha$ et son incertitude. Les contributions de $I$, $R$ et $\alpha$ sont malheureusement mal connues, aussi nous n'en tenons pas compte dans les tableaux qui suivent. 
\item Pour les pulsars dont la distance est calculée à partir du modèle NE2001 \citep{NE2001}, une incertitude sur $D$ de 20\% est utilisée.
\end{itemize}

\end{itemize}

\newpage

\begin{center}
\begin{scriptsize}

\tablefirsthead{\hline
\hline
JName & BName & Binaire ? & l & b & $P$ & $\dot P$ & Distance & $\dot E$ & $\tau$ & $B_S$ & $B_{LC}$ \\
 & & & ($^\circ$) & ($^\circ$) & (ms) & (10$^{-21}$) & (kpc) & (10$^{33}$ erg/s) & (10$^{8}$ yr) & (10$^8$ G) & (10$^2$ G) \\ \hline}
\tablehead{\hline
\hline
JName & BName & Binaire ? & l & b & $P$ & $\dot P$ & Distance & $\dot E$ & $\tau$ & $B_S$ & $B_{LC}$ \\
 & & & ($^\circ$) & ($^\circ$) & (ms) & (10$^{-21}$) & (kpc) & (10$^{33}$ erg/s) & (10$^{8}$ yr) & (10$^8$ G) & (10$^2$ G) \\ \hline}
\tabletail{Se poursuit sur une autre page...}
\bottomcaption{Table des pulsars milliseconde du champ galactique\label{tableauMSPsGAL}}
\tabletail{\hline}
\begin{supertabular}{cccccccccccc}
J0030+0451 & -- & N & 113,14 & -57,61 & 4,865 & 10,05(4)  & 0,300$_{-0,065}^{+0,110}$ & 3,44(1)  & 76,7(3)  & 2,237(7)  & 176,6(5)  \\
J0034$-$0534 & -- & O & 111,49 & -68,07 & 1,877 & 3(1)  & 0,540$ \pm$ 0,108* & 15(6)  & 11(5)  & 0,7(2)  & 10(3)  \\
J0218+4232 & -- & O & 139,51 & -17,53 & 2,323 & 77,0(6)  & 2,670$ \pm$ 0,534* & 243(2)  & 4,78(4)  & 4,28(3)  & 310(2)  \\
J0407+1607 & -- & O & 176,62 & -25,66 & 25,702 & 79(3)  & 1,330$ \pm$ 0,266* & 0,184(7)  & 52(2)  & 14,4(4)  & 7,7(2)  \\
J0437$-$4715 & -- & O & 253,39 & -41,96 & 5,757 & 13,9(4)  & 0,1563$ \pm$ 0,0013 & 2,87(7)  & 66(2)  & 2,86(5)  & 136(3)  \\
J0610$-$2100 & -- & O & 227,75 & -18,18 & 3,861 & 7(2)  & 3,540$ \pm$ 0,708* & 5(1)  & 9(3)  & 1,6(4)  & 26(6)  \\
J0613$-$0200 & -- & O & 210,41 & -9,30 & 3,062 & 9,2(2)  & 0,48$_{-0,11}^{+0,19}$ & 12,6(2)  & 53(1)  & 1,7(2)  & 537(7)  \\
J0621+1002 & -- & O & 200,57 & -2,01 & 28,854 & 46,2(3)  & 1,360$ \pm$ 0,272* & 0,0758(5)  & 99,1(6)  & 11,68(5)  & 4,42(2)  \\
J0711$-$6830 & -- & N & 279,53 & -23,28 & 5,491 & 9(2)  & 0,860$ \pm$ 0,172* & 2,2(4)  & 9(2)  & 2,3(3)  & 13(1)  \\
J0737$-$3039A & -- & O & 245,24 & -4,50 & 22,699 & 1758,8(3)  & 1,15$_{-0,16}^{+0,22}$ & 5,937(1)  & 2,0448(3)  & 63,939(7)  & 49,701(6)  \\
J0751+1807 & -- & O & 202,73 & 21,09 & 3,479 & 7,6(2)  & 0,625$_{-0,210}^{+0,625}$ & 7,1(2)  & 73(2)  & 1,64(3)  & 355(7)  \\
J0900$-$3144 & -- & O & 256,16 & 9,49 & 11,110 & 49,1(1)  & 0,540$ \pm$ 0,108* & 1,414(3)  & 35,84(9)  & 7,48(1)  & 49,56(9)  \\
J1012+5307 & -- & O & 160,35 & 50,86 & 5,256 & 13,8(7)  & 0,410$ \pm$ 0,082* & 3,7(2)  & 60(3)  & 2,72(9)  & 171(6)  \\
J1022+1001 & -- & O & 231,79 & 51,10 & 16,453 & 43,34(4)  & 0,40$_{-0,10}^{+0,19}$ & 0,3842(4)  & 60,15(6)  & 8,545(6)  & 17,44(1)  \\
J1024$-$0719 & -- & N & 251,70 & 40,52 & 5,162 & 18,531(6)  & 0,52$_{-0,15}^{+0,39}$ & 5,318(2)  & 44,14(1)  & 3,1298(7)  & 206,85(5)  \\
J1038+0032 & -- & N & 247,15 & 48,47 & 28,852 & 67(7)  & 1,180$ \pm$ 0,236* & 0,11(1)  & 68(7)  & 14(1)  & 5,3(4)  \\
J1045$-$4509 & -- & O & 280,85 & 12,25 & 7,474 & 14(1)  & 1,960$ \pm$ 0,392* & 1,3(1)  & 87(8)  & 3,2(2)  & 70(4)  \\
J1125$-$6014 & -- & O & 292,50 & 0,89 & 2,630 & 4,01(9)  & 1,5$ \pm$ 0,3* & 8,7(2)  & 104(2)  & 1,04(2)  & 519(8)  \\
J1216$-$6410 & -- & O & 299,10 & -1,56 & 3,539 & 1,6(2)  & 1,330$ \pm$ 0,266* & 1,4(2)  & 35(4)  & 0,77(6)  & 16(1)  \\
J1300+1240 & B1257+12 & O & 311,31 & 75,41 & 6,219 & -- & 0,77$_{-0,18}^{+0,34}$ & -- & -- & -- & -- \\
J1435$-$6100 & -- & O & 315,19 & -0,64 & 9,348 & 24,5(5)  & 2,160$ \pm$ 0,432* & 1,18(2)  & 60(1)  & 4,84(7)  & 53,9(8)  \\
J1439$-$5501 & -- & O & 318,10 & 4,63 & 28,635 & 142(1)  & 0,60$ \pm$ 0,12* & 0,238(2)  & 32,0(2)  & 20,4(1)  & 7,9(4)  \\
J1453+1902 & -- & N & 23,39 & 60,81 & 5,792 & 10,6(6)  & 1,15$ \pm$ 0,23* & 2,2(1)  & 87(5)  & 2,5(1)  & 117(5)  \\
J1455$-$3330 & -- & O & 330,72 & 22,56 & 7,987 & 18(5)  & 0,530$ \pm$ 0,106* & 1,4(4)  & 7(2)  & 3,8(7)  & 7(1)  \\
J1600$-$3053 & -- & O & 344,09 & 16,45 & 3,598 & 8,8(1)  & 1,630$ \pm$ 0,326* & 7,5(1)  & 65(1)  & 1,8(2)  & 352(4)  \\
J1603$-$7202 & -- & O & 316,63 & -14,50 & 14,842 & 13,9(5)  & 1,170$ \pm$ 0,234* & 0,168(7)  & 169(7)  & 4,6(1)  & 12,8(4)  \\
J1614$-$2230 & -- & O & 352,54 & 20,30 & 3,151 & 4(2)  & 1,290$ \pm$ 0,258* & 5(2)  & 12(6)  & 1,1(4)  & 3(1)  \\
J1618$-$39 & -- & O & 340,77 & 7,89 & 11,987 & -- & 2,730$ \pm$ 0,546* & -- & -- & -- & -- \\
J1629$-$6902 & -- & N & 320,37 & -13,93 & 6,001 & 10,0(3)  & 0,960$ \pm$ 0,192* & 1,83(5)  & 95(3)  & 2,48(5)  & 104(2)  \\
J1640+2224 & -- & O & 41,05 & 38,27 & 3,163 & 1,7(2)  & 1,160$ \pm$ 0,232* & 2,1(3)  & 30(4)  & 0,73(7)  & 21(2)  \\
J1643$-$1224 & -- & O & 5,67 & 21,22 & 4,622 & 16(2)  & 2,410$ \pm$ 0,482* & 6,5(7)  & 45(5)  & 2,8(2)  & 26(2)  \\
J1709+2313 & -- & O & 44,52 & 32,21 & 4,631 & 2,0(4)  & 1,410$ \pm$ 0,282* & 0,8(2)  & 37(7)  & 1,0(1)  & 9(1)  \\
J1713+0747 & -- & O & 28,75 & 25,22 & 4,570 & 8,07(3)  & 1,05$_{-0,07}^{+0,06}$ & 3,34(1)  & 89,8(3)  & 1,943(5)  & 185,1(5)  \\
J1721$-$2457 & -- & N & 0,39 & 6,75 & 3,497 & 6(1)  & 1,290$ \pm$ 0,258* & 5,4(9)  & 9(2)  & 1,5(2)  & 31(4)  \\
J1730$-$2304 & -- & N & 3,14 & 6,02 & 8,123 & 20,21(1)  & 0,530$ \pm$ 0,106* & 1,4887(7)  & 63,68(3)  & 4,1(1)  & 69,55(2)  \\
J1732$-$5049 & -- & O & 340,03 & -9,45 & 5,313 & 13,8(9)  & 1,410$ \pm$ 0,282* & 3,6(2)  & 61(4)  & 2,7(1)  & 166(8)  \\
J1738+0333 & -- & O & 27,72 & 17,74 & 5,850 & 23,1(6)  & 1,430$ \pm$ 0,286* & 4,6(1)  & 40(1)  & 3,72(7)  & 169(3)  \\
J1741+1354 & -- & O & 37,89 & 21,64 & 3,747 & -- & 0,90$ \pm$ 0,18* & -- & -- & -- & -- \\
J1744$-$1134 & -- & N & 14,79 & 9,18 & 4,075 & 6,9(5)  & 0,47$_{-0,07}^{+0,12}$ & 4,0(3)  & 93(7)  & 1,7(9)  & 23(1)  \\
J1745$-$0952 & -- & O & 16,37 & 9,89 & 19,376 & 95(4)  & 1,830$ \pm$ 0,366* & 0,52(2)  & 32(1)  & 13,7(4)  & 17,2(5)  \\
J1751$-$2857 & -- & O & 0,65 & -1,12 & 3,915 & 11,26(4)  & 1,10$ \pm$ 0,22* & 7,41(3)  & 55,1(2)  & 2,125(5)  & 321,9(8)  \\
J1756$-$2251 & -- & O & 6,50 & 0,95 & 28,462 & 1017,1(2)  & 2,480$ \pm$ 0,496* & 1,7416(3)  & 4,4337(9)  & 54,445(8)  & 21,469(3)  \\
J1757$-$5322 & -- & O & 339,64 & -13,98 & 8,870 & 26,3(4)  & 0,960$ \pm$ 0,192* & 1,49(2)  & 53,4(8)  & 4,89(5)  & 63,7(7)  \\
J1801$-$1417 & -- & N & 14,55 & 4,16 & 3,625 & 5,3(1)  & 1,520$ \pm$ 0,304* & 4,4(9)  & 108(2)  & 1,4(2)  & 268(4)  \\
J1802$-$2124 & -- & O & 8,38 & 0,61 & 12,648 & 72(1)  & 2,940$ \pm$ 0,588* & 1,4(2)  & 27,8(4)  & 9,66(9)  & 43,4(4)  \\
J1804$-$2717 & -- & O & 3,50 & -2,74 & 9,343 & 40,89(8)  & 0,780$ \pm$ 0,156* & 1,979(4)  & 36,2(7)  & 6,255(9)  & 69,7(1)  \\
J1841+0130 & -- & O & 33,12 & 2,94 & 29,773 & 817(5)  & 3,590$ \pm$ 0,718* & 12,22(7)  & 0,577(4)  & 157,8(7)  & 54,4(2)  \\
J1843$-$1113 & -- & N & 22,05 & -3,40 & 1,846 & 9,59(5)  & 1,690$ \pm$ 0,338* & 60,2(3)  & 30,5(2)  & 1,346(5)  & 1947(7)  \\
J1853+1303 & -- & O & 44,88 & 5,37 & 4,092 & 8,9(1)  & 2,090$ \pm$ 0,418* & 5,1(6)  & 73,3(8)  & 1,93(2)  & 256(2)  \\
J1857+0943 & B1855+09 & O & 42,29 & 3,06 & 5,362 & 17,3(2)  & 0,91$_{-0,20}^{+0,35}$ & 4,42(4)  & 49,2(5)  & 3,08(2)  & 182(1)  \\
J1903+0327 & -- & O & 37,32 & -0,99 & 2,150 & 18,79(2)  & 6,360$ \pm$ 1,272* & 74,65(8)  & 18,13(2)  & 2,034(2)  & 1861(1)  \\
J1905+0400 & -- & N & 38,09 & -1,29 & 3,784 & 4,86(6)  & 1,710$ \pm$ 0,342* & 3,54(4)  & 123(2)  & 1,37(1)  & 230(2)  \\
J1909$-$3744 & -- & O & 359,73 & -19,60 & 2,947 & 2,7(5)  & 1,14$ \pm$ 0,05 & 4,1(8)  & 17(3)  & 0,9(1)  & 32(4)  \\
J1910+1256 & -- & O & 46,56 & 1,79 & 4,984 & 9,77(7)  & 2,330$ \pm$ 0,466* & 3,12(2)  & 80,8(6)  & 2,23(1)  & 164,0(8)  \\
J1911$-$1114 & -- & O & 25,14 & -9,58 & 3,626 & 8(5)  & 1,220$ \pm$ 0,244* & 7(4)  & 7(4)  & 1,7(7)  & 3(1)  \\
J1911+1347 & -- & N & 47,52 & 1,81 & 4,626 & 17,1(2)  & 2,070$ \pm$ 0,414* & 6,83(8)  & 42,8(5)  & 2,85(2)  & 262(2)  \\
J1918$-$0642 & -- & O & 30,03 & -9,12 & 7,646 & 24(3)  & 1,240$ \pm$ 0,248* & 2,1(3)  & 50(6)  & 4,3(4)  & 88(8)  \\
J1933$-$6211 & -- & O & 334,43 & -28,63 & 3,543 & 3,7(1)  & 0,520$ \pm$ 0,104* & 3,28(9)  & 152(4)  & 1,16(2)  & 237(5)  \\
J1939+2134 & B1937+21 & N & 57,51 & -0,29 & 1,558 & 105,1(4)  & 8,33$_{-3,30}^{+17,0}$ & 1097,6(4)  & 2,3484(9)  & 4,095(1)  & 9847(3)  \\
J1944+0907 & -- & N & 47,16 & -7,36 & 5,185 & 6(3)  & 1,790$ \pm$ 0,358* & 1,8(9)  & 13(6)  & 1,8(6)  & 12(4)  \\
J1955+2908 & B1953+29 & O & 65,84 & 0,44 & 6,133 & 28,7(2)  & 4,640$ \pm$ 0,928* & 4,92(4)  & 33,8(3)  & 4,25(2)  & 167,4(9)  \\
J1959+2048 & B1957+20 & O & 59,20 & -4,70 & 1,607 & 8(2)  & 2,490$ \pm$ 0,498* & 7(2)  & 32(7)  & 1,1(2)  & 25(4)  \\
J2010$-$1323 & -- & N & 29,45 & -23,54 & 5,223 & 4,82(7)  & 1,020$ \pm$ 0,204* & 1,34(2)  & 172(2)  & 1,61(2)  & 102(1)  \\
J2019+2425 & -- & O & 64,75 & -6,62 & 3,935 & 7,024(1)  & 1,490$ \pm$ 0,298* & 4,5525(8)  & 88,75(2)  & 1,6822(2)  & 251,1(3)  \\
J2033+17 & -- & O & 60,86 & -13,12 & 5,949 & 11(1)  & 2,0$ \pm$ 0,4* & 2,1(2)  & 86(8)  & 2,6(2)  & 112(7)  \\
J2051$-$0827 & -- & O & 39,19 & -30,41 & 4,509 & 12,4(1)  & 1,040$ \pm$ 0,208* & 5,35(5)  & 57,5(5)  & 2,39(2)  & 238(2)  \\
J2124$-$3358 & -- & N & 10,93 & -45,44 & 4,931 & 12(8)  & 0,25$_{-0,08}^{+0,25}$ & 4(3)  & 6(4)  & 3(1)  & 19(9)  \\
J2129$-$5721 & -- & O & 338,00 & -43,57 & 3,726 & 29,0(5)  & 1,360$ \pm$ 0,272* & 22,1(3)  & 20,3(3)  & 3,33(4)  & 585(6)  \\
J2145$-$0750 & -- & O & 47,78 & -42,08 & 16,052 & 26(2)  & 0,50$_{-0,11}^{+0,21}$ & 0,25(2)  & 98(6)  & 6,5(3)  & 14,3(6)  \\
J2229+2643 & -- & O & 87,69 & -26,28 & 2,978 & 1,46(2)  & 1,45$ \pm$ 0,29* & 2,18(3)  & 323(4)  & 0,667(6)  & 230(2)  \\
J2317+1439 & -- & O & 91,36 & -42,36 & 3,445 & 2,0(3)  & 0,830$ \pm$ 0,166* & 1,9(3)  & 28(5)  & 0,8(1)  & 19(2)  \\
J2322+2057 & -- & N & 96,52 & -37,31 & 4,808 & 4(2)  & 0,80$ \pm$ 0,16* & 1,4(5)  & 20(8)  & 1,4(4)  & 11(3)  \\
\hline
\end{supertabular}

\end{scriptsize}
\end{center}

Les deux premières colonnes donnent les noms J2000 et éventuellement B1950. La troisième indique les pulsars appartenant à des systèmes binaires. On donne également leur position en coordonnées galactiques, leur période de rotation $P$ et leur taux de ralentissement $\dot P$, intrinsèque si possible, apparent sinon. Les colonnes suivantes donnent les valeurs de $d$, $\dot E$, $\tau$, $B_S$ et $B_{LC}$. Les chiffres placés entre parenthèses indiquent l'incertitude sur la dernière décimale citée. Les distances calculées avec le modèle NE2001 sont indiquées avec des astérisques. Les valeurs de $P$, $\dot P$ ainsi que les quantités dérivées pour le pulsar J1614$-$2230 sont différentes de celles du catalogue ATNF. Elles sont issues de l'éphéméride utilisée pour la détection avec le LAT. Un article traitant de la découverte et de la chronométrie de PSR J1614$-$2230 est par ailleurs en cours de rédaction (M. Roberts, S. Ransom, communication privée). 

%% file: annexes/annexeB.tex
\chapter{Table des pulsars milliseconde des amas globulaires}
\label{annexeB}

La Table qui suit donne les noms des pulsars milliseconde répertoriés (non nécessairement confirmés) situés dans des amas globulaires. Les distances citées ont été obtenues à partir du catalogue de Harris \citep{Harris1996}, régulièrement mis à jour\footnote{\emph{cf.} http://physwww.mcmaster.ca/~harris/mwgc.dat}. Lorsque possible les valeurs de $\dot P$ et des quantités dérivées tiennent compte de l'effet Shklovskii (pour le détail, \emph{cf.} Annexe \ref{annexeA}). Néanmoins les valeurs citées sont à manipuler avec prudence, car elles ne tiennent pas compte des effets locaux d'accélération gravitationnelle, interférant sur la valeur de $\dot P$ observée.

\begin{center}
\begin{scriptsize}

\tablefirsthead{\hline
\hline
JName & BName & Amas & Binaire ? & l & b & $P$ & $\dot P$ & Distance & $\dot E$ & Notes \\
 & & & & ($^\circ$) & ($^\circ$) & (ms) & (10$^{-21}$) & (kpc) & (10$^{33}$ erg/s) & \\ 
\hline}
\tablehead{\hline
\hline
JName & BName & Amas & Binaire ? & l & b & $P$ & $\dot P$ & Distance & $\dot E$ & Notes \\
 & & & & ($^\circ$) & ($^\circ$) & (ms) & (10$^{-21}$) & (kpc) & (10$^{33}$ erg/s) & \\ 
\hline}
\tabletail{Se poursuit sur une autre page...}
\bottomcaption{Table des pulsars milliseconde des amas globulaires\label{tableauMSPsGC}} 
\tabletail{\hline}
\begin{supertabular}{ccccccccccc}
J0024$-$7204C & B0021$-$72C & 47Tuc (NGC104) & N & 305,92 & -44,89 & 5,757 &   & 4,5 &   & * \\
J0024$-$7204D & B0021$-$72D & 47Tuc (NGC104) & N & 305,88 & -44,89 & 5,358 &   & 4,5 &   & * \\
J0024$-$7204E & B0021$-$72E & 47Tuc (NGC104) & O & 305,88 & -44,88 & 3,536 & 96,7(2)  & 4,5 & 86,3(2)  & * \\
J0024$-$7204F & B0021$-$72F & 47Tuc (NGC104) & N & 305,90 & -44,89 & 2,624 & 63,7(1)  & 4,5 & 139,2(3)  & * \\
J0024$-$7204G & B0021$-$72G & 47Tuc (NGC104) & N & 305,89 & -44,89 & 4,040 &   & 4,5 &   & * \\
J0024$-$7204H & B0021$-$72H & 47Tuc (NGC104) & O & 305,90 & -44,90 & 3,210 &   & 4,5 &   & * \\
J0024$-$7204I & B0021$-$72I & 47Tuc (NGC104) & O & 305,89 & -44,89 & 3,485 &   & 4,5 &   & * \\
J0024$-$7204J & B0021$-$72J & 47Tuc (NGC104) & O & 305,91 & -44,90 & 2,101 &   & 4,5 &   & * \\
J0024$-$7204L & B0021$-$72L & 47Tuc (NGC104) & N & 305,90 & -44,89 & 4,346 &   & 4,5 &   & * \\
J0024$-$7204M & B0021$-$72M & 47Tuc (NGC104) & N & 305,91 & -44,88 & 3,677 &   & 4,5 &   & * \\
J0024$-$7204N & B0021$-$72N & 47Tuc (NGC104) & N & 305,89 & -44,90 & 3,054 &   & 4,5 &   & * \\
J0024$-$7204O & -- & 47Tuc (NGC104) & O & 305,90 & -44,89 & 2,643 & 29,5(2)  & 4,5 & 63,1(5)  & * \\
J0024$-$7204P & -- & 47Tuc (NGC104) & O & 305,90 & -44,90 & 3,643 &   & 4,5 &   & * \\
J0024$-$7204Q & -- & 47Tuc (NGC104) & O & 305,88 & -44,90 & 4,033 & 34,01(2)  & 4,5 & 20,47(1)  & * \\
J0024$-$7204R & -- & 47Tuc (NGC104) & O & 305,90 & -44,90 & 3,480 &   & 4,5 &   & * \\
J0024$-$7204S & -- & 47Tuc (NGC104) & O & 305,90 & -44,89 & 2,830 &   & 4,5 &   & * \\
J0024$-$7204T & -- & 47Tuc (NGC104) & O & 305,89 & -44,89 & 7,588 & 293,7(1)  & 4,5 & 26,54(1)  & * \\
J0024$-$7204U & -- & 47Tuc (NGC104) & O & 305,89 & -44,91 & 4,343 & 93,0(3)  & 4,5 & 44,8(2)  & * \\
J0024$-$7204V & -- & 47Tuc (NGC104) & N & 305,90 & -44,90 & 4,810 &   & 4,5 &   & * \\
J0024$-$7204W & -- & 47Tuc (NGC104) & O & 305,90 & -44,90 & 2,352 &   & 4,5 &   & * \\
J0024$-$7204X & -- & 47Tuc (NGC104) & N & 305,90 & -44,90 & 4,771 &   & 4,5 &   & * \\
J0024$-$7204Y & -- & 47Tuc (NGC104) & O & 305,90 & -44,90 & 2,197 &   & 4,5 &   & * \\
J0024$-$7204Z & -- & 47Tuc (NGC104) & N & 305,90 & -44,90 & 4,554 &   & 4,5 &   & Non publié \\
J0514$-$4002A & -- & NGC1851 & O & 244,51 & -35,04 & 4,991 & 1,2(1)  & 12,1 & 0,37(4)  & * \\
J1342+2822A & -- & M3 (NGC5272) & N & 42,21 & 78,71 & 2,545 &   & 10,4 &   & * \\
J1342+2822B & -- & M3 (NGC5272) & O & 42,22 & 78,71 & 2,389 & 18,58(4)  & 10,4 & 53,8(1)  & * \\
J1342+2822C & -- & M3 (NGC5272) & N & 42,21 & 78,71 & 2,166 &   & 10,4 &   & Incertain \\
J1342+2822D & -- & M3 (NGC5272) & O & 42,22 & 78,71 & 5,443 &   & 10,4 &   & * \\
J1518+0204A & B1516+02A & M5 (NGC5904) & N & 3,87 & 46,80 & 5,554 & 41,22(6)  & 7,5 & 9,5(1)  & * \\
J1518+0204B & B1516+02B & M5 (NGC5904) & O & 3,86 & 46,81 & 7,947 &   & 7,5 &   & * \\
J1518+0204C & -- & M5 (NGC5904) & O & 3,86 & 46,80 & 2,484 &   & 7,5 &   & * \\
J1518+0204D & -- & M5 (NGC5904) & O & 3,86 & 46,80 & 2,988 &   & 7,5 &   & * \\
J1518+0204E & -- & M5 (NGC5904) & O & 3,86 & 46,80 & 3,182 &   & 7,5 &   & * \\
J1623$-$2631 & B1620$-$26 & M4 (NGC6121) & O & 350,98 & 15,96 & 11,076 & 62(1)  & 2,2 & 18,1(3)  & * \\
J1641+3627A & B1639+36A & M13 (NGC6205) & N & 59,00 & 40,91 & 10,378 &   & 7,7 &   & * \\
J1641+3627B & B1639+36B & M13 (NGC6205) & O & 59,01 & 40,91 & 3,528 &   & 7,7 &   & * \\
J1641+3627C & -- & M13 (NGC6205) & N & 59,01 & 40,91 & 3,722 &   & 7,7 &   & * \\
J1641+3627D & -- & M13 (NGC6205) & O & 59,01 & 40,91 & 3,118 &   & 7,7 &   & * \\
J1641+3627E & -- & M13 (NGC6205) & O & 59,01 & 40,91 & 2,487 &   & 7,7 &   & * \\
J1701$-$3006A & -- & M62 (NGC6266) & O & 353,58 & 7,32 & 5,242 &   & 6,9 &   & * \\
J1701$-$3006B & -- & M62 (NGC6266) & O & 353,57 & 7,32 & 3,594 &   & 6,9 &   & * \\
J1701$-$3006C & -- & M62 (NGC6266) & O & 353,57 & 7,32 & 7,613 &   & 6,9 &   & * \\
J1701$-$3006D & -- & M62 (NGC6266) & O & 353,58 & 7,33 & 3,418 &   & 6,9 &   & * \\
J1701$-$3006E & -- & M62 (NGC6266) & O & 353,58 & 7,33 & 3,234 &   & 6,9 &   & * \\
J1701$-$3006F & -- & M62 (NGC6266) & O & 353,58 & 7,33 & 2,295 &   & 6,9 &   & * \\
J1740$-$5340 & -- & NGC6397 & O & 338,17 & -11,97 & 3,650 & 168(7)  & 2,3 & 136(6)  & * \\
J1748$-$2021B & -- & NGC6440 & O & 7,73 & 3,80 & 16,760 &   & 8,4 &   & * \\
J1748$-$2021C & -- & NGC6440 & N & 7,72 & 3,80 & 6,227 &   & 8,4 &   & * \\
J1748$-$2021D & -- & NGC6440 & O & 7,73 & 3,81 & 13,496 & 586,8(2)  & 8,4 & 9,424(3)  & * \\
J1748$-$2021E & -- & NGC6440 & N & 7,73 & 3,80 & 16,264 & 312,4(4)  & 8,4 & 2,867(4)  & * \\
J1748$-$2021F & -- & NGC6440 & O & 7,73 & 3,80 & 3,794 &   & 8,4 &   & * \\
J1748$-$2446A & B1744$-$24A & Terzan5 & O & 3,84 & 1,70 & 11,563 &   & 10,3 &   & * \\
J1748$-$2446aa & -- & Terzan5 & N & 3,84 & 1,69 & 5,788 &   & 10,3 &   & Non publié \\
J1748$-$2446ab & -- & Terzan5 & N & 3,84 & 1,69 & 5,120 &   & 10,3 &   & Non publié \\
J1748$-$2446ac & -- & Terzan5 & N & 3,84 & 1,69 & 5,087 &   & 10,3 &   & Non publié \\
J1748$-$2446ad & -- & Terzan5 & O & 3,84 & 1,69 & 1,396 &   & 10,3 &   & * \\
J1748$-$2446ae & -- & Terzan5 & O & 3,84 & 1,69 & 3,659 &   & 10,3 &   & Non publié \\
J1748$-$2446af & -- & Terzan5 & N & 3,84 & 1,69 & 3,304 &   & 10,3 &   & Non publié \\
J1748$-$2446ag & -- & Terzan5 & N & 3,84 & 1,69 & 4,448 &   & 10,3 &   & Non publié \\
J1748$-$2446ah & -- & Terzan5 & N & 3,84 & 1,69 & 4,965 &   & 10,3 &   & Non publié \\
J1748$-$2446C & -- & Terzan5 & N & 3,84 & 1,69 & 8,436 &   & 10,3 &   & * \\
J1748$-$2446D & -- & Terzan5 & N & 3,84 & 1,69 & 4,714 &   & 10,3 &   & * \\
J1748$-$2446E & -- & Terzan5 & O & 3,84 & 1,69 & 2,198 &   & 10,3 &   & * \\
J1748$-$2446F & -- & Terzan5 & N & 3,84 & 1,69 & 5,540 &   & 10,3 &   & * \\
J1748$-$2446G & -- & Terzan5 & N & 3,84 & 1,69 & 21,672 &   & 10,3 &   & * \\
J1748$-$2446H & -- & Terzan5 & N & 3,84 & 1,69 & 4,926 &   & 10,3 &   & * \\
J1748$-$2446I & -- & Terzan5 & O & 3,84 & 1,69 & 9,570 &   & 10,3 &   & * \\
J1748$-$2446K & -- & Terzan5 & N & 3,84 & 1,69 & 2,970 &   & 10,3 &   & * \\
J1748$-$2446L & -- & Terzan5 & N & 3,84 & 1,69 & 2,245 &   & 10,3 &   & * \\
J1748$-$2446M & -- & Terzan5 & O & 3,84 & 1,69 & 3,570 &   & 10,3 &   & * \\
J1748$-$2446N & -- & Terzan5 & O & 3,84 & 1,69 & 8,667 &   & 10,3 &   & * \\
J1748$-$2446O & -- & Terzan5 & O & 3,84 & 1,69 & 1,677 &   & 10,3 &   & * \\
J1748$-$2446P & -- & Terzan5 & O & 3,84 & 1,69 & 1,729 &   & 10,3 &   & * \\
J1748$-$2446Q & -- & Terzan5 & O & 3,84 & 1,69 & 2,812 &   & 10,3 &   & * \\
J1748$-$2446R & -- & Terzan5 & N & 3,84 & 1,69 & 5,029 &   & 10,3 &   & * \\
J1748$-$2446S & -- & Terzan5 & N & 3,84 & 1,69 & 6,117 &   & 10,3 &   & * \\
J1748$-$2446T & -- & Terzan5 & N & 3,84 & 1,69 & 7,085 &   & 10,3 &   & * \\
J1748$-$2446U & -- & Terzan5 & O & 3,84 & 1,69 & 3,289 &   & 10,3 &   & * \\
J1748$-$2446V & -- & Terzan5 & O & 3,84 & 1,69 & 2,073 &   & 10,3 &   & * \\
J1748$-$2446W & -- & Terzan5 & O & 3,84 & 1,69 & 4,205 &   & 10,3 &   & * \\
J1748$-$2446X & -- & Terzan5 & O & 3,84 & 1,69 & 2,999 &   & 10,3 &   & * \\
J1748$-$2446Y & -- & Terzan5 & O & 3,84 & 1,69 & 2,048 &   & 10,3 &   & * \\
J1748$-$2446Z & -- & Terzan5 & O & 3,84 & 1,69 & 2,463 &   & 10,3 &   & * \\
J1750$-$3703B & -- & NGC6441 & O & 353,53 & -5,01 & 6,075 & 19,2(5)  & 11,7 & 3,38(9)  & * \\
J1750$-$3703C & -- & NGC6441 & N & 353,53 & -5,01 & 26,569 &   & 11,7 &   & * \\
J1750$-$3703D & -- & NGC6441 & N & 353,53 & -5,01 & 5,140 & 492,8(3)  & 11,7 & 143,27(9)  & * \\
J1801$-$0857A & -- & NGC6517 & N & 19,23 & 6,76 & 7,176 &   & 10,8 &   & Non publié \\
J1801$-$0857B & -- & NGC6517 & N & 19,23 & 6,76 & 28,961 &   & 10,8 &   & Non publié \\
J1801$-$0857C & -- & NGC6517 & N & 19,23 & 6,76 & 3,739 &   & 10,8 &   & Non publié \\
J1803$-$3002B & -- & NGC6522 & N & 1,02 & -3,93 & 4,397 &   & 7,8 &   & Non publié \\
J1803$-$3002C & -- & NGC6522 & N & 1,02 & -3,93 & 5,840 &   & 7,8 &   & Non publié \\
J1803$-$30A & -- & NGC6522 & N & 1,02 & -3,93 & 7,101 &   & 7,8 &   & * \\
J1804$-$0735 & B1802$-$07 & NGC6539 & O & 20,79 & 6,77 & 23,101 & 467,1(3)  & 8,4 & 1,496(1)  & * \\
J1807$-$2459A & -- & NGC6544 & O & 5,84 & -2,20 & 3,059 &   & 2,7 &   & * \\
J1807$-$2459B & -- & NGC6544 & N & 5,84 & -2,20 & 4,186 &   & 2,7 &   & * \\
J1823$-$3021A & B1820$-$30A & NGC6624 & N & 2,79 & -7,91 & 5,440 & 3384,14(3)  & 7,9 & 829,871(7)  & * \\
J1823$-$3021D & -- & NGC6624 & N & 2,79 & -7,91 & 3,020 &   & 7,9 &   & Non publié \\
J1823$-$3021E & -- & NGC6624 & N & 2,79 & -7,91 & 4,394 &   & 7,9 &   & Non publié \\
J1823$-$3021F & -- & NGC6624 & N & 2,79 & -7,91 & 4,850 &   & 7,9 &   & Non publié \\
J1824$-$2452 & B1821$-$24 & M28 (NGC6626) & N & 7,80 & -5,58 & 3,054 & 1617,8(5)  & 5,6 & 2241,6(7)  & * \\
J1824$-$2452B & -- & M28 (NGC6626) & N & 7,80 & -5,58 & 6,547 &   & 5,6 &   & Non publié \\
J1824$-$2452C & -- & M28 (NGC6626) & O & 7,80 & -5,58 & 4,159 &   & 5,6 &   & Non publié \\
J1824$-$2452E & -- & M28 (NGC6626) & N & 7,80 & -5,58 & 5,420 &   & 5,6 &   & Non publié \\
J1824$-$2452F & -- & M28 (NGC6626) & N & 7,80 & -5,58 & 2,451 &   & 5,6 &   & Non publié \\
J1824$-$2452G & -- & M28 (NGC6626) & O & 7,80 & -5,58 & 5,909 &   & 5,6 &   & Non publié \\
J1824$-$2452H & -- & M28 (NGC6626) & O & 7,80 & -5,58 & 4,629 &   & 5,6 &   & Non publié \\
J1824$-$2452I & -- & M28 (NGC6626) & O & 7,80 & -5,58 & 3,932 &   & 5,6 &   & Non publié \\
J1824$-$2452J & -- & M28 (NGC6626) & O & 7,80 & -5,58 & 4,039 &   & 5,6 &   & Non publié \\
J1824$-$2452K & -- & M28 (NGC6626) & O & 7,80 & -5,58 & 4,461 &   & 5,6 &   & Non publié \\
J1824$-$2452L & -- & M28 (NGC6626) & O & 7,80 & -5,58 & 4,100 &   & 5,6 &   & Non publié \\
J1836$-$2354A & -- & M22 (NGC6656) & O & 9,89 & -7,55 & 3,354 &   & 3,2 &   & Non publié \\
J1836$-$2354B & -- & M22 (NGC6656) & N & 9,89 & -7,55 & 3,232 &   & 3,2 &   & Non publié \\
J1905+0154A & -- & NGC6749 & O & 36,21 & -2,20 & 3,193 &   & 7,9 &   & * \\
J1905+0154B & -- & NGC6749 & N & 36,21 & -2,20 & 4,968 &   & 7,9 &   & Incertain \\
J1910$-$5959A & -- & NGC6752 & O & 336,52 & -25,73 & 3,266 & 2,18(7)  & 4 & 2,48(7)  & * \\
J1910$-$5959B & -- & NGC6752 & N & 336,49 & -25,63 & 8,358 &   & 4 &   & * \\
J1910$-$5959C & -- & NGC6752 & N & 336,46 & -25,66 & 5,277 &   & 4 &   & * \\
J1910$-$5959D & -- & NGC6752 & N & 336,49 & -25,63 & 9,035 & 964,31(6)  & 4 & 51,612(3)  & * \\
J1910$-$5959E & -- & NGC6752 & N & 336,49 & -25,63 & 4,572 &   & 4 &   & * \\
J1911+0101A & B1908+00 & NGC6760 & O & 36,11 & -3,92 & 3,619 &   & 7,4 &   & * \\
J1911+0101B & -- & NGC6760 & N & 36,11 & -3,93 & 5,384 &   & 7,4 &   & * \\
J1953+1846A & -- & M71 (NGC6838) & O & 56,74 & -4,56 & 4,888 &   & 4 &   & * \\
J2129+1210D & B2127+11D & M15 (NGC7078) & N & 65,01 & -27,31 & 4,803 &   & 10,3 &   & * \\
J2129+1210E & B2127+11E & M15 (NGC7078) & N & 65,01 & -27,31 & 4,651 & 178(7)  & 10,3 & 70(3)  & * \\
J2129+1210F & B2127+11F & M15 (NGC7078) & N & 65,01 & -27,31 & 4,027 & 32(8)  & 10,3 & 19(5)  & * \\
J2129+1210H & B2127+11H & M15 (NGC7078) & N & 65,01 & -27,31 & 6,743 & 2(1)  & 10,3 & 3(2)  & * \\
J2140$-$2310A & -- & M30 (NGC7099) & O & 27,18 & -46,84 & 11,019 &   & 8 &   & * \\
J2140$-$2310B & -- & M30 (NGC7099) & O & 27,18 & -46,84 & 12,986 &   & 8 &   & * \\
\hline
\end{supertabular}

\end{scriptsize}
\end{center}

Les deux premières colonnes donnent les noms J2000 et éventuellement B1950. La troisième colonne donne le nom de l'amas globulaire hôte. La quatrième indique les pulsars appartenant à des systèmes binaires. Puis les positions en coordonnées galactiques sont données, ainsi que $P$ et $\dot P$. La distance de l'amas (issue de \citet{Harris1996}) est donnée ensuite, ainsi que la valeur de $\dot E$. Enfin la colonne « Notes » apporte des renseignements complémentaires ; indiquant par exemple les pulsars non publiés jusqu'à présent. Les chiffres placés entre parenthèses indiquent l'incertitude sur la dernière décimale citée.

%% file: annexes/annexeC.tex
\chapter{Courbes de lumière des huit MSPs détectés}
\label{annexeC}

Les courbes de lumière présentées dans le chapitre V (\emph{cf.} Figures \ref{phasos1} et \ref{phasos2}) correspondent à des coupures angulaires indépendantes de l'énergie : $\rho =$ 0,5$^\circ$ ou 1$^\circ$. Or la résolution angulaire de l'instrument est variable en fonction de l'énergie : elle est de plusieurs degrés à 100 MeV et elle décroît lorsque l'énergie augmente. Ainsi, la sélection d'une région étroite autour d'une source $\gamma$ a pour conséquence l'élimination d'une grande partie des photons de basse énergie. Ces phasogrammes sont donc biaisés vers les hautes énergies. 

Dans cette annexe nous montrons les courbes de lumière des huit MSPs pour lesquels des pulsations ont été détectées, en appliquant cette fois des coupures angulaires adaptées au profil de la PSF en fonction de l'énergie. Cette méthode présente l'avantage de ne privilégier aucune partie du spectre des pulsars, ce qui est utile pour l'interprétation de l'émission des pulsars à différentes gammes d'énergie. 

Nous avons retenu les photons enregistrés depuis la mise en service du LAT, le 30 juin 2008, jusqu'au 2 juin 2009. Les événements appartenant à la classe \emph{Diffuse} ont été sélectionnés, et ceux dont la direction forme un angle de plus de 105$^\circ$ avec le zénith ont été exclus. 

La coupure angulaire est déterminée de la manière suivante : soit un photon d'énergie $E$, d'inclinaison et d'azimut ($\theta$,$\phi$) par rapport à l'axe du télescope et de direction reconstruite formant un angle $\rho$ avec la source considérée. Pour cette configuration ($E$, $\theta$, $\phi$, $\rho$), nous calculons la fraction de la PSF qui inclut ce photon (la fonction de réponse de l'instrument utilisée ici est \emph{P6\_V3\_diffuse}). Enfin nous fixons une limite sur la fraction de PSF désirée : 68\% ou 33\% pour les pulsars situés dans des régions bruitées ou lorsque des sources brillantes sont à proximité, comme pour J0218+4232 (\emph{cf.} Table \ref{coupuresC}). La PSF de l'instrument à 68\% est représentée dans la Figure \ref{AngRes}. Pour une fraction de 33\% et en incidence normale, elle est de 1,8$^\circ$ à 100 MeV, 0,75$^\circ$ à 300 MeV et 0,3$^\circ$ à 1 GeV. 

\tab[Coupures et niveaux de fond pour les courbes de lumière de l'annexe C]{coupuresC.tex}{Fractions de PSF retenues pour les Figures \ref{phasosC1}, \ref{phasosC2}, \ref{phasosC3}, \ref{phasosC4}, \ref{phasosC5}, \ref{phasosC6}, \ref{phasosC7} et \ref{phasosC8}. Les quatre dernières colonnes indiquent les niveaux de fond attendus dans les différentes gammes d'énergie. La première valeur est le nombre de photons, la valeur entre parenthèses est le nombre de photons de fond moyenné sur le nombre de subdivisions du phasogramme.}{coupuresC}

Les courbes de lumière $\gamma$ au-dessus de 100 MeV, entre 100 et 300 MeV, entre 300 MeV et 1 GeV et au-dessus de 1 GeV sont montrées en Figures \ref{phasosC1}, \ref{phasosC2}, \ref{phasosC3}, \ref{phasosC4}, \ref{phasosC5}, \ref{phasosC6}, \ref{phasosC7} et \ref{phasosC8}. Les lignes pointillées horizontales indiquent les niveaux de fond, pour les bandes en énergie et les coupures angulaires considérées. Ces niveaux ont été calculés en utilisant 16 régions voisines de dimensions identiques, situées à 1$^\circ$ du pulsar (les points centraux de ces régions forment un cercle autour du pulsar). Pour ces photons, on calcule la phase rotationnelle du pulsar, et on rejette les photons appartenant à la région pulsée du phasogramme. On obtient ainsi 16 valeurs dont on retient la médiane. Contrairement à la valeur moyenne, la médiane est moins sensible aux perturbations des valeurs extrémales, introduites par exemple lorsque des sources brillantes sont situées à proximité. La Table \ref{coupuresC} donne les valeurs calculées pour le niveau de fond dans chaque bande d'énergie. Les Figures \ref{nrj_phaseC1}, \ref{nrj_phaseC2}, \ref{nrj_phaseC3}, \ref{nrj_phaseC4}, \ref{nrj_phaseC5}, \ref{nrj_phaseC6}, \ref{nrj_phaseC7} et \ref{nrj_phaseC8} montrent des cartes du ciel $\gamma$ au-dessus de 100 MeV dans un rayon de 5$^\circ$ autour des pulsars, et l'énergie des photons $\gamma$ en fonction de la phase rotationnelle. 

\begin{twocolumn}

\begin{figure}[bt]
\begin{center}
\includegraphics[scale=0.67]{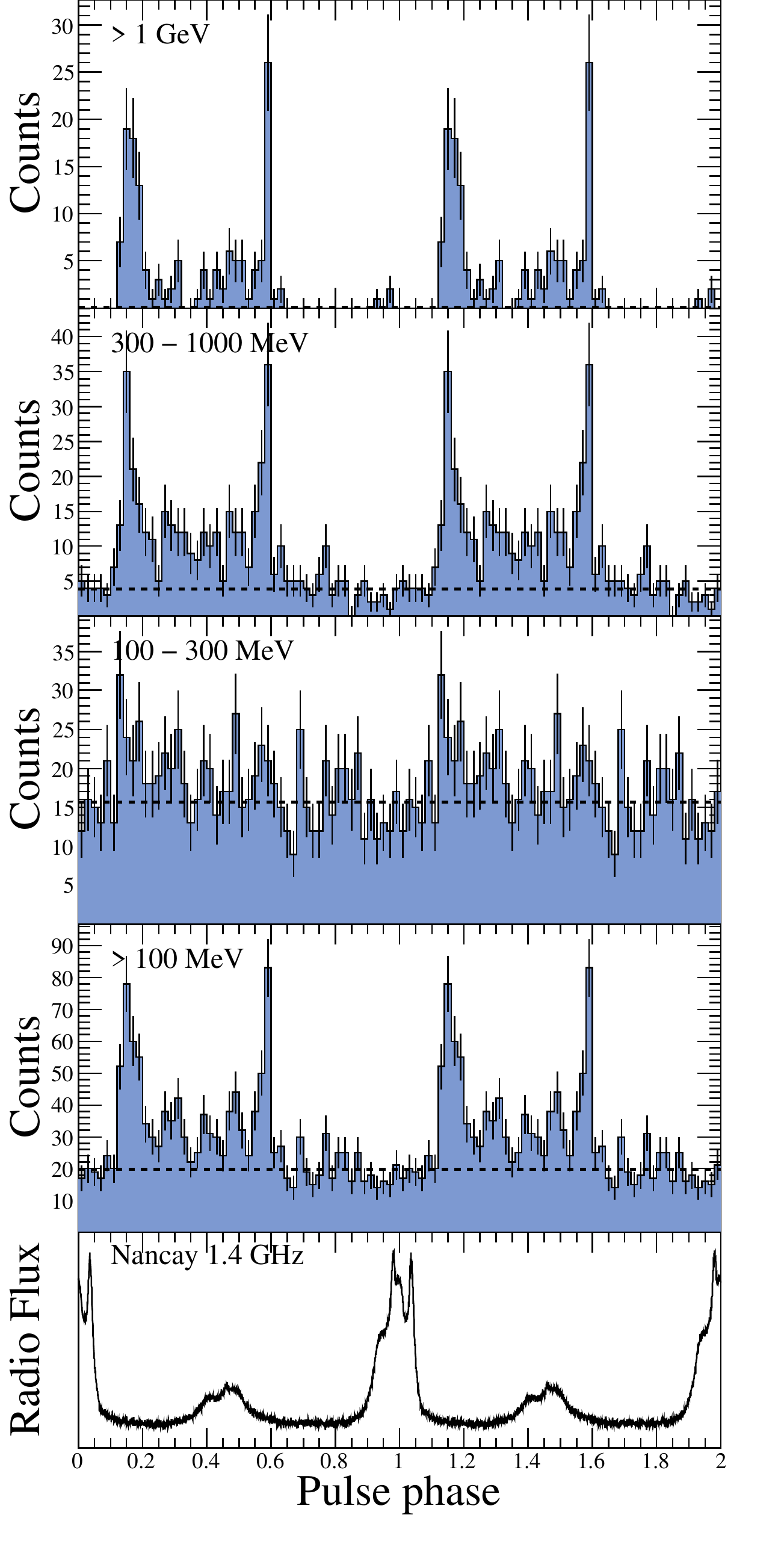}
\caption{Courbes de lumière pour J0030+0451 (rayon dépendant de l'énergie)}
\label{phasosC1}
\end{center}
\end{figure}

\begin{figure}[bt]
\begin{center}
\includegraphics[scale=0.4]{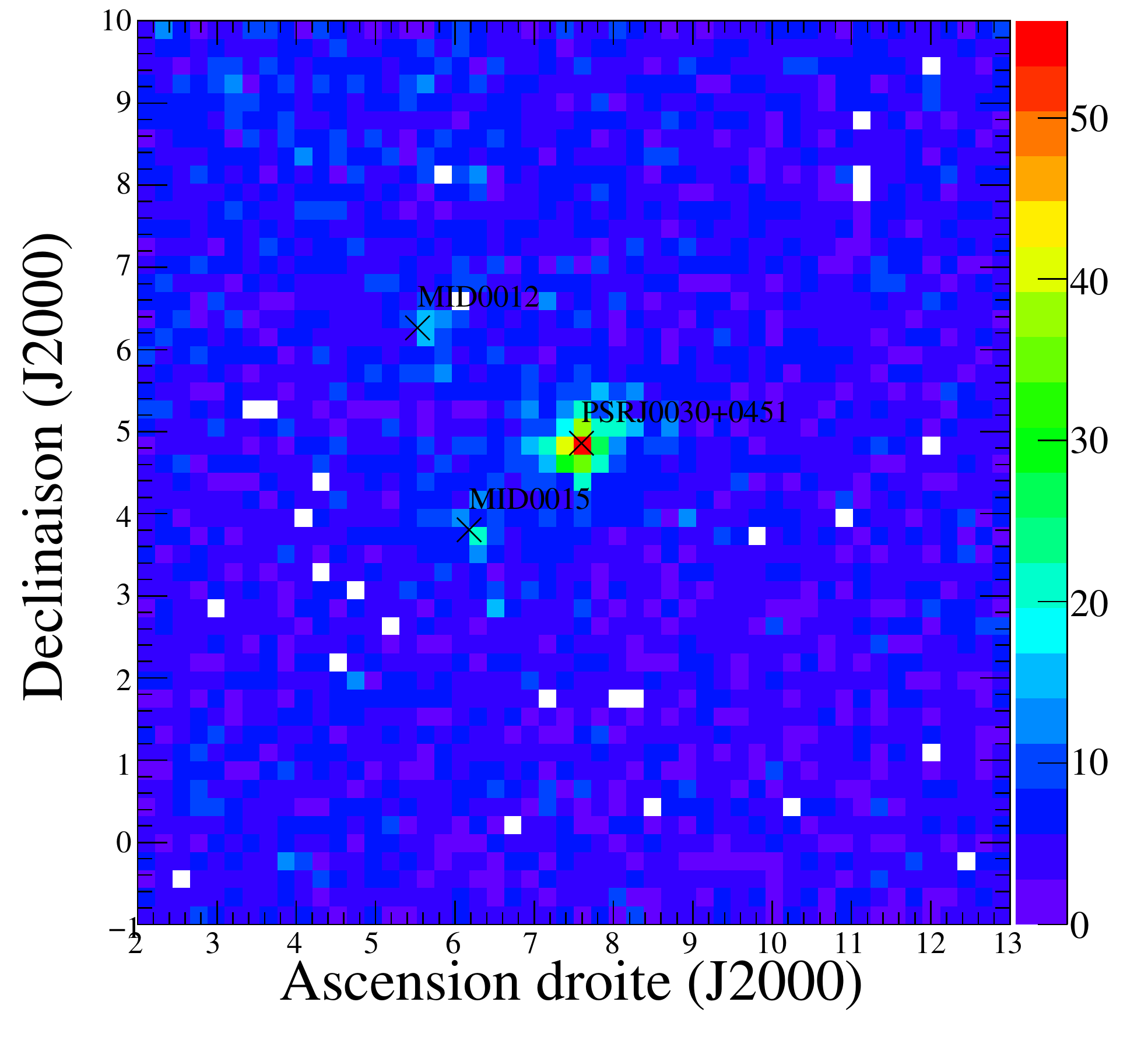}
\includegraphics[scale=0.4]{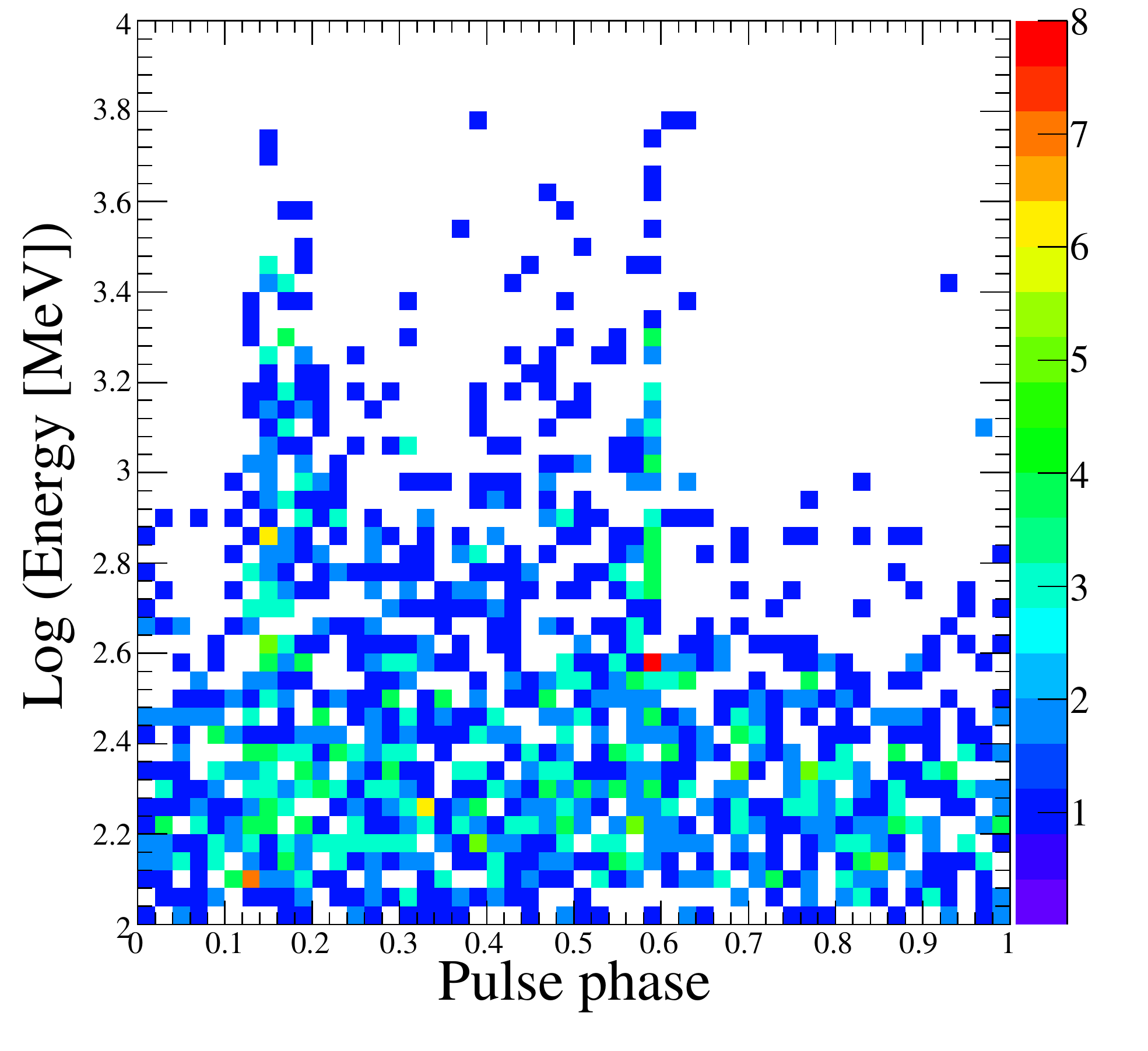}
\caption{Carte du ciel autour de J0030+0451 et énergie des photons en fonction de la phase rotationnelle}
\label{nrj_phaseC1}
\end{center}
\end{figure}

\end{twocolumn}
\onecolumn

\begin{twocolumn}

\begin{figure}[bt]
\begin{center}
\includegraphics[scale=0.67]{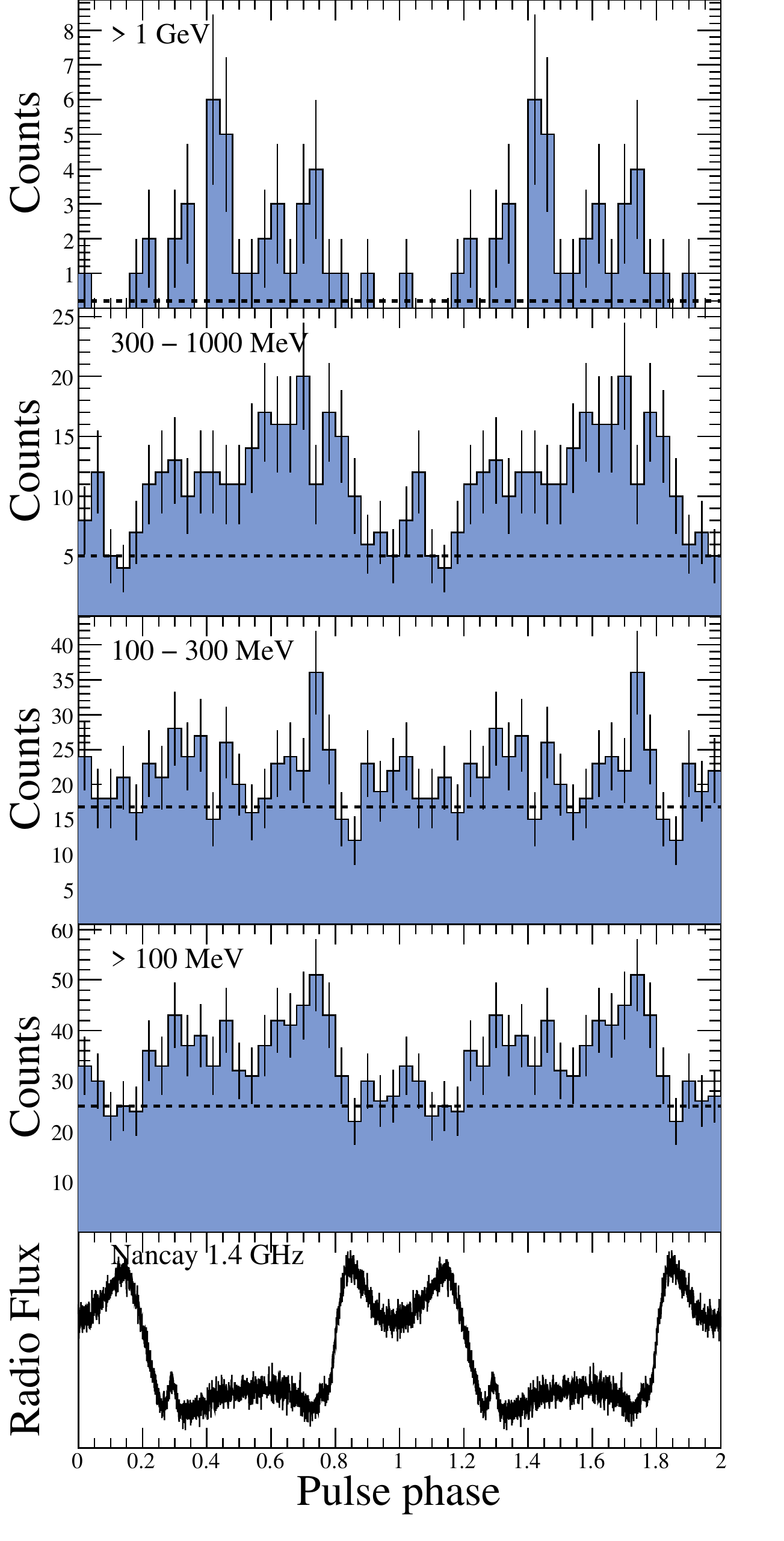}
\caption{Courbes de lumière pour J0218+4232 (rayon dépendant de l'énergie)}
\label{phasosC2}
\end{center}
\end{figure}

\begin{figure}[bt]
\begin{center}
\includegraphics[scale=0.4]{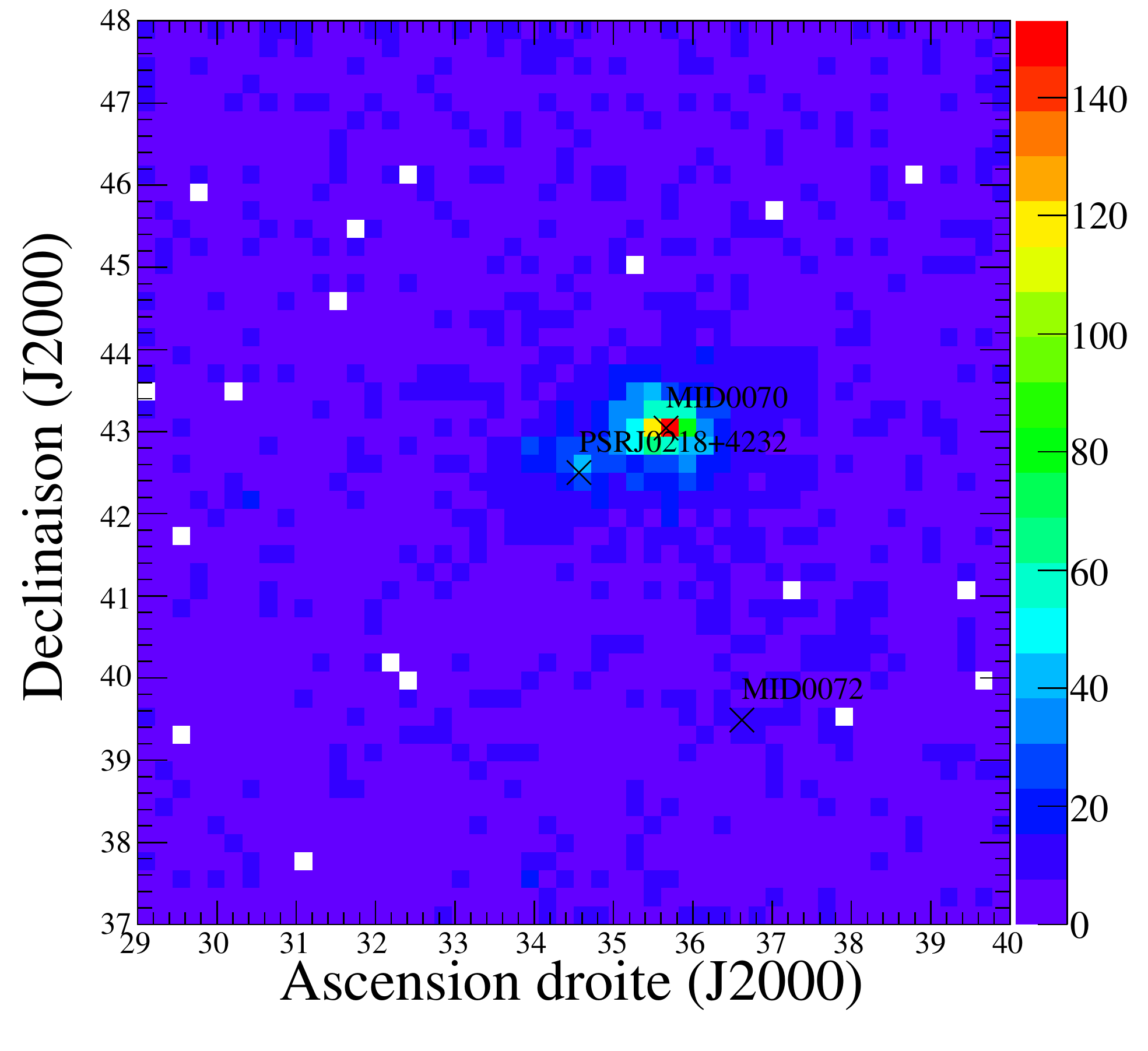}
\includegraphics[scale=0.4]{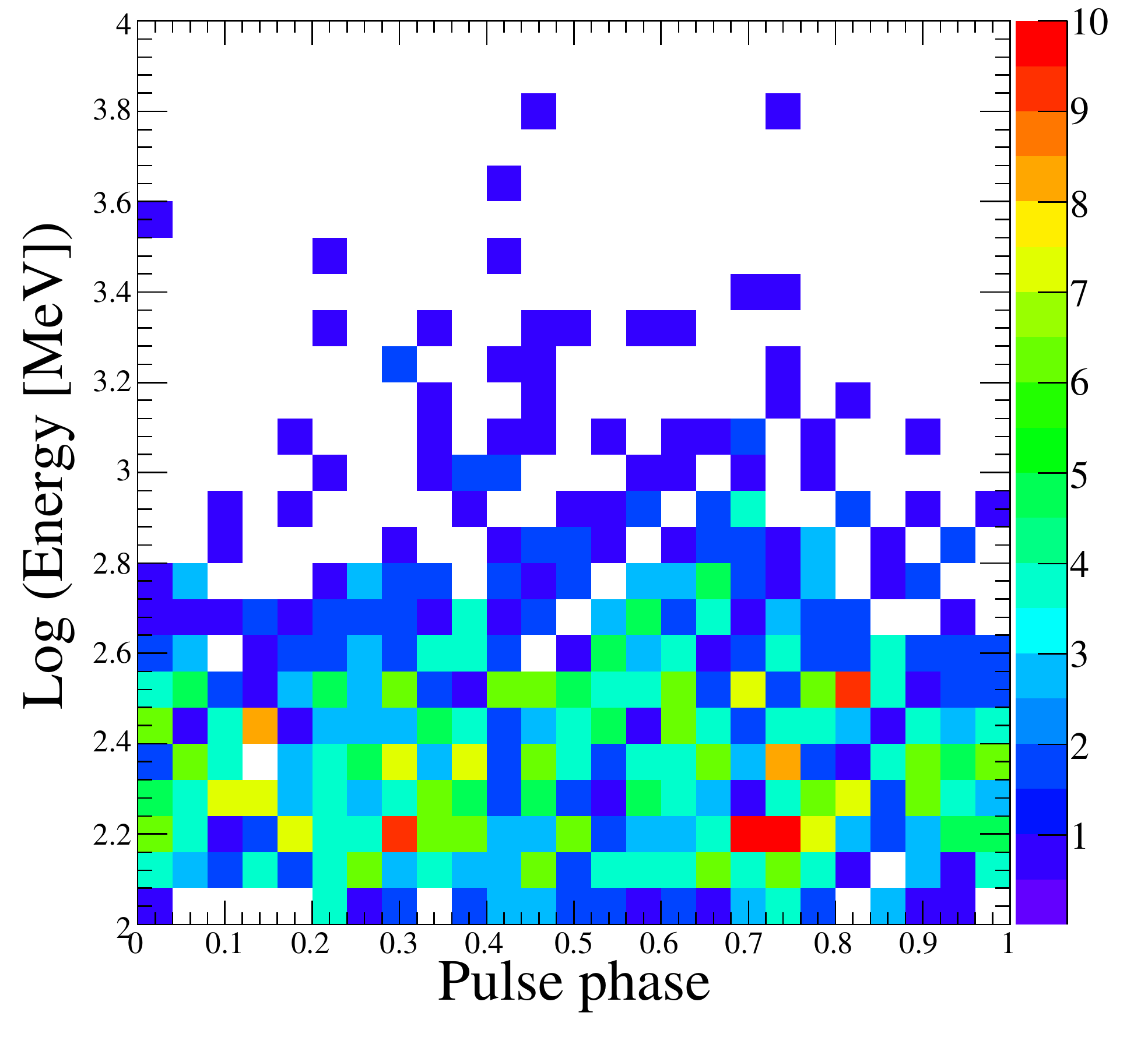}
\caption{Carte du ciel autour de J0218+4232 et énergie des photons en fonction de la phase rotationnelle}
\label{nrj_phaseC2}
\end{center}
\end{figure}

\end{twocolumn}
\onecolumn

\begin{twocolumn}

\begin{figure}[bt]
\begin{center}
\includegraphics[scale=0.67]{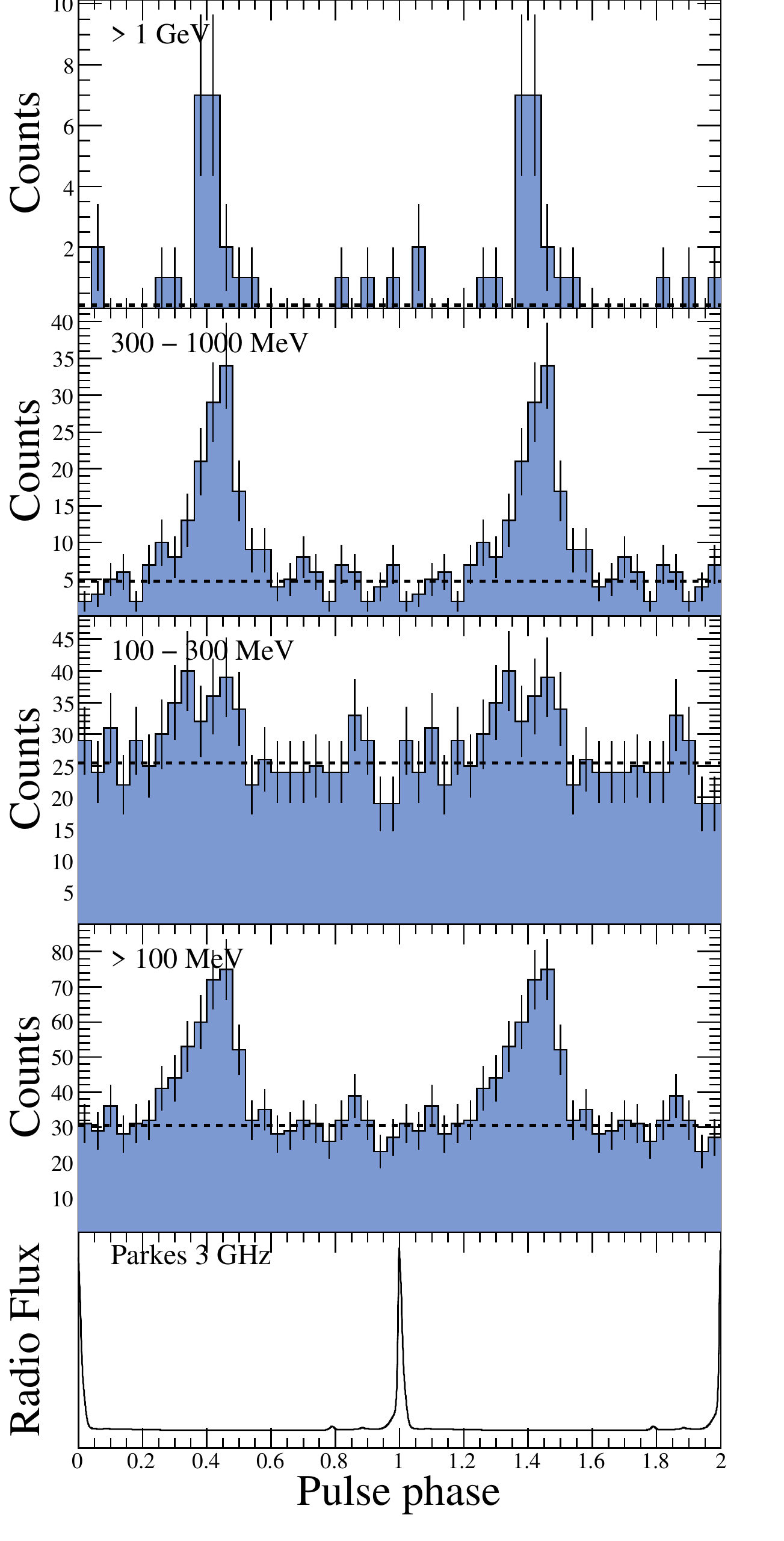}
\caption{Courbes de lumière pour J0437$-$4715 (rayon dépendant de l'énergie)}
\label{phasosC3}
\end{center}
\end{figure}

\begin{figure}[bt]
\begin{center}
\includegraphics[scale=0.4]{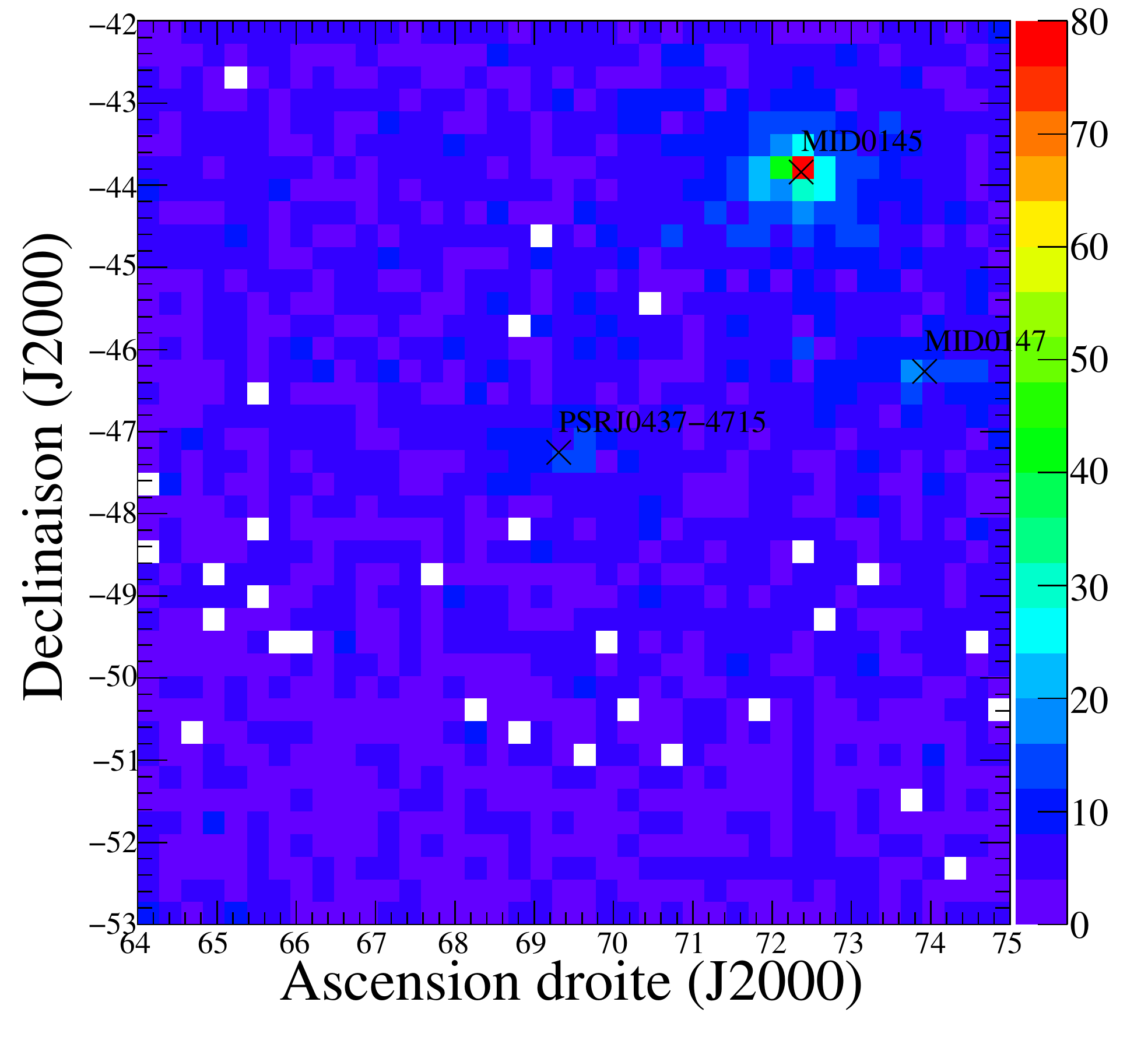}
\includegraphics[scale=0.4]{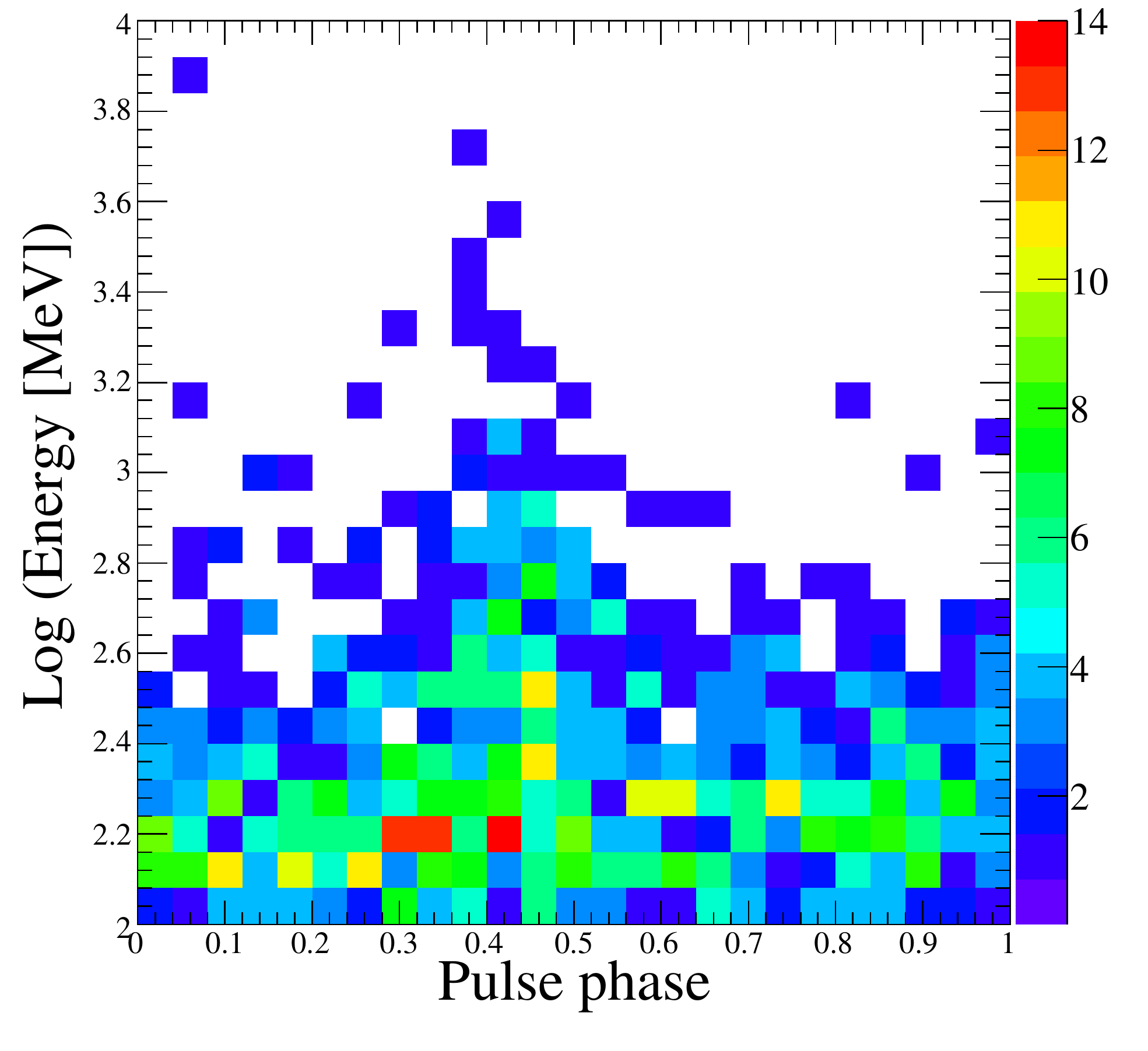}
\caption{Carte du ciel autour de J0437$-$4715 et énergie des photons en fonction de la phase rotationnelle}
\label{nrj_phaseC3}
\end{center}
\end{figure}

\end{twocolumn}
\onecolumn

\begin{twocolumn}

\begin{figure}[bt]
\begin{center}
\includegraphics[scale=0.67]{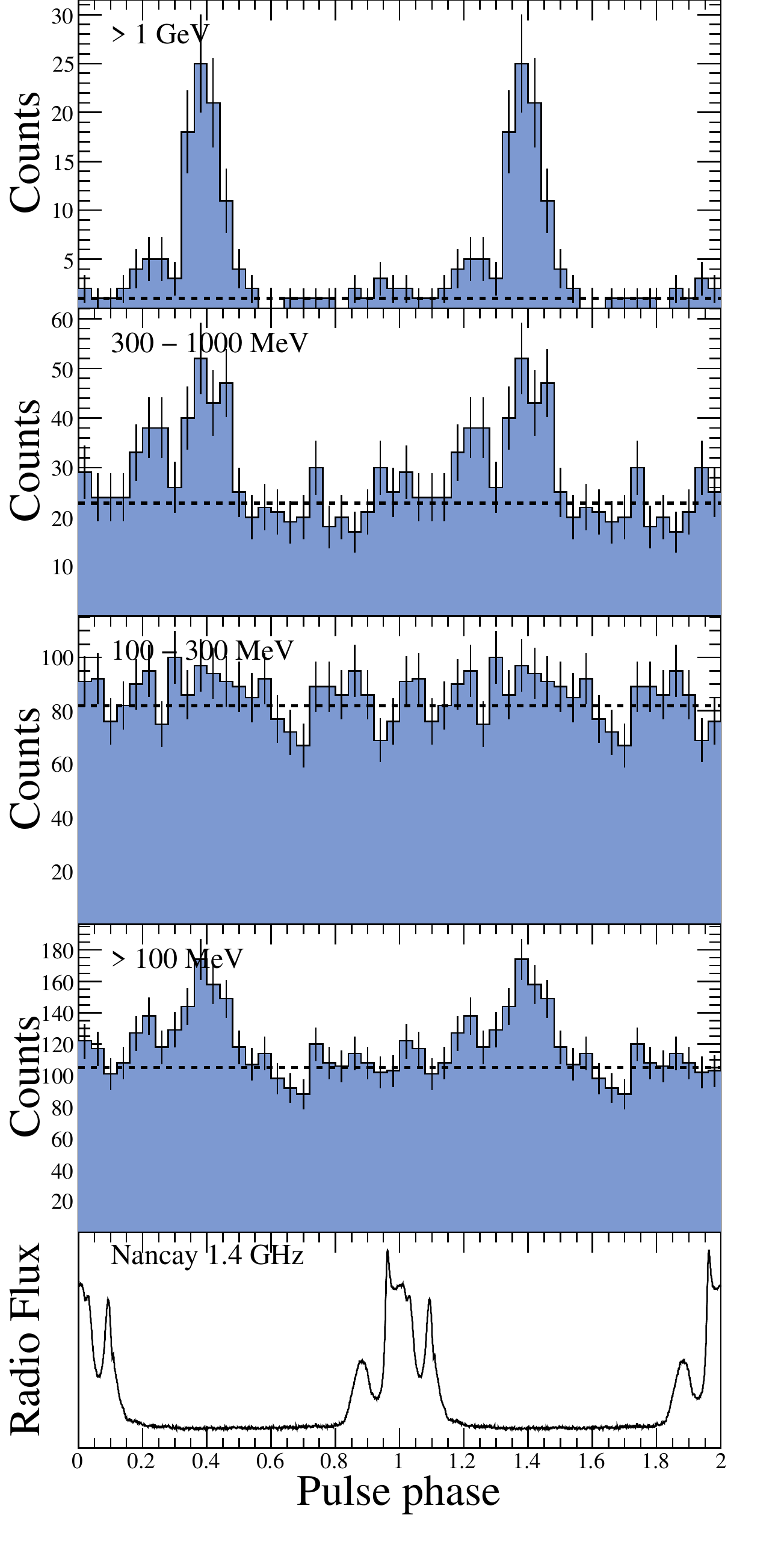}
\caption{Courbes de lumière pour J0613$-$0200 (rayon dépendant de l'énergie)}
\label{phasosC4}
\end{center}
\end{figure}

\begin{figure}[bt]
\begin{center}
\includegraphics[scale=0.4]{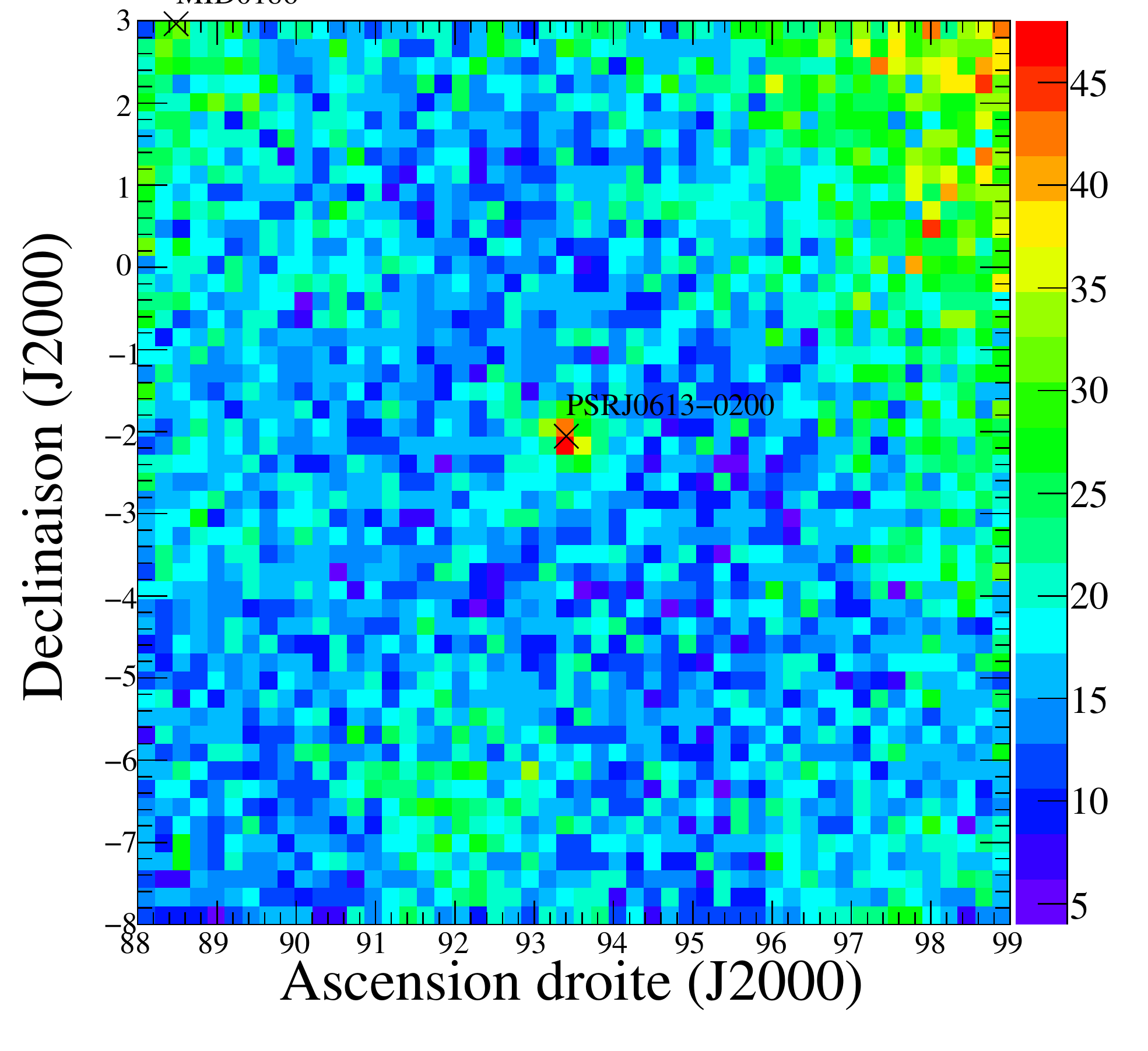}
\includegraphics[scale=0.4]{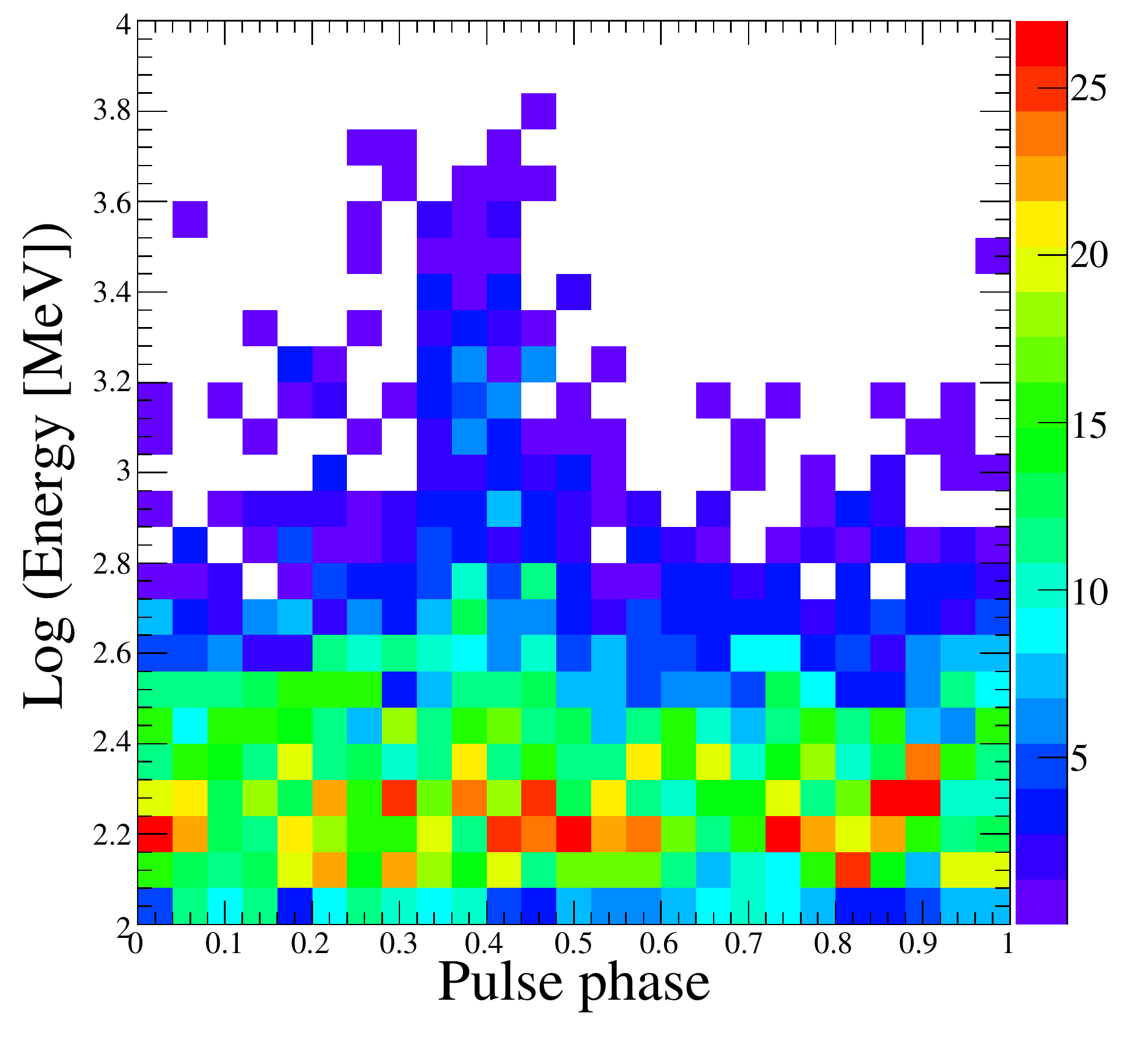}
\caption{Carte du ciel autour de J0613$-$0200 et énergie des photons en fonction de la phase rotationnelle}
\label{nrj_phaseC4}
\end{center}
\end{figure}

\end{twocolumn}
\onecolumn

\begin{twocolumn}

\begin{figure}[bt]
\begin{center}
\includegraphics[scale=0.67]{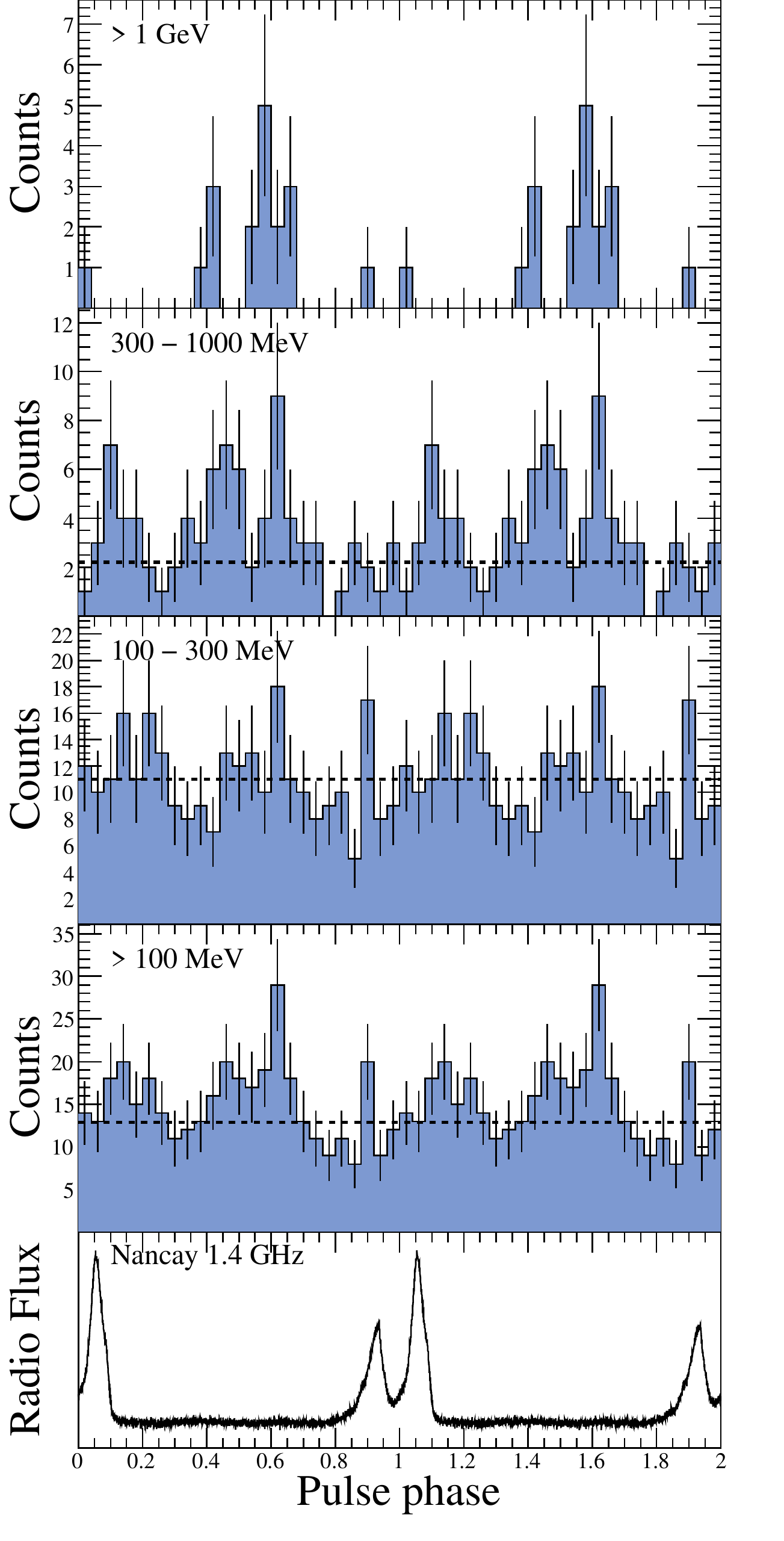}
\caption{Courbes de lumière pour J0751+0751 (rayon dépendant de l'énergie)}
\label{phasosC5}
\end{center}
\end{figure}

\begin{figure}[bt]
\begin{center}
\includegraphics[scale=0.4]{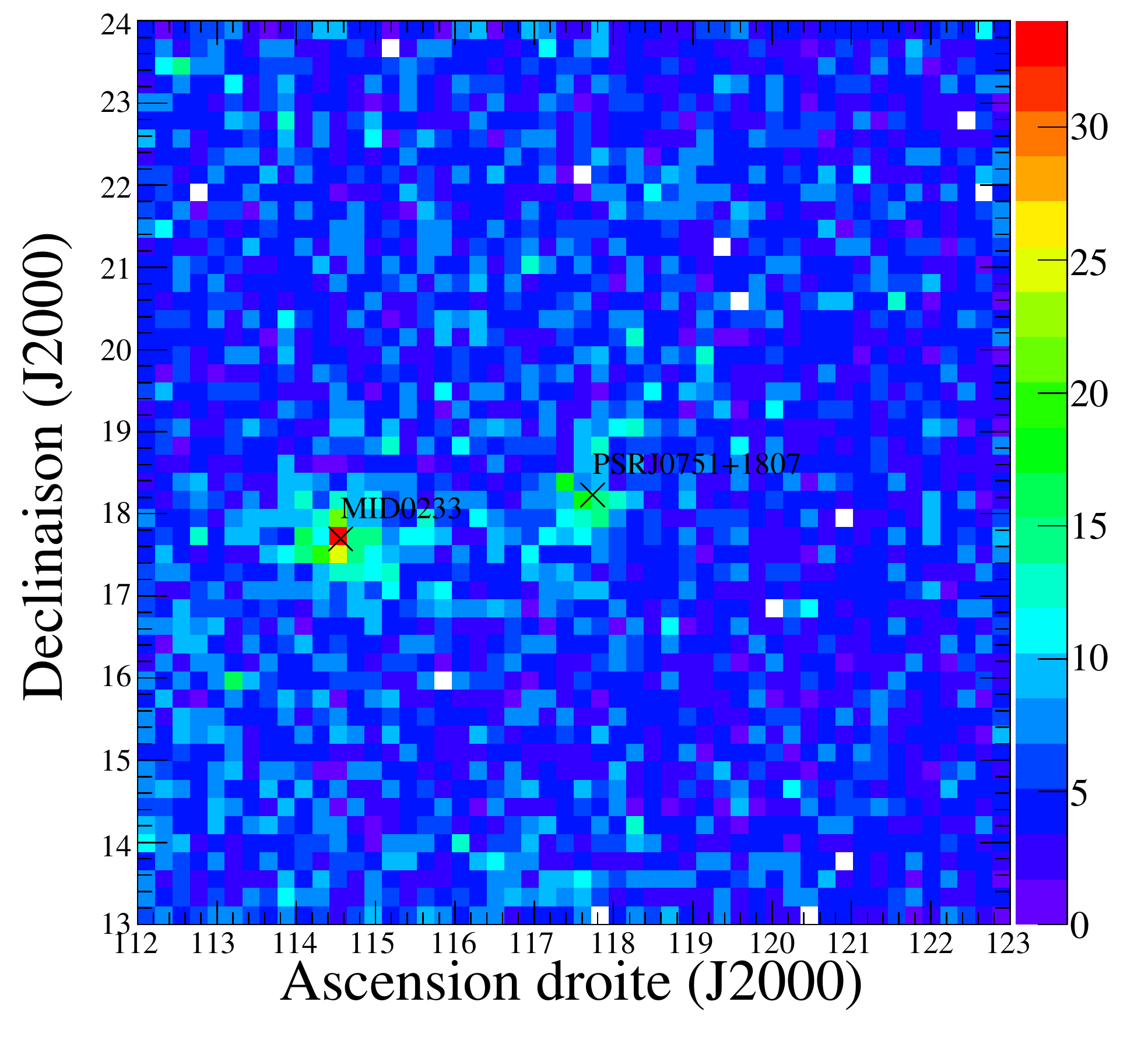}
\includegraphics[scale=0.4]{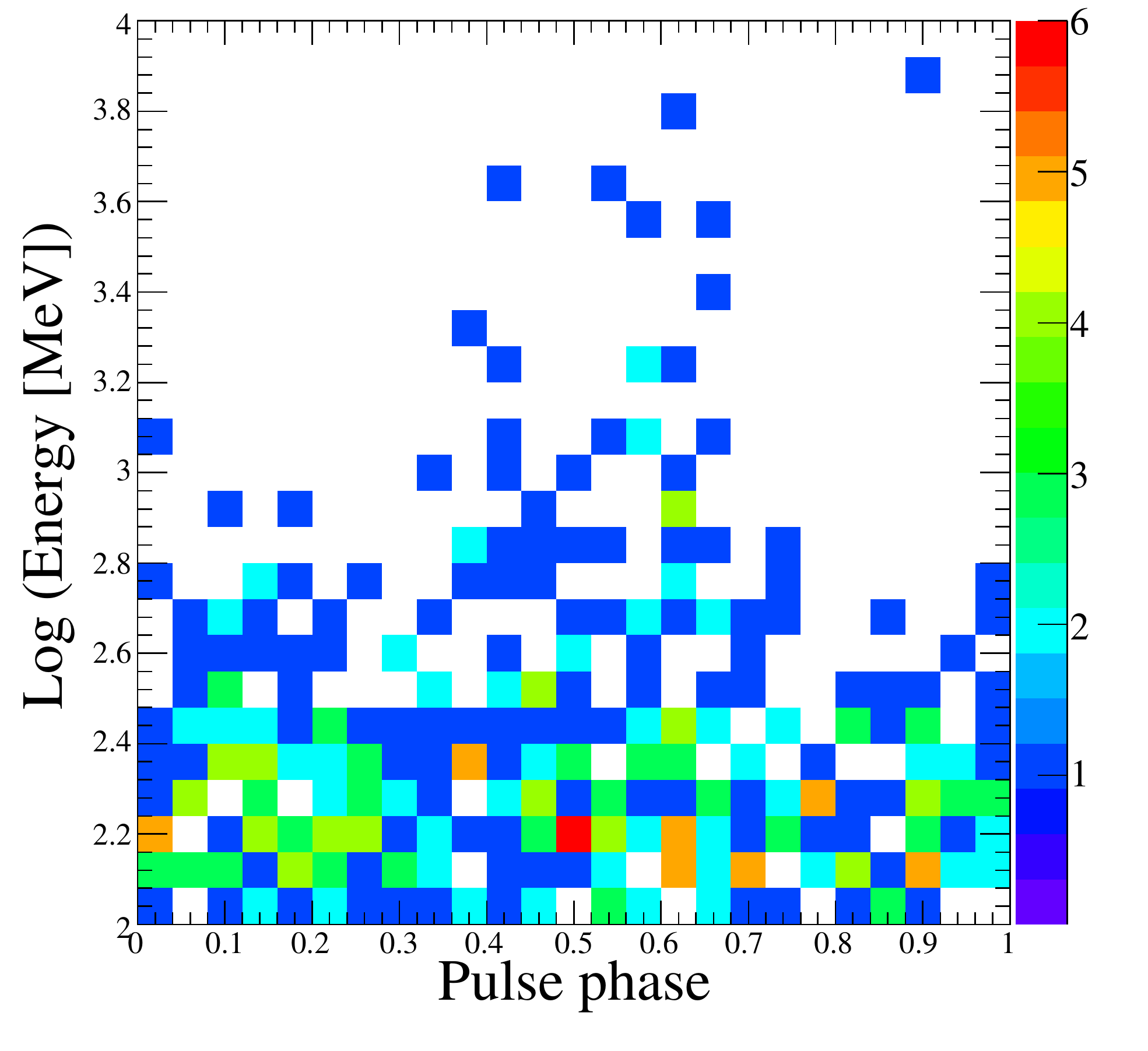}
\caption{Carte du ciel autour de J0751+1807 et énergie des photons en fonction de la phase rotationnelle}
\label{nrj_phaseC5}
\end{center}
\end{figure}

\end{twocolumn}
\onecolumn

\begin{twocolumn}

\begin{figure}[bt]
\begin{center}
\includegraphics[scale=0.67]{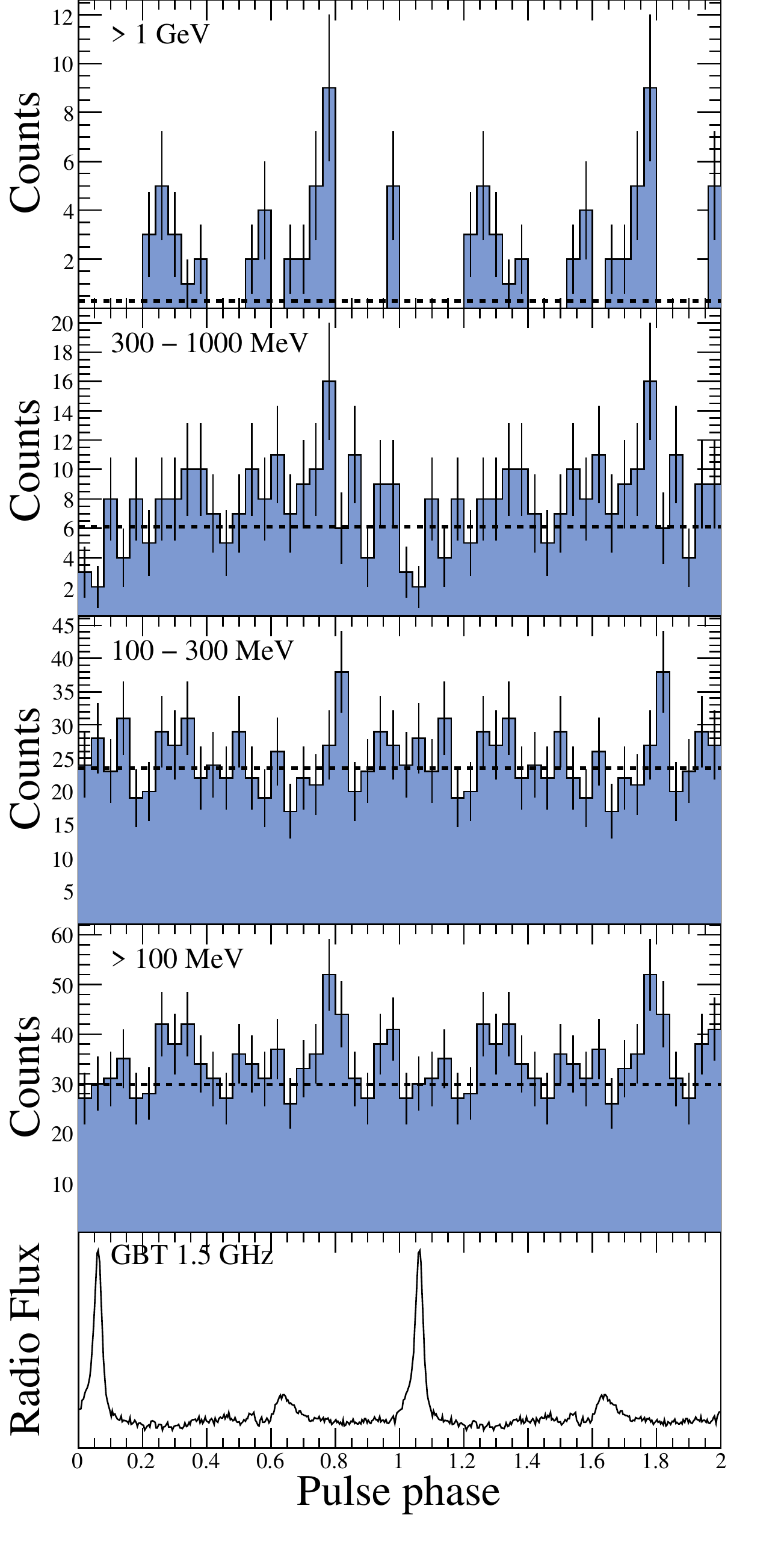}
\caption{Courbes de lumière pour J1614$-$2230 (rayon dépendant de l'énergie)}
\label{phasosC6}
\end{center}
\end{figure}

\begin{figure}[bt]
\begin{center}
\includegraphics[scale=0.4]{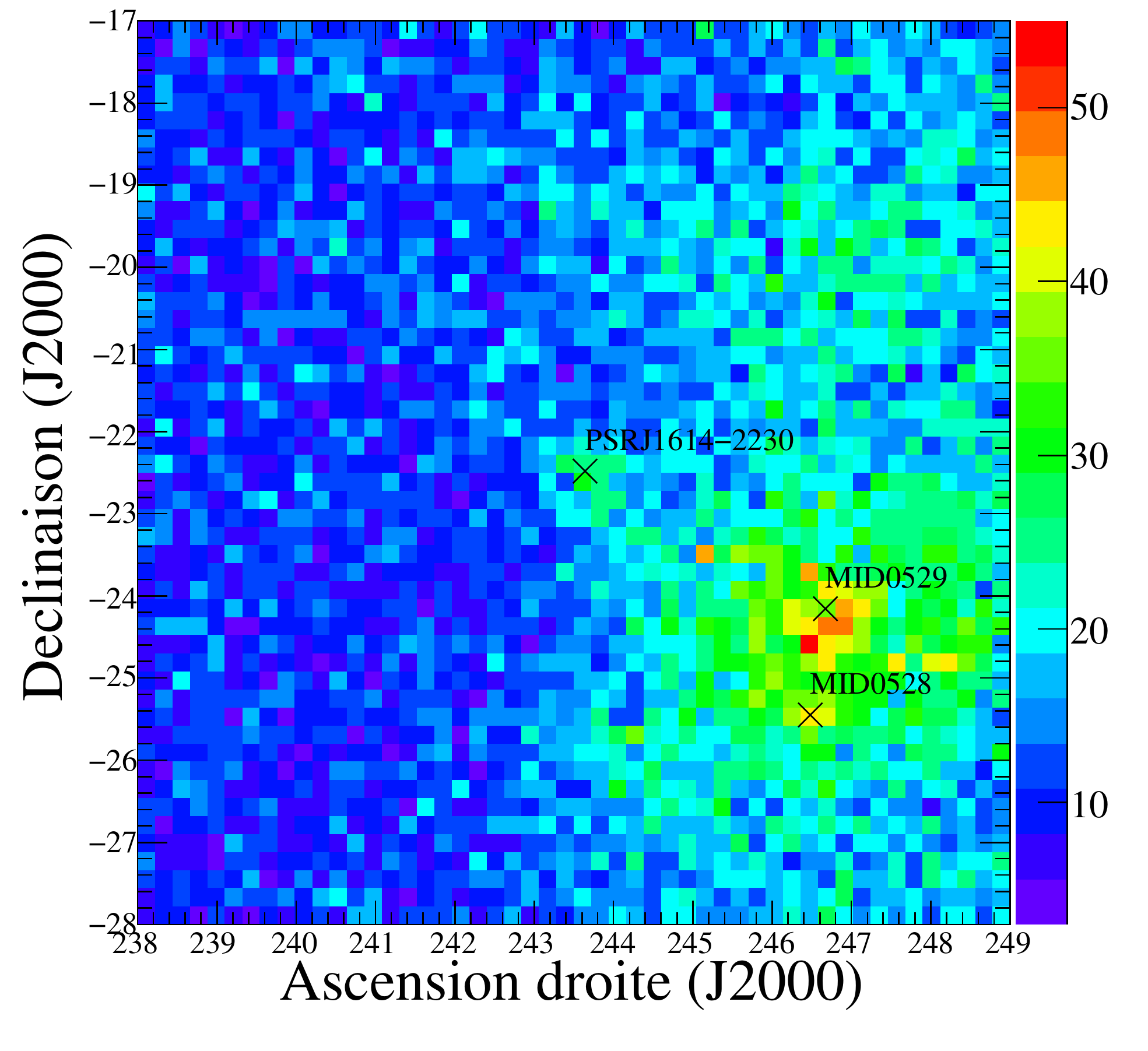}
\includegraphics[scale=0.4]{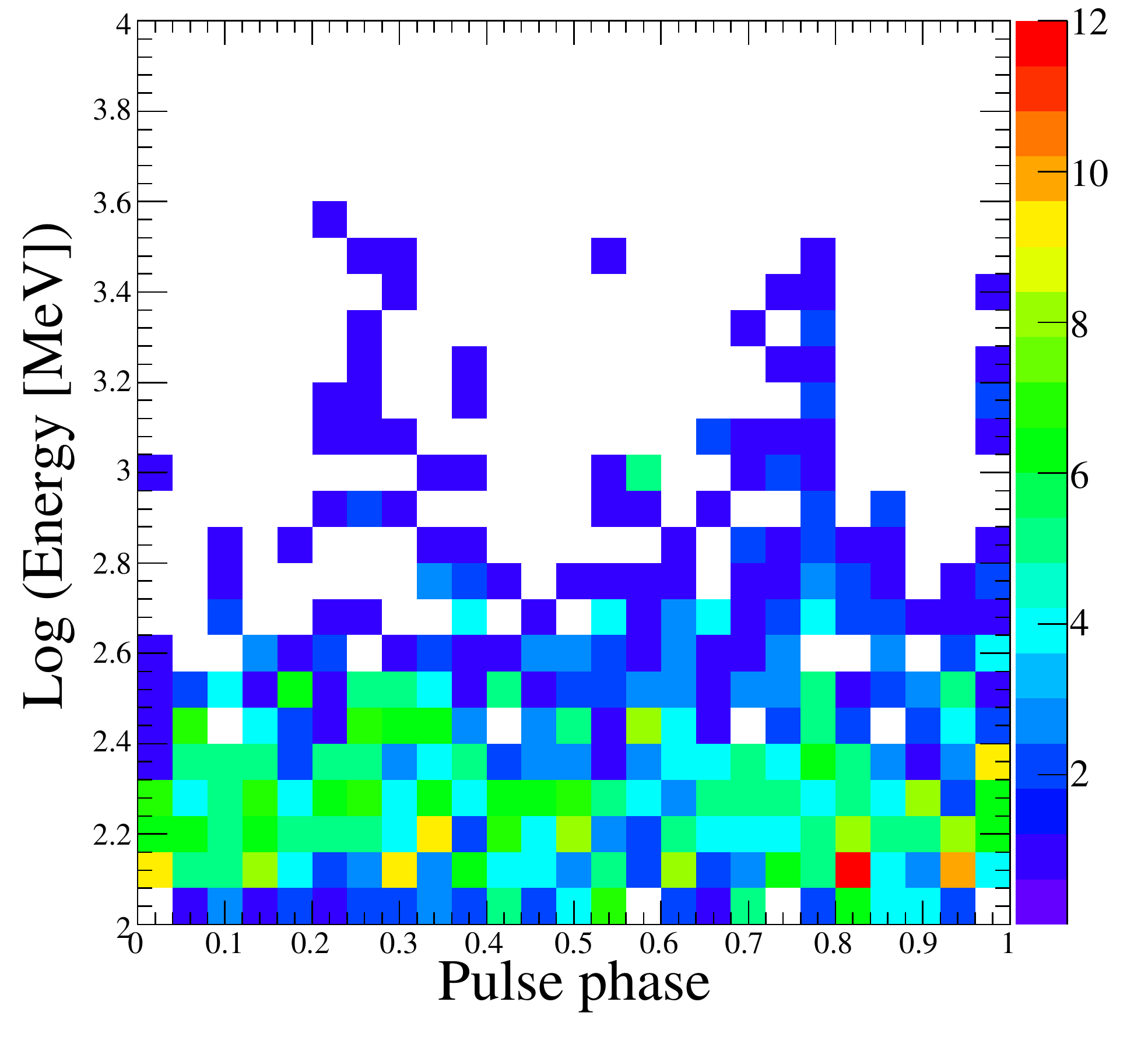}
\caption{Carte du ciel autour de J1614$-$2230 et énergie des photons en fonction de la phase rotationnelle}
\label{nrj_phaseC6}
\end{center}
\end{figure}

\end{twocolumn}
\onecolumn

\begin{twocolumn}

\begin{figure}[bt]
\begin{center}
\includegraphics[scale=0.67]{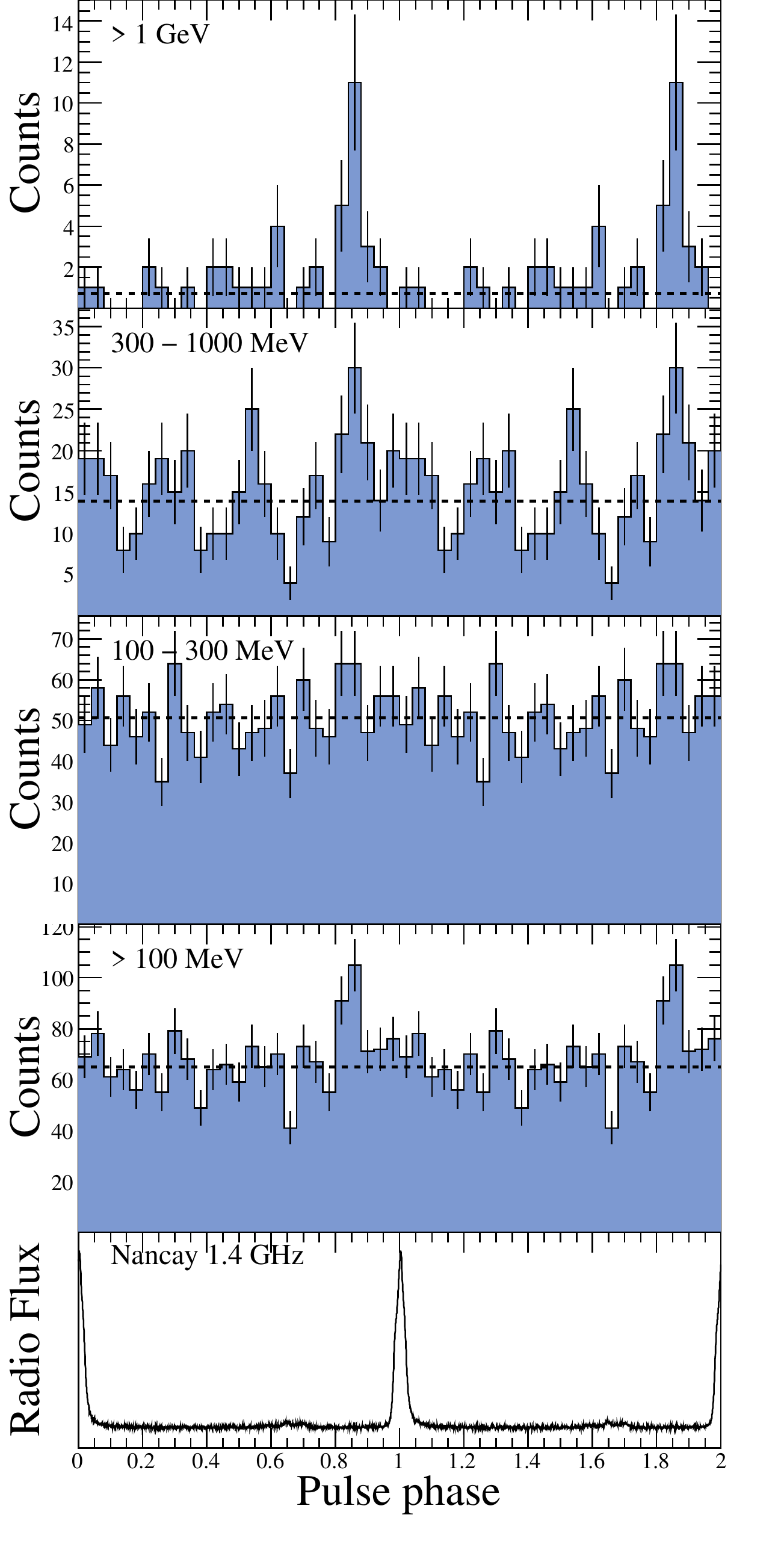}
\caption{Courbes de lumière pour J1744$-$1134 (rayon dépendant de l'énergie)}
\label{phasosC7}
\end{center}
\end{figure}

\begin{figure}[bt]
\begin{center}
\includegraphics[scale=0.4]{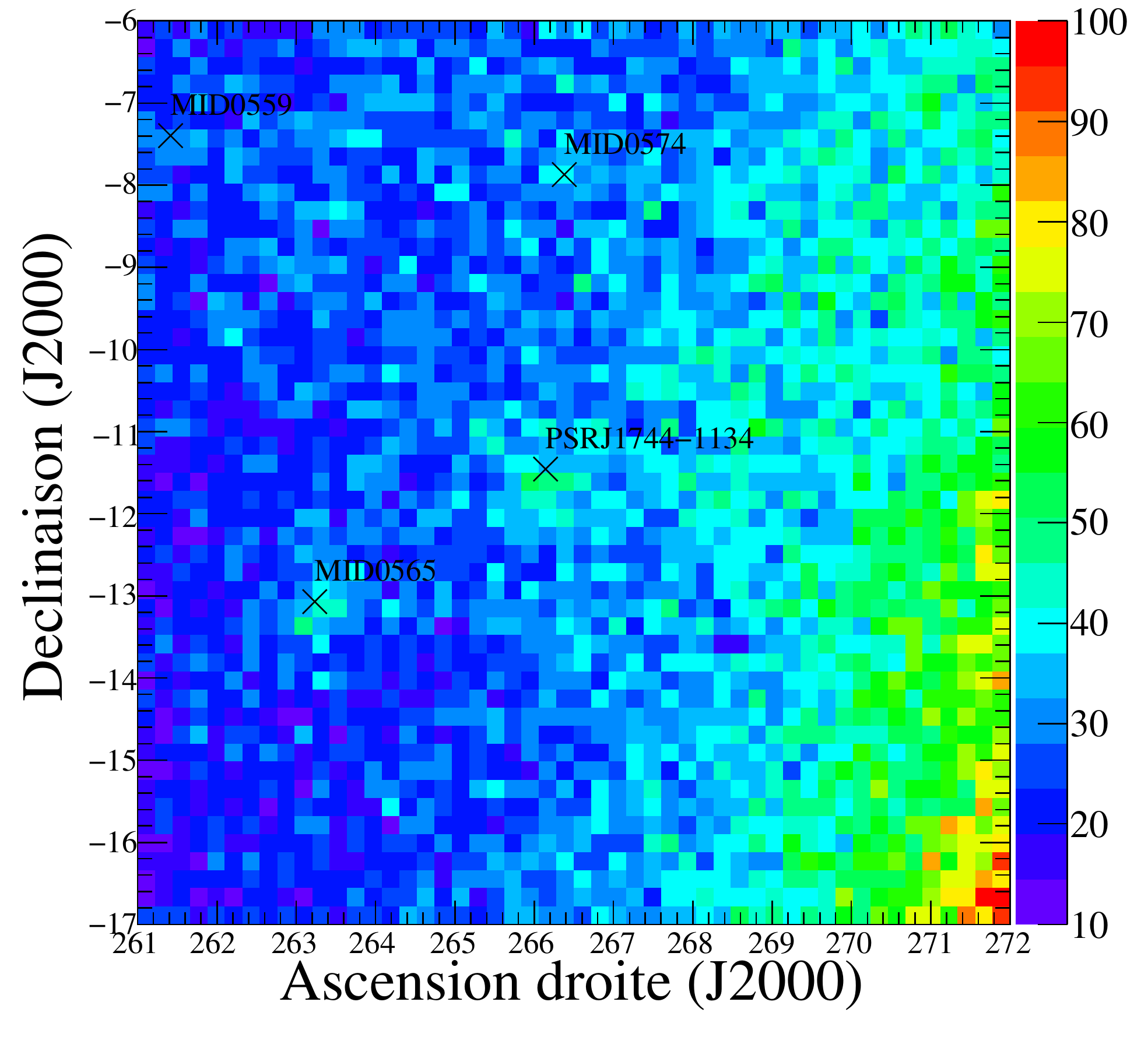}
\includegraphics[scale=0.4]{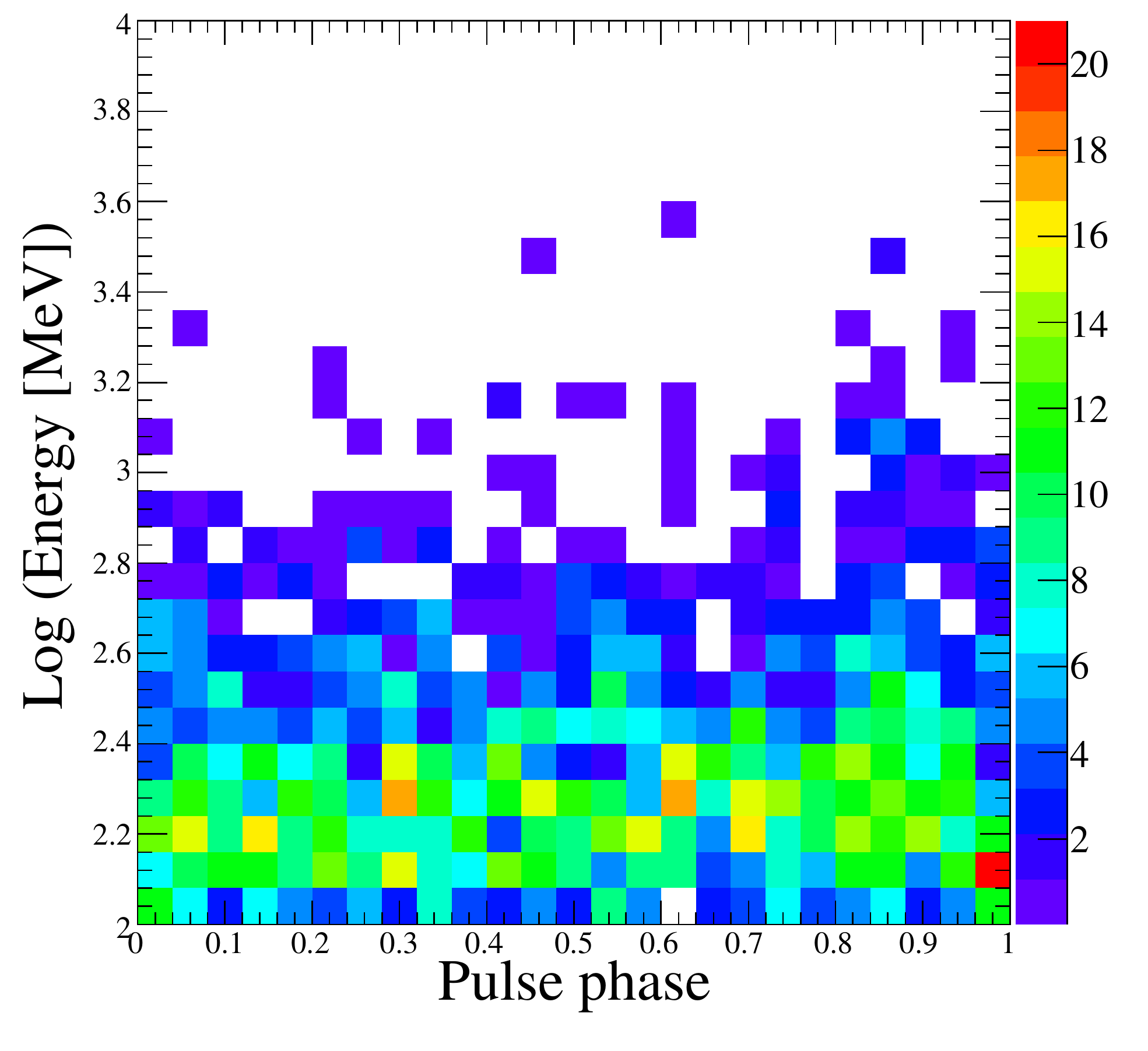}
\caption{Carte du ciel autour de J1744$-$1134 et énergie des photons en fonction de la phase rotationnelle}
\label{nrj_phaseC7}
\end{center}
\end{figure}

\end{twocolumn}
\onecolumn

\begin{twocolumn}

\begin{figure}[bt]
\begin{center}
\includegraphics[scale=0.67]{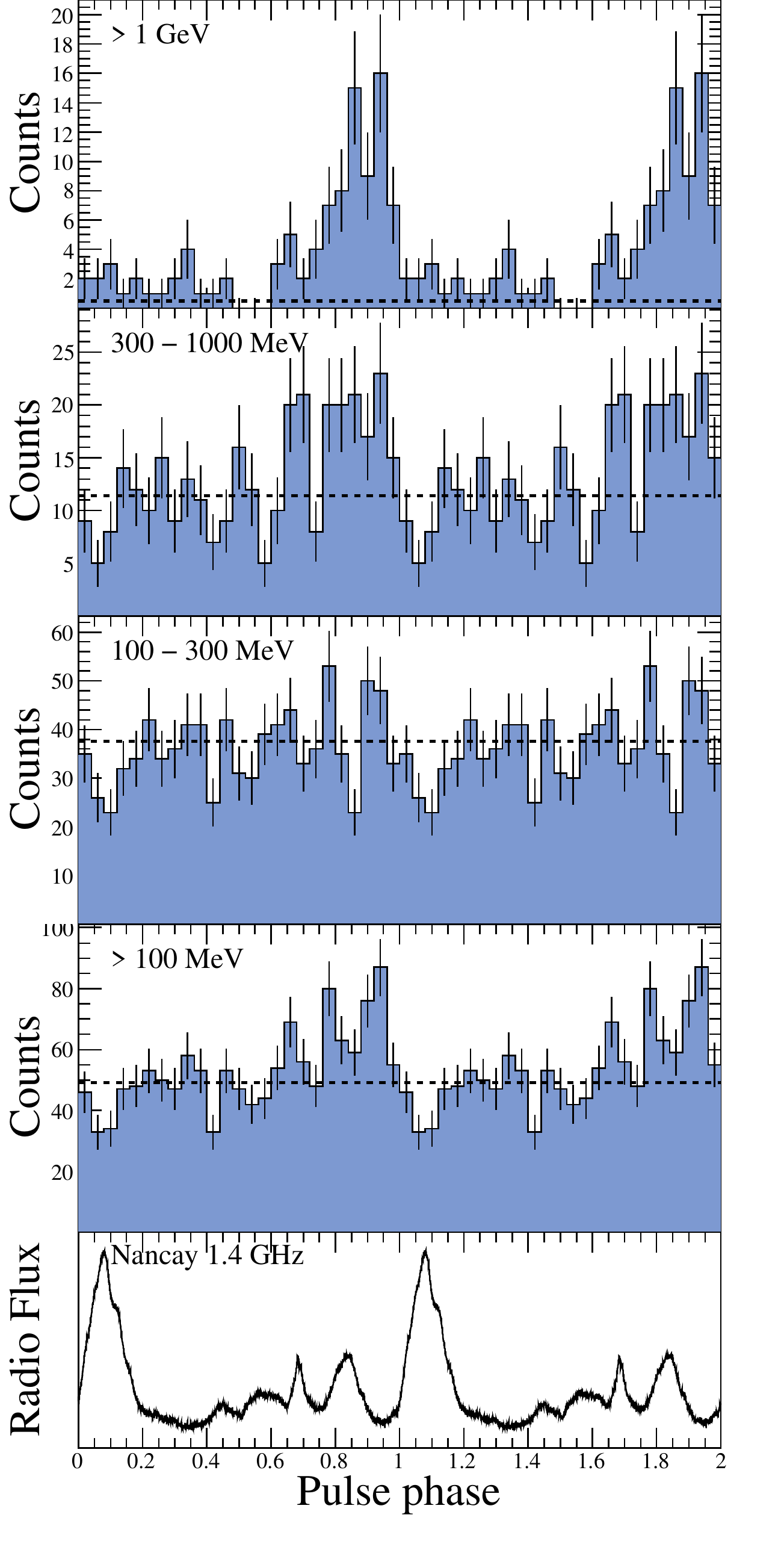}
\caption{Courbes de lumière pour J2124$-$3358 (rayon dépendant de l'énergie)}
\label{phasosC8}
\end{center}
\end{figure}

\begin{figure}[bt]
\begin{center}
\includegraphics[scale=0.4]{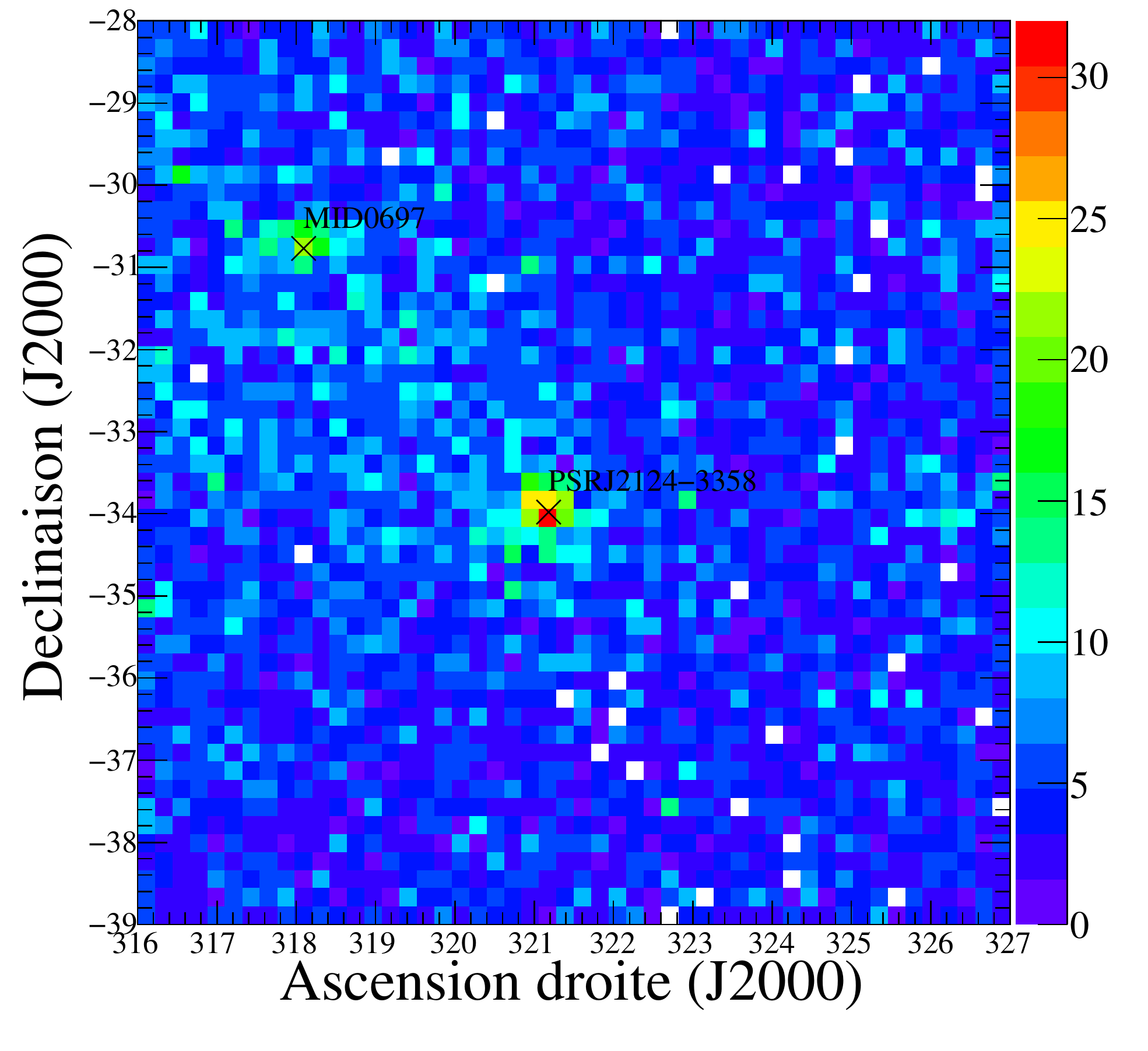}
\includegraphics[scale=0.4]{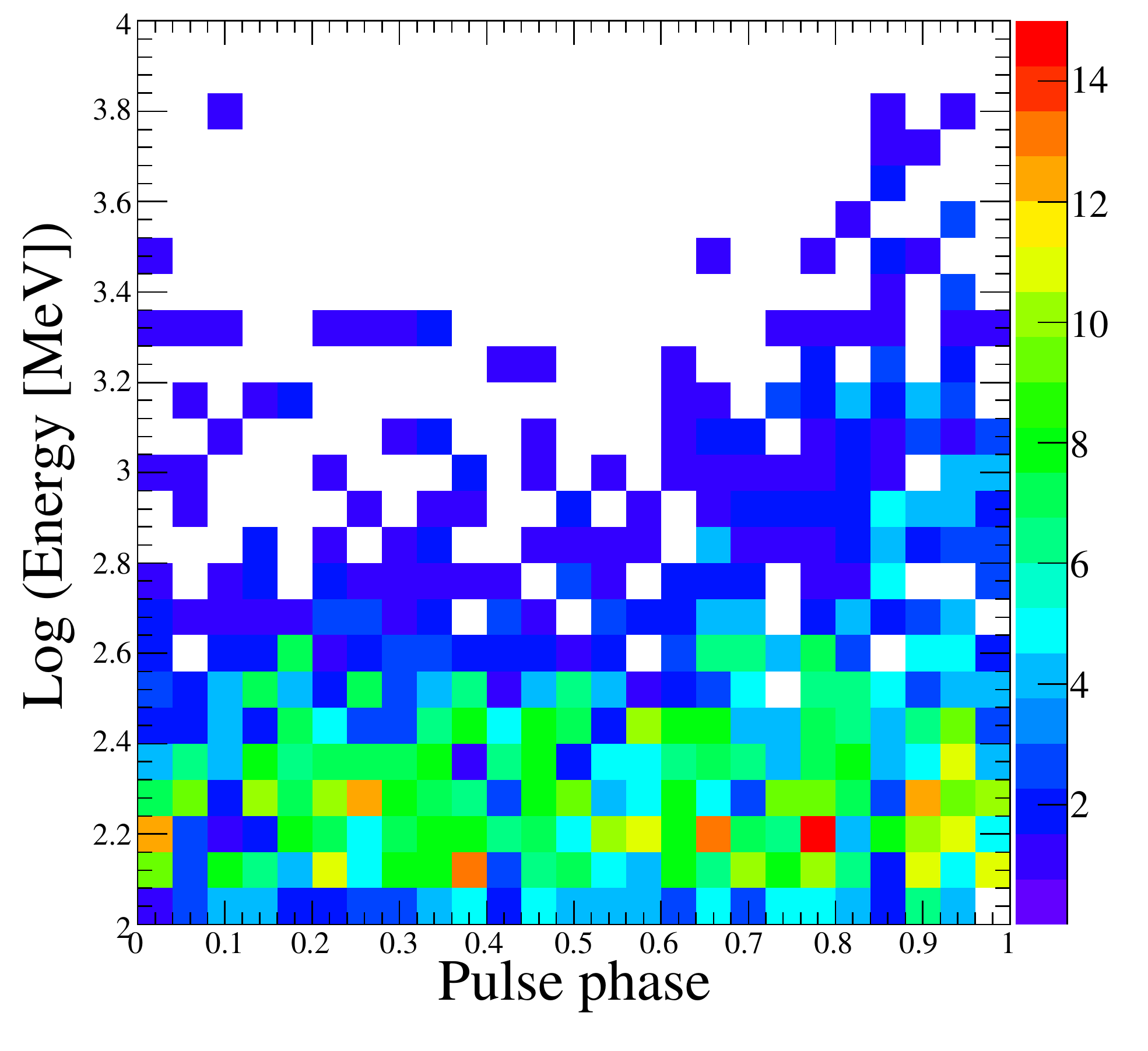}
\caption{Carte du ciel autour de J2124$-$3358 et énergie des photons en fonction de la phase rotationnelle}
\label{nrj_phaseC8}
\end{center}
\end{figure}

\end{twocolumn}
\onecolumn

%


%% file: frontmatter/acronymes.tex
\chapter*{Acronymes et abréviations}

La liste ci-dessous regroupe les acronymes et abréviations fréquemment utilisées dans ce manuscrit :\\

\vspace*{0.5cm}

\begin{center}
\begin{tabular}{l | l}
ACD & \emph{Anti Coincidence Detector}, bouclier anti-co\"incidences\\
AGILE & \emph{Astro-rivelatore Gamma a Immagini Leggero}\\
ATNF & \emph{Australia Telescope National Facility} \\
CAL & \emph{Calorimeter}, calorimètre\\
CGRO & \emph{Compton Gamma Ray Observatory} \\
DM & \emph{Dispersion Measure}, mesure de dispersion\\
EGRET & \emph{Energetic Gamma Ray Experiment Telescope}\\
FGST & \emph{Fermi Gamma-ray Space Telescope}, anciennement GLAST\\
FSSC & \emph{Fermi Science Support Center}\\
GBM & \emph{Gamma-ray Burst Monitor}\\
GLAST & \emph{Gamma-ray Large Area Space Telescope}\\
GPS & \emph{Global Positioning System}\\
GRP & \emph{Giant Radio Pulse}, pulsations radio géantes\\
IRF & \emph{Instrument Response Function}, fonction de réponse instrumentale\\
LAT & \emph{Large Area Telescope}\\
LC & \emph{Light Cylinder}, cylindre de lumière\\
MET & \emph{Mission Elapsed Time}\\
MJD & \emph{Modified Julian Date}, date julienne modifiée\\
MSP & \emph{Millisecond Pulsar}, pulsar milliseconde \\
OG & \emph{Outer Gap}, cavité externe\\
PC & \emph{Polar Cap}, calotte polaire\\
PSF & \emph{Point Spread Function}, fonction d'étalement du point\\
PWN & \emph{Pulsar Wind Nebula}, nébuleuse à vent de pulsar\\
SG & \emph{Slot Gap}, cavité à fentes\\
SNR & \emph{Supernova Remnant}, vestige de supernova\\
SSB & \emph{Solar System Barycenter}, barycentre du système solaire\\
TAI & Temps Atomique International\\
TDB & Temps Dynamique Barycentrique\\ 
TKR & \emph{Tracker}, trajectographe\\
TOA & \emph{Time Of Arrival}, temps d'arrivée\\
TPC & \emph{Two-Pole Caustic}\\
TT & Temps Terrestre\\
UTC & \emph{Coordinated Universal Time}, temps universel coordonné \\
\end{tabular}
\end{center}

%% file: frontmatter/resume.tex
\addstarredchapter{Résumé - Abstract} 

\thispagestyle{reallyempty}

\paragraph{Résumé\newline}

Le satellite Fermi a été lancé le 11 juin 2008, avec à son bord le \emph{Large Area Telescope} (LAT). Le LAT est un télescope sensible au rayonnement $\gamma$ de 20 MeV à plus de 300 GeV. Au début de l'activité de Fermi, neuf pulsars jeunes et énergétiques étaient connus dans le domaine $\gamma$. Le nombre de détections de pulsars par le LAT prédit avant lancement était de plusieurs dizaines au moins. Le LAT permettait également l'étude des pulsars milliseconde (MSPs), jamais détectés avec certitude à très haute énergie jusqu'alors.

Cette thèse aborde dans un premier temps la campagne de chronométrie des pulsars émetteurs radio et/ou X, candidats à la détection par le LAT, en collaboration avec les grands radiotélescopes et télescopes X. Cette campagne a permis la recherche de signaux $\gamma$ pulsés avec une grande sensibilité. En outre, la plupart des MSPs galactiques ont été suivis dans le cadre de cette campagne, sans biais de sélection \emph{a priori} sur cette population d'étoiles. 

Pour la première fois, des pulsations ont été détectées pour huit MSPs galactiques au-dessus de 100 MeV. Quelques bons candidats à une détection prochaine apparaissent. Une recherche similaire a été conduite pour des MSPs d'amas globulaires, sans succès à présent. L'analyse des courbes de lumière et des propriétés spectrales des huit MSPs détectés révèle que leur rayonnement $\gamma$ est relativement similaire à celui des pulsars ordinaires, et est vraisemblablement produit dans la magnétosphère externe. Cette découverte suggère que certaines sources non identifiées sont des MSPs, pour l'instant inconnus.

\paragraph{Mots-clés:}{\it Astronomie gamma, Fermi, Large Area Telescope (LAT), chronométrie des pulsars, pulsars milliseconde.}

\paragraph{Abstract\newline}

The Fermi observatory was launched on June 11, 2008. It hosts the \emph{Large Area Telescope} (LAT), sensitive to $\gamma$-ray photons from 20 MeV to over 300 GeV. When the LAT began its activity, nine young and energetic pulsars were known in $\gamma$ rays. At least several tens of pulsar detections by the LAT were predicted before launch. The LAT also allowed the study of millisecond pulsars (MSPs), never firmly detected in $\gamma$ rays before Fermi. 

This thesis first presents the pulsar timing campaign for the LAT, in collaboration with large radiotelescopes and X-ray telescopes, allowing for high sensitivity pulsed searches. Furthermore, it lead to quasi-homogeneous coverage of the galactic MSPs, so that the search for pulsations in LAT data for this population of stars was not affected by an \emph{a priori} bias. 

We present a search for pulsations from these objects in LAT data. For the first time, eight galactic MSPs have been detected as sources of pulsed $\gamma$-ray emission over 100 MeV. In addition, a couple of good candidates for future detection are seen. A similar search for globular cluster MSPs has not succeeded so far. Comparison of the phase-aligned $\gamma$-ray and radio light curves, as well as the spectral shapes, leads to the conclusion that their $\gamma$-ray emission is similar to that of normal pulsars, and is probably produced in the outer-magnetosphere. This discovery suggests that many unresolved $\gamma$-ray sources are unknown MSPs. 

\paragraph{Keywords:}{\it Gamma-ray astronomy, Fermi, Large Area Telescope (LAT), pulsar timing, millisecond pulsars.}